\documentclass[iop]{emulateapj}
\usepackage{multirow}
\usepackage{booktabs}
\usepackage[figuresright]{rotating}
\usepackage{subfigure}
\usepackage{epic,eepic}
\usepackage{graphicx}
\usepackage{graphics}
\usepackage{hyperref}
\usepackage{threeparttable}
\usepackage{amsmath}

\shorttitle{STELLAR X-RAY ACTIVITY}
\shortauthors{Wang et al.}

\begin{document}
\title{Stellar X-ray activity across the Hertzsprung-Russell diagram. I. Catalogs}

\author{Song Wang\altaffilmark{1}, Yu Bai\altaffilmark{1}, Lin He\altaffilmark{2}, Jifeng Liu\altaffilmark{1,3,4}}

\altaffiltext{1}{Key Laboratory of Optical Astronomy, National Astronomical Observatories,
Chinese Academy of Sciences, Beijing 100101, China; songw@bao.ac.cn}
\altaffiltext{2}{Hebei University of Science and Technology, Shijiazhuang 050018, China}
\altaffiltext{3}{College of Astronomy and Space Sciences,
University of Chinese Academy of Sciences, Beijing 100049, China}
\altaffiltext{4}{WHU-NAOC Joint Center for Astronomy, Wuhan University, Wuhan, Hubei 430072, China}

\begin{abstract}

Stellar magnetic activity provides substantial information on the magnetic dynamo and the coronal heating process.
We present a catalog of X-ray activity for about 6000 stars, based on the $Chandra$ and $Gaia$ DR2 data.
We also classified more than 3000 stars as young stellar objects, dwarf stars, or giant stars.
By using the stars with valid stellar parameters and classifications, we studied the distribution of X-ray luminosity ($L_X$) and the ratio of X-ray-to-bolometric luminosities ($R_X$), the positive relation between $L_X$, $R_X$, and hardness ratio, and the long-term X-ray variation.
This catalog can be used to investigate some important scientific topics, including the activity-rotation relation, the comparison between different activity indicators, and the activities of interesting objects (e.g., A-type stars and giants).
As an example, we use the catalog to study the activity-rotation relation, and find that the young stellar objects, dwarfs, and giants fall on a single sequence in the relation $R_X$ versus Rossby number, while the giants do not follow the relation $R_X$ versus $P_{\rm rot}^{-2}R^{-4}$ valid for dwarfs.

\end{abstract}


\section{INTRODUCTION}
\label{intro.sec}

The detailed physical process responsible for coronal heating remains one major open problem in astrophysics  \citep{Rosner1980,Parker1988,Hudson1991,Klimchuk2006,
vanBallegooijen2011}.
The most popular heating mechanisms include
dissipation of magnetic stresses as in the ``nanoflare" model with the twisting of magnetic field lines and subsequent magnetic reconnection \citep[e.g.,][]{Parker1988,Priest2002}, and
dissipation of magnetohydrodynamic (Alfv{\'e}n) waves \citep[e.g.,][]{vanBallegooijen2011,vanBallegooijen2014}.
Studies of stellar magnetic activity can help advance our understanding of the characteristics of magnetic field, and of the process that heats the outer atmosphere \citep{Testa2015}.

One key question about the magnetic activity is the magnetic dynamo in different type stars.
In the solar-type dynamo mechanism ($\alpha$-$\Omega$ dynamo or tachocline dynamo), the magnetic field is generated in the deep convection zones because of the interior radial differential rotation and amplified by the interaction between magnetic flux tubes and the convection;
the magnetic field then rises to the stellar surface and produce chromospheric heating through interaction with the uppermost convection zones \citep[e.g.,][]{Parker1975, Reid2000}.
Fully convective stars (e.g., T Tauri stars, very late M dwarfs) do not have a tachocline and their dynamo mechanism is expected to be very different.
However, there is no precipitate decline or change in coronal properties between fully convective stars and stars with tachoclines \citep{Testa2015}. Some studies even found that fully convective stars also operate a solar-type dynamo \citep[e.g.,][]{Wright2016}.
It therefore suggests that the tachocline is not vital in the generation of the magnetic field and the dynamo may originate throughout the convection zone \citep{Wright2016}.

The magnetic dynamo can be observationally tested from stellar activities in different parts of stellar atmospheres,
such as star spots, flares, and X-ray emission.
Stellar activity is ubiquitous in late-type stars, and strongly depends on stellar parameters (e.g., stellar mass, age, metallicity).
Significant progress have been made in recent years, including
the activity-rotation relationship\citep{Pizzolato2003,Wright2011},
the evolution of activity with stellar age \citep{Mamajek2008,Pace2013,vanSaders2013,Reinhold2015},
the Gyrochronology \citep{Barnes2003,Barnes2007,Barnes2010,McQuillan2014,Meibom2015},
flaring activity \citep{Shibayama2013,Hawley2014,Balona2015,Yang2019},
and stellar cycles \citep{Mathur2014, Ferreira2015}.
On the other hand, the growing number of observations and data raise new questions about stellar magnetic activity.
Giants, which are expected to harbor weak surface magnetic fields due to slow rotation, show clear activities \citep{Auriere2015,He2019b};
A-type stars, with shallow convective envelopes, also show substantial activities \citep{Balona2012, Balona2013, Balona2017}.
However, some studies argued that these stars belong to binary systems and the activities are attributed to their unresolved low-mass companions \citep[e.g.,][]{Pedersen2017,Ozdarcan2018}.

These advances in observations and theory can also help us explore the solar-stellar connection \citep[e.g.,][]{Rosner1985,Peres1997,Peres2000,Brun2015}.
Although the solar and stellar activity show close similarity, many differences have been observed.
Active stars tend to have large polar spots \citep[e.g.,][]{Berdyugina1998,Hussain2007}, quite unlike the sun.
A typical flare on the sun lasts several minutes with energy ranging from 10$^{29}$ to 10$^{32}$ erg, which can be
well supported by the classical model of the solar magnetic reconnection.
However, for stellar flares, their energies are much larger than those of solar flares \citep{Walkowicz2011,Maehara2012,Yang2019},
and their durations are much shorter than expected values deduced from solar flares \citep{Namekata2017}.
There could be some differences between solar and stellar coronae,
such as the flaring activity and coronal plasma densities \citep[e.g.,][]{Testa2004}.
In addition, it is still on debate whether the candidate heating mechanism for stellar coronae (e.g., ``nanoflares")
is effective for the heating of the solar corona.
Most of our understanding on stellar activity and atmospheric structure are from the Sun,
and detailed studies of stellar activity can also help re-examine the solar physics.

A large sample covering different type of stars, with stellar activities estimated from a uniform procedure,
can help us doing detailed investigation of the physical properties of stellar activity,
and provide potential diagnostics of heating mechanisms.
Stellar activity has been explored in the X-ray regime for nearly forty years
\citep{Vaiana1981,Schmitt1985,Pizzolato2003,Mamajek2008}.
Benefited from new X-ray missions with high spatial resolution and low background noise, our knowledge of stellar coronae and activity can be continually refined.
In this paper, we use the $Chandra$ data archive and the {\it Gaia} DR2
to provide a large sample of stars with accurately estimated X-ray activities.
In Section \ref{analysis.sec}, we describe the sample selection and data reduction process.
Section \ref{activity.sec} presents the stellar activity investigations,
including the X-ray activity in different stellar types,
the relation between X-ray activity and hardness ratio,
and the X-ray flux variation.
In Section \ref{discussion.sec}, we show some analyses and application examples of our catalog.

\section{SAMPLE SELECTION AND DATA REDUCTION}
\label{analysis.sec}

\subsection{Sample Selection}
\label{sample.sec}

The unprecedented subarcsecond spatial resolution of {\it Chandra} telescope allows X-ray sources to be unambiguously matched to their optical counterparts.
Our primary database is the $Chandra$ point source catalog \citep{Wang2016},
which includes 217,828 distinct X-ray sources with 363,530 detections.
The $Gaia$ DR2 released about 1.33 billion celestial objects with estimated distances \citep{Bailer-Jones2018}, which can be used to evaluate X-ray luminosity and bolometric luminosity of the stars.
Firstly, we cross-matched the $Chandra$ point source catalog and the $Gaia$ DR2  catalog, using a radius of 3$^{\prime\prime}$.
In some cases there are multiple matches, and we only selected the closest neighbour as the counterpart.
This led to more than 60 thousand unique sources with $Chandra$ detections.

\begin{figure}[!htb]
\center
\includegraphics[width=0.46\textwidth]{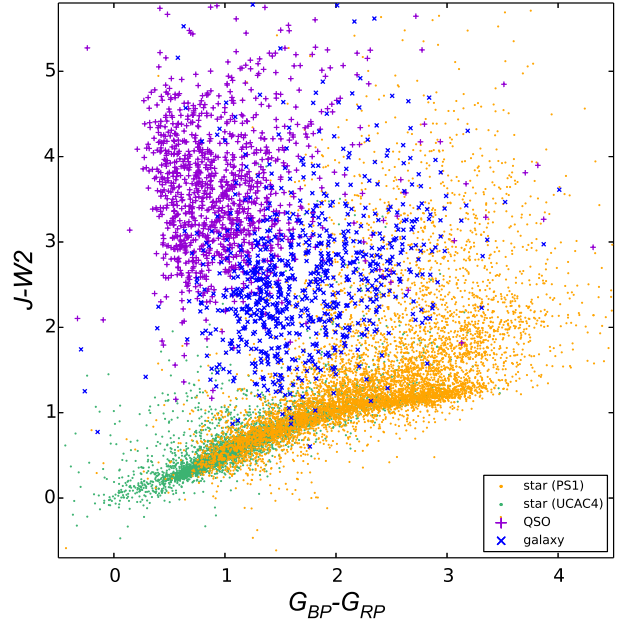}
\caption[]{Color-color diagram of the initial sample. The yellow and green points represent stars with PS1 and UCAC4 magnitudes, respectively. The purple pluses and blue crosses show the distinguished QSOs and galaxies, respectively.}
\label{galaxyqso.fig}
\end{figure}

\begin{figure}[!htb]
\center
\includegraphics[width=0.49\textwidth]{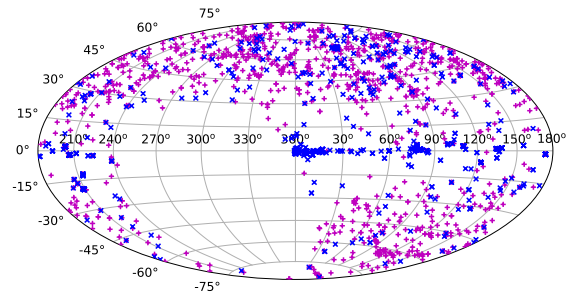}
\caption[]{Sky distribution of the classified QSOs (plus) and galaxies (cross), in Galactic coordinates.}
\label{galactic.fig}
\end{figure}

Secondly, we used a machine learning method \citep{Bai2019} to classify possible contamination from non-stellar objects, such as galaxy and quasi-stellar object (QSO).
The classifier uses nine colors (i.e., $g-r$, $r-i$, $i-J$, $J-H$, $H-K_{\rm s}$, $K_{\rm s}-W1$, $W$1$-W$2, ${\rm w1mag\_1-w1mag\_3}$, ${\rm w2mag\_1-w2mag\_3}$), spanning from optical to far-infrared bands.
In order to obtain these colors, we cross-matched the $\sim$60 thousand sources with
the Pan-STARRS DR1 \citep[hereafter PS1;][]{Chambers2016}, UCAC4 \citep{Zacharias2013}, and WISE All-Sky \citep{Cutri2012} catalogs, by using the TOPCAT\footnote{http://www.star.bris.ac.uk/\%7Embt/topcat} and CasJobs\footnote{http://skyserver.sdss.org/CasJobs}.
The matching radius is 3$^{\prime\prime}$, and only the closest counterpart was selected for the multiple coincidences.
We first collected the $g$, $r$, and $i$ magnitudes from the PS1.
For objects without PS1 observations, we derived the magnitudes from the UCAC4 catalog.
The WISE All-Sky data release includes the 2MASS magnitudes and WISE magnitudes.
The $W$1 and $W$2 magnitudes are the w1(2)mpro magnitudes in the WISE catalog;
the w1(2)mag\_1 and w1(2)mag\_3 are magnitudes measured with circular apertures of radii of
5.5$^{\prime\prime}$ and 11$^{\prime\prime}$,
and the differential aperture magnitude is expected to  have different distributions for point and extended sources \citep{Bilicki2014, Krakowski2016}.
Furthermore, we set the saturation thresholds for these different surveys:
$W$2 $=$ 6.7 mag for WISE\footnote{http://wise2.ipac.caltech.edu/docs/release/allsky/expsup/sec6\_
3d.html};
$K_{\rm S}$ $=$ 8 mag for 2MASS\footnote{https://old.ipac.caltech.edu/2mass/releases/allsky/doc/sec1\_
6b.html};
$g$ $=$ 14 mag,  $r$ $=$ 14 mag, and $i$ $=$ 14 mag for PS1\footnote{https://panstarrs.stsci.edu};
$V$ $=$ 7 mag for UCAC4\footnote{https://www.aavso.org/apass}.
We also set the detection limits using the $g$, $r$, and $i$ magnitudes ($\sim$23.2 mag) for PS1 observations,
and the $V$ magnitude ($\sim$17 mag) for UCAC4 observations, respectively.
This method led to more than 11000 stars, $\sim$800 galaxies, and $\sim$1100 QSOs (Figure \ref{galaxyqso.fig}).
Figure \ref{galactic.fig} shows the sky distribution of the extragalactic sources in Galactic coordinates. About 490 galaxies lie across the Galactic plane ($-10^{\circ} \leq l \leq 10^{\circ}$). Among these sources, about 150/250 sources have a neighbour within $3^{\prime\prime}$/5$^{\prime\prime}$, indicating their photometry may suffer from a contamination by nearby objects. 
Considering that the crowding can result in inaccurate photometry that may move the color index towards the blue band \citep{Gaia2018a,Gaia2018b}, the classification of these galaxies may be unreliable.

Thirdly, we collected the extinction from the Pan-STARRS 3D dust map, which is constructed with the high-quality stellar photometry of 800 million stars from PS1 and 2MASS \citep{Green2015}.
The reddening values $E(B-V)$ were used to estimate the absolute optical magnitude and the unabsorbed X-ray flux.

Fourthly,  in order to have accurate distance estimations, we excluded the objects with relative parallax uncertainties larger than 0.2, although this may introduce biases towards bright nearby sources in our sample \citep{Luri2018}.
Furthermore, we removed those objects with SIMBAD classification as white dwarf, planet nebula, binary, galaxy, and QSO.
Finally, there are more than 5900 stars left in our sample.

\begin{figure}[!htb]
\center
\includegraphics[width=0.49\textwidth]{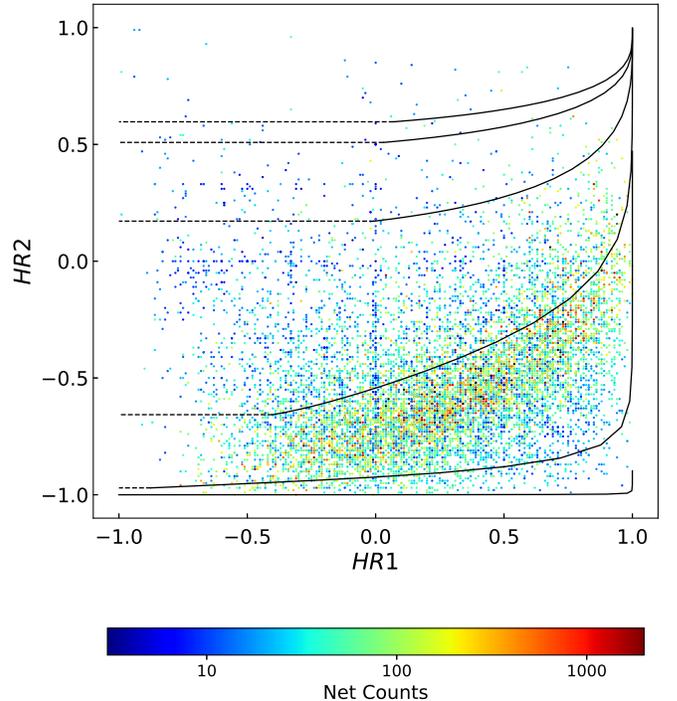}
\caption[]{X-ray $HR$ diagrams for all $Chandra$ detections. Overplotted are examples of thermal spectra (APEC model, red lines) with temperatures of log$T$(K) $=$ 6.0 (bottom), 6.5, 7.0, 7.5, 8.0, 8.5 (top).
These lines are obtained with $N_{\rm H}$ varying from 0 to 10$^{23}$ cm$^{-2}$. The blue horizontal lines are extension of the red lines.}
\label{apec.fig}
\end{figure}

\subsection{$Chandra$ Data Analysis}
\label{chandra.sec}

We extracted the X-ray information (i.e., net counts and count rates in different bands) for each star in each detection from the $Chandra$ point source catalog \citep{Wang2016}.
The hardness ratios ($HR$1 and $HR$2) for each detection were then calculated with the net photon counts for the three bands defined by \citet{Prestwich2003}, i.e., the soft band (S: 0.3--1.0 keV), the medium band (M: 1.0--2.0 keV), and the hard band (H: 2.0--8.0 keV). The $HR$1 and $HR$2 were defined as ($M$-$S$)/($M$+$S$) and  ($H$-$M$)/($H$+$M$), respectively.

We converted the net count rate (0.3--8 keV) into unabsorbed flux using PIMMS\footnote{http://cxc.harvard.edu/toolkit/pimms.jsp} with an APEC model.
APEC is a model of emission spectrum from collisionally ionized diffuse gas in XSPEC and is often used to describe the X-ray emission of stars.
First, we constructed spectral tracks of absorbed APEC spectra with the coronal temperature covering from ${\rm log}T {\rm (K)} =$ 6 to 8.5 with $\Delta~{\rm log}T =$ 0.1, and the hydrogen column density $N_{\rm H}\ {\rm (cm^{-2})}$ covering from 0 to $10^{23}$ with $\Delta~{\rm log}N_{\rm H} =$ 0.1.
The $HR$s of each model in the resulting grid was calculated.
Second,
we derived the coronal temperature, for each detection of each object in our catalog, by selecting the model with closest $HR$s (Figure \ref{apec.fig}).
The models (solid lines) can not predict temperatures of the sources with low $HR$1 values (in the left part of Figure \ref{apec.fig}).
The low $HR$1 values can be mostly due to statistical fluctuation of the photons, especially for low-count detections.
A few sources show a soft excess (i.e., an additional thermal component) in their spectra, which can also reduce $HR$1 values (see Appendix A).
Here we simply extended each APEC model (dashed line) to derive the temperatures of those sources.
Third, we determined the observation cycle (or AO) for each detection, which was used in PIMMS
to convert the count rate to flux for {\it Chandra} archive data.
Finally, for each detection, we calculated the flux using the APEC model ($Z=0.5\ {\rm Z_{\odot}}$) with individual coronal temperature and $N_{\rm H}$, the latter of which was estimated from individual optical extinction \citep{Zhu2017}:
\begin{equation}
N_{\rm H} {\rm (cm ^{-2})} = 2.19 \times 10^{21} A_V {\rm (mag)}.
\label{nh.eq}
\end{equation}

Before we estimated an averaged quiescent flux for each star, we removed those detections with X-ray flares.
We first used the
standard nonparametric Kolmogorov--Smirnov (K--S) test
to quantitatively test the source variability during each an
observation.
The observations with variable light curve were identified with K-S probability $P_{\rm K-S} <$ 0.1.
Then we applied Bayesian block analysis \citep[{\it astropy.stats.bayesian\_blocks};][]{Scargle2013} to these observations to detect flares.
By operating on unbinned events, this method can
detect and characterize local variance in the count rate,
and works well in flare identification \citep[e.g.,][]{Neilsen2013, Ponti2015}.
The false alarm probability is set as $p0 =$ 0.01.
A detailed study of the X-ray flares in our sample is currently in preparation as an independent work.
The vignetting-corrected net count rate and unabsorbed flux for each observation are listed in Table \ref{chandra.tab}.
Finally, we calculated an exposure-weighted averaged X-ray flux for each star.
The X-ray luminosity ($L_X$) was determined from the unabsorbed averaged flux ($f_X$) and distance (Table \ref{lx2lbol.tab}).

\begin{figure}[tbp]
\center
\includegraphics[width=0.48\textwidth]{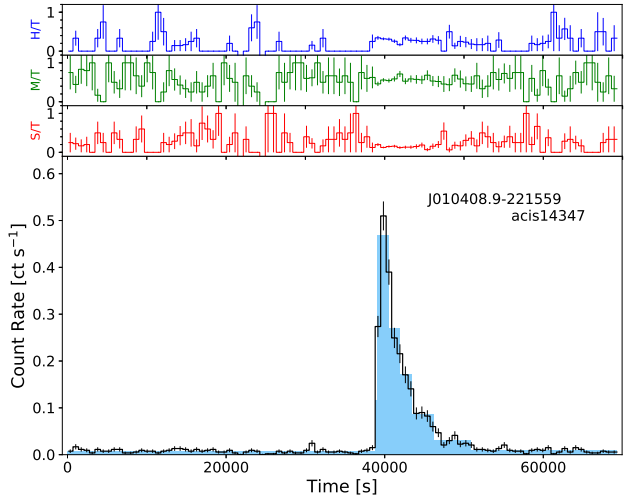}
\caption[]{An example of the light curve of stellar X-ray flare. The $S$, $M$, and $H$ represent the soft (0.3--1 keV), medium (1--2 keV), and hard (2--8 keV) bands, respectively. The blue histograms show the Bayesian blocks.}
\label{flarelc.fig}
\end{figure}

\begin{table}
\begin{center}
\renewcommand{\arraystretch}{1.5}
\scriptsize
\setlength{\tabcolsep}{1.5pt}
\caption[]{Individual {\it Chandra} observation for stars in our sample.}
\label{chandra.tab}
\begin{tabular}{lcccc}
\hline\noalign{\smallskip}
Object  &   ObsID    &   MJD   &   Count Rate & Flux  \\
        &             &  (day) &   ks$^{-1}$  & (10$^{-15}$ erg cm$^{-2}$ s$^{-1}$)       \\
  (1)   &     (2)  &    (3)      &         (4)                        &      (5)        \\
\hline\noalign{\smallskip}
J000012.8+622947 & acis2810 & 52531.383  & 0.92$\pm$0.21 & 11$\pm$2\\
J000100.1-245742 & acis13394 & 55784.056  & 0.32$\pm$0.11 & 2.8$\pm$1.0\\
J000101.9-250431 & acis13394 & 55784.056  & 0.21$\pm$0.09 & 1.8$\pm$0.8\\
J000136.1+130639 & acis6978 & 54075.017  & 23$\pm$1 & 175$\pm$7\\
J000136.1+130639 & acis8491 & 54124.292  & 12$\pm$1 & 110$\pm$8\\
J000238.8+255219 & acis5610 & 53394.483  & 10$\pm$2 & 40$\pm$6\\
J000611.4+725929 & acis3835 & 52742.424  & 1.8$\pm$0.3 & 17$\pm$2\\
J000645.6+730635 & acis3835 & 52742.424  & 0.31$\pm$0.14 & 2.4$\pm$1.1\\
J000753.8+512400 & acis8942 & 54544.217  & 0.67$\pm$0.15 & 3.5$\pm$0.8\\
J000759.3+512655 & acis8942 & 54544.217  & 1.8$\pm$0.3 & 7.3$\pm$1.0\\
J000833.8+512412 & acis8942 & 54544.217  & 0.46$\pm$0.15 & 2.1$\pm$0.7\\
J000835.3+512142 & acis8942 & 54544.217  & 0.98$\pm$0.21 & 3.8$\pm$0.8\\
J000836.5+512616 & acis8942 & 54544.217  & 0.29$\pm$0.14 & 1.6$\pm$0.8\\
J000849.5+512514 & acis8942 & 54544.217  & 5.7$\pm$0.5 & 24$\pm$2\\
J001051.6-120543 & acis15061 & 56448.701  & 1.4$\pm$0.5 & 9.6$\pm$3.7\\
J001116.8-151526 & acis6105 & 53549.484  & 1.3$\pm$0.3 & 14$\pm$3\\
J001123.9-151541 & acis6105 & 53549.484  & 0.56$\pm$0.19 & 4.1$\pm$1.4\\
J001144.7+522859 & acis15318 & 56453.998  & 1.4$\pm$0.3 & 17$\pm$3\\
J001145.8-285501 & acis5797 & 53610.327  & 0.64$\pm$0.29 & 7.0$\pm$3.2\\
J001147.4-152319 & acis6105 & 53549.484  & 2.0$\pm$0.3 & 13$\pm$2\\ 
\noalign{\smallskip}\hline
\end{tabular}
\end{center}
(This table is available in its entirety in machine-readable and Virtual Observatory (VO) forms in the online journal. A portion is shown here for guidance regarding its form and content.)
\end{table}

\subsection{Stellar parameters and classification}
\label{classify.sec}

To derive individual stellar parameters (i.e, effective temperature, surface gravity, and metallicity), our primary choice is the Large Sky Area Multi-Object Fiber Spectroscopic Telescope (hereafter LAMOST, also called the Guo Shou Jing Telescope) data base.
LAMOST a reflecting Schmidt telescope, with an effective aperture of 4 m and a field of view of 5 degrees \citep{Cui2012, Zhao2012}.
The LAMOST DR7 dataset released more than 14 million spectra,
and presented parameter estimations for more than 6 million A, F, G, and K stars.
For stars not included in the LAMOST DR7 catalog, we adopted the parameters ({\it teff50}, {\it logg50}, {\it met50}) from  \citet{Anders2019}, who derived Bayesian stellar parameters and distances for 265 million stars using the code StarHorse, based on the combination of {\it Gaia} DR2 and the photometric catalogs of PS1, 2MASS, and AllWISE.
There are 447 and 3550 objects in common between our sample and LAMOST DR7 catalog and \citet{Anders2019}, respectively.
We defined the stars with parameter estimations as the ``{\it parameter}" sample, and other stars as the ``{\it non-parameter}" sample.
Using the  effective temperatures, we classified the ``{\it parameter}" sample into different subclasses:
O type with $T_{\rm eff}$ $\geq$ 30000 K;
B type with 10000 K $\leq$ $T_{\rm eff}$ $<$ 30000 K;
A type with 7500 K $\leq$ $T_{\rm eff}$ $<$ 10000 K;
F type with 6000 K $\leq$ $T_{\rm eff}$ $<$ 7500 K;
G type with 5200 K $\leq$ $T_{\rm eff}$ $<$ 6000 K;
K type with 3700 K $\leq$ $T_{\rm eff}$ $<$ 5200 K;
M type with $T_{\rm eff}$ $<$ 3700 K.

We further classified the sample into giants, YSOs, and dwarfs.
First, a star was considered as a giant if the log$g$ value is smaller than that specified in the following algorithm \citep{Ciardi2011}: 
\begin{equation}
{\rm log}g < \left \{
\begin{array}{lcl}
3.5 & & {\rm if}\ T_{\rm eff} \geq 6000, \\
4.0 & & {\rm if}\ T_{\rm eff} \leq 4250, \\
5.2-2.8\times~10^{-4}T_{\rm eff} & & {\rm if}\ 4250 \leq T_{\rm eff} \leq 6000.
\end{array}
\right.
\label{eq:tefflogg}
\end{equation}
This leads to 685 giants and 3312 dwarfs and YSOs.
We then cross-matched our sample with the catalogs in \citet{Marton2016, Marton2019}.
By using the 2MASS and WISE photometric data, combined with {\it Planck} dust opacity values, \citet{Marton2016} presented a catalog of Class I/II and III YSO candidates with the support vector machine method.
By adding the {\it Gaia} data base, \citet{Marton2019}  presented a new catalog, classifying more than 100 million sources into four classes: YSOs, extragalactic objects, main-sequence stars, and evolved stars.
There are 1100 objects in common between our sample and these catalogs, including 68 giants, 900 YSOs, and 132 dwarfs. 
For sources in \citet{Marton2019}, only those with reliable classification probability ($P$ $>$ 90\%) were selected.
Sometimes one star has different classifications from above methods, we preferentially adopted the classification by using the log$g$ value from LAMOST DR7 and \citet{Anders2019}), followed by the classification of \citet{Marton2019}. 
We classified a total of 697 giants.

Second, we defined some criteria to classify more new YSOs.
As \citet{Marton2019} reported, 99\% of the known YSOs are located in the regions where the dust opacity value is higher than 1.3$\times$10$^{-5}$. 
Thus we labeled as YSO candidates those sources that meet the requirements: (1) $\tau >$ 1.3$\times$10$^{-5}$ and (2)
$J-H > 1 - (H-K_{\rm S})$ or $W1-W2 > $ 0.04
(See Appendix B for more details).
In addition, considering that there may be many AGB stars with $W2-W3 < $1 \citep{Koenig2014}, we add one criterion $W2-W3 \geq $1 for the ``{\it non-parameter}" sample.
With the constraints of colors and dust opacity,
we classified additional $\approx$300 stars to be YSOs.
We finally categorized 1196 YSOs.

Third, for the ``{\it parameter}" sample, if they were not classified as YSOs or giants from above steps and if their dust opacity values are lower than 1.3$\times$10$^{-5}$, we flagged them as main-sequence dwarfs.

\begin{figure}[!htb]
\center
\includegraphics[width=0.49\textwidth]{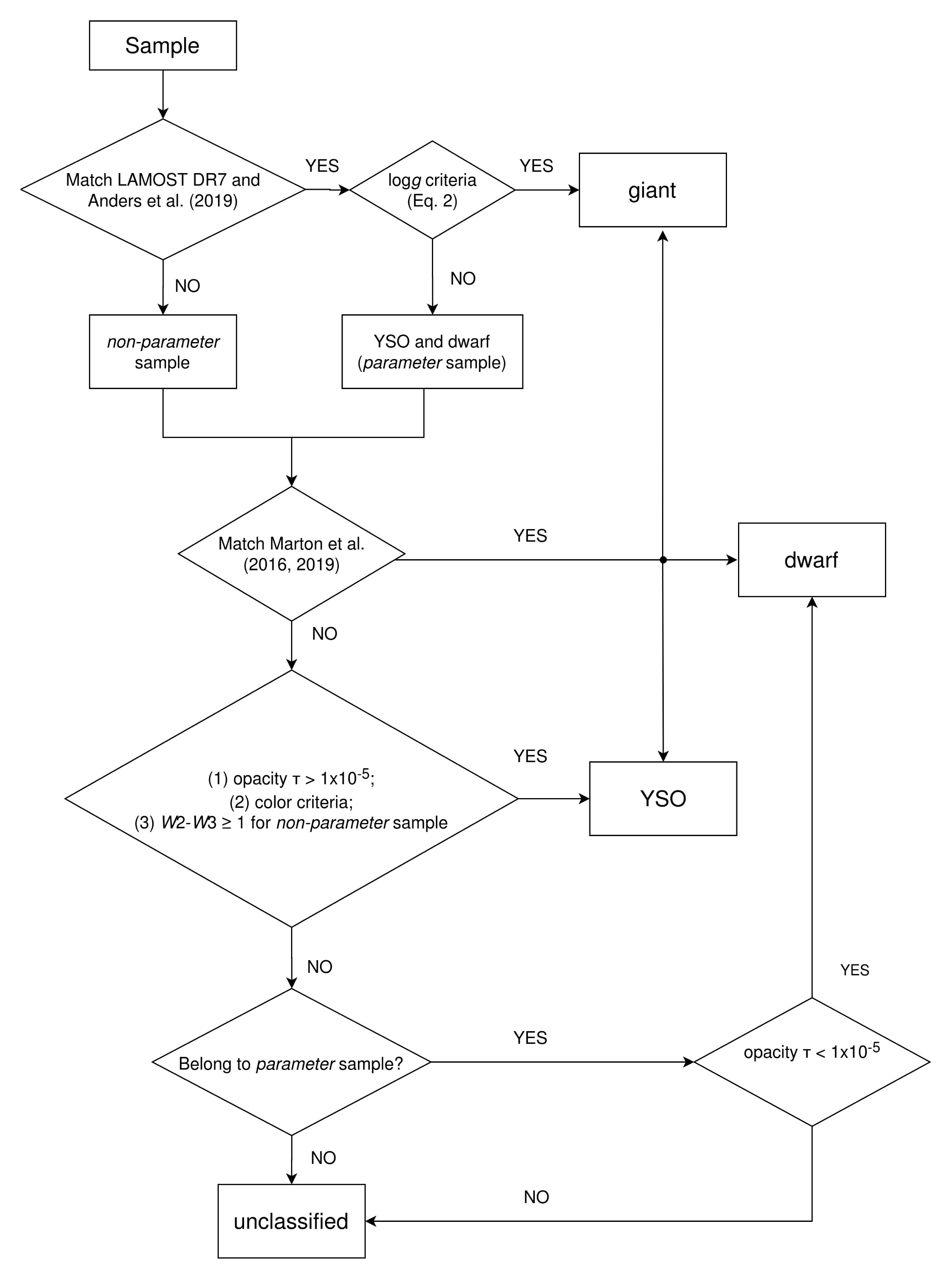}
\caption[]{Summary flowchart of the stellar classification process in this paper.}
\label{process.fig}
\end{figure}

In summary, the giants were collected from the ``{\it parameter}" sample and \citet{Marton2019}; the YSOs were classified from previous studies \citep{Marton2016, Marton2019} and our criteria;
for the dwarfs, one part was selected from \citet{Marton2019}, and the other part was classified from the ``{\it parameter}" sample. Figure \ref{process.fig} summarizes the steps in a flowchart for reference.

At the end, we presented stellar classifications for 3005 sources in our sample, including 1196 YSOs, 1112 dwarfs, and 697 giants.
Among these sources, 432 YSOs, 1111 dwarfs, and 695 giants have parameter estimations. We defined these 2238 sources as the ``{\it A-class}" sample, which will be used for following analysis.

\begin{figure*}[!htb]
\center
\includegraphics[width=0.48\textwidth]{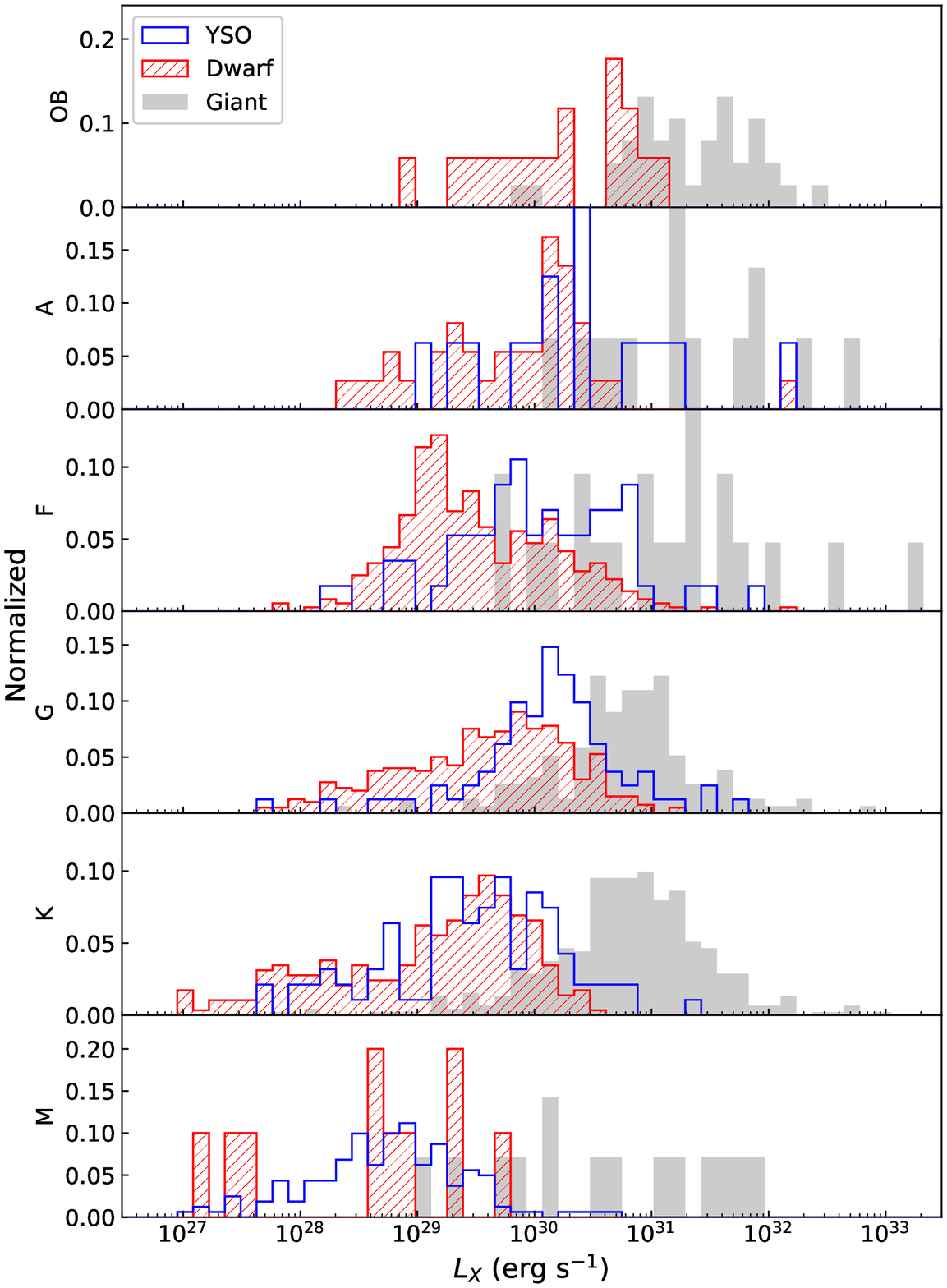}
\includegraphics[width=0.48\textwidth]{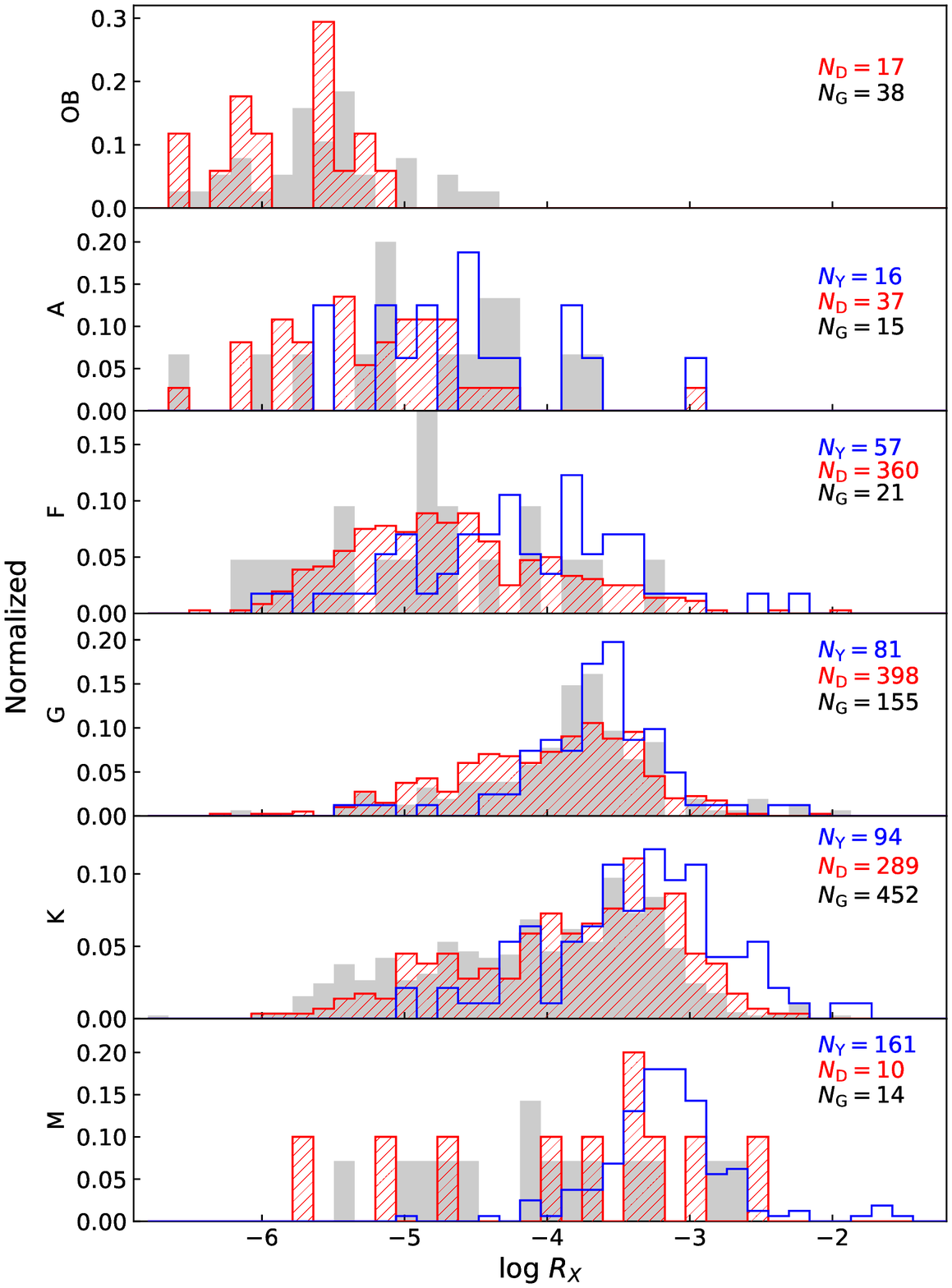}
\caption[]{Left panel: Histograms showing the distributions of $L_X$ for stars from OB to M type. Right panel: Histograms showing the distributions of $R_X$ for stars of spectral type OB through M.}
\label{subclass.fig}
\end{figure*}

\section{Stellar X-ray Activity}
\label{activity.sec}

\subsection{X-ray Activity in Different Stellar Types}
\label{types.sec}

X-ray emission is prevalent among almost all stellar classes \citep[e.g.,][]{Stocke1991},
although with different mechanisms \citep[see][for a review]{Gudel2004}.
For the hot and massive early-type stars,
the emitted X-rays arise from either small-scale shocks in their winds or collisions between the wind and circumstellar material \citep{Lucy1980, Parkin2009}.
For late-type stars, the X-ray emission is attributed to the presence of a magnetic corona \citep{Vaiana1981}.
The X-ray emission of YSOs may be from accretion shock, star-disk interaction, or solar-like corona \citep[e.g.,][]{Preibisch2005}.
In this paper, we used the ratio of X-ray-to-bolometric luminosities ($R_X$ $= L_X/L_{\rm bol}$) as the X-ray activity indicators.
We used the PARSEC theoretical models to determine the bolometric luminosity (see Appendix C).

Figure \ref{subclass.fig} displays the distributions of $L_X$ and $R_X$ for different stellar types in our ``{\it A-class}" sample.
For each stellar type, giants have the highest luminosities up to 10$^{32}$--10$^{33}$ erg s$^{-1}$.
For G and K stars, we ran a K-S test ({\it scipy.stats.ks\_2samp}) to check the similarity of $L_X$ distributions. Since the YSOs, dwarfs, and giants have different numbers, we randomly selected $\approx$100 objects for each class and calculated the $P_{\rm K-S}$ values. The selection and calculation was repeated 10$^4$ times, allowing us to have an averaged $P_{\rm K-S}$ estimation.
For giants and dwarfs, the $P_{\rm K-S}$ values are smaller than 10$^{-10}$ for both G and K stars, indicating a different distribution; for YSOs and dwarfs, the $P_{\rm K-S}$ values are $\approx$6$\times$10$^{-4}$ (G stars) and $\approx$0.06 (K stars), respectively.

In previous studies, YSOs were found to have X-ray luminosities 10--10$^4$ above those typically seen in main-sequence stars \citep[e.g,][]{Feigelson2002,Preibisch2005}.
However, no such notable difference can be seen in Figure \ref{subclass.fig}.
This may be explained by the incompleteness of our dwarf sample.
Take K stars as an example, previous observations ({\it ROSAT}; 0.1--2.4 keV) showed that the field dwarfs have X-ray luminosities ranging from $\approx$6$\times$10$^{26}$ erg/s to $\approx$2$\times$10$^{30}$ erg/s, while  
the dwarfs in open clusters Pleiades and Hyades have X-ray luminosities spanning from $\approx$3$\times$10$^{28}$ erg/s to $\approx$4$\times$10$^{30}$ erg/s \citep{Wright2011}.
The X-ray luminosities ({\it Chandra}; 0.5--8 keV) of pre-main-sequence K5-7 stars in the Orion nebula are from $\approx$3$\times$10$^{29}$ erg/s to $\approx$2$\times$10$^{31}$ erg/s \citep{Wolk2005}.
It seems that most of the K dwarfs in our catalog are in the high-luminosity tail of the $L_X$ distribution of Galactic K stars. 
This may be due to the {\it Chandra} observation mode and our sample selection methods.
For example, the requirement of {\it WISE} photometry may remove a number of dwarfs.

In general, the $R_X$ distributions are similar among different classes.
For giants and dwarfs, the $P_{\rm K-S}$ values are about 0.14 and 0.20, respectively.
For YSOs and dwarfs, the $P_{\rm K-S}$ values are smaller than 0.01 for both G and K stars.
More YSOs have higher $R_X$ values than the dwarfs in the same stellar type.

We noted that there is a large difference between the number of M-type YSOs and dwarfs. Among the 161 M-type YSOs with parameter estimations, about half is from \citet{Marton2016, Marton2019} and the other half is picked out with our criteria. One caveat is that about 50 ones have $W2 - W3 < 1$, some of which may be wrongly classified (i.e., contaminated by dwarfs or giants). However, the statistical distributions will not be much affected.

\begin{table*}
\begin{center}
\renewcommand{\arraystretch}{1.5}
\scriptsize
\setlength{\tabcolsep}{2pt}
\caption[]{Main properties for the stars in our sample.}
\label{lx2lbol.tab}
\begin{tabular}{lcccccccccc}
\hline\noalign{\smallskip}
Object  &     RA  &   Dec  &   Class &  $D$ & $E(B-V)$ &     $f_X$   & $L_X$  & log$R_X$  &        $HR$1   &   $HR$2   \\
        &   ($^{\rm o}$)      &    ($^{\rm o}$)     &            & (pc)  &  (mag)        & (10$^{-15}$ erg cm$^{-2}$ s$^{-1}$)  &   (erg s$^{-1}$)  &      &       & \\
  (1)   &     (2)  &    (3)      &         (4)                        &      (5)          &   (6)     &     (7)    & (8)    &   (9)          &   (10)     &     (11)     \\
\hline\noalign{\smallskip}
J000012.8+622947 & 0.05374 & 62.49657 & G & 899$^{+21}_{-20}$ & 0.26 & 11$\pm$2 & 1.0e+30$\pm$2.4e+29 & -4.0$\pm$0.1 & 0.59$\pm$0.22 & -0.84$\pm$0.29\\
J000100.1-245742 & 0.25066 & -24.96192 & M & 376$^{+21}_{-19}$ & 0.02 & 2.8$\pm$1.0 & 4.7e+28$\pm$1.6e+28 & -3.52$\pm$0.15 & 0.15$\pm$0.4 & -0.7$\pm$0.52\\
J000101.9-250431 & 0.25818 & -25.07535 & Kd & 242$^{+2}_{-2}$ & 0.01 & 1.8$\pm$0.8 & 1.3e+28$\pm$5.7e+27 & -5.0$\pm$0.2 & 0.0$\pm$0.57 & -0.38$\pm$0.74\\
J000136.1+130639 & 0.40042 & 13.11095 & Fd & 381$^{+7}_{-6}$ & 0.08 & 149$\pm$5 & 2.6e+30$\pm$9.6e+28 & -3.74$\pm$0.02 & 0.21$\pm$0.04 & -0.66$\pm$0.04\\
J000238.8+255219 & 0.66206 & 25.87221 & K & 703$^{+30}_{-27}$ & 0.03 & 40$\pm$6 & 2.4e+30$\pm$3.6e+29 & -2.66$\pm$0.07 & 0.09$\pm$0.17 & -0.71$\pm$0.18\\
J000611.4+725929 & 1.54788 & 72.99162 & G & 1680$^{+138}_{-119}$ & 0.37 & 17$\pm$2 & 5.7e+30$\pm$7.8e+29 & -3.09$\pm$0.06 & -0.01$\pm$0.15 & -0.42$\pm$0.2\\
J000645.6+730635 & 1.69003 & 73.10997 & K & 368$^{+3}_{-3}$ & 0.33 & 2.4$\pm$1.1 & 3.9e+28$\pm$1.8e+28 & -4.4$\pm$0.2 & -0.1$\pm$0.52 & -0.5$\pm$0.83\\
J000753.8+512400 & 1.97436 & 51.40010 & Ky & 597$^{+28}_{-26}$ & 0.12 & 2.1$\pm$0.9 & 1.5e+29$\pm$3.4e+28 & -3.48$\pm$0.1 & -0.27$\pm$0.25 & -0.47$\pm$0.43\\
J000759.3+512655 & 1.99728 & 51.44876 & My & 290$^{+10}_{-10}$ & 0.05 & 3.8$\pm$1.8 & 7.4e+28$\pm$1.0e+28 & -3.15$\pm$0.06 & -0.19$\pm$0.15 & -0.59$\pm$0.24\\
J000833.8+512412 & 2.14095 & 51.40359 & Ad & 434$^{+8}_{-8}$ & 0.09 & 11$\pm$1 & 4.7e+28$\pm$1.6e+28 & -6.53$\pm$0.15 & -0.14$\pm$0.38 & -0.76$\pm$0.61\\
J000835.3+512142 & 2.14724 & 51.36168 & Kd & 50$^{+3}_{-2}$ & 0.01 & 3.5$\pm$0.8 & 1.1e+27$\pm$2.5e+26 & -4.81$\pm$0.09 & 0.07$\pm$0.24 & -0.83$\pm$0.29\\
J000836.5+512616 & 2.15234 & 51.43779 & G & 1497$^{+115}_{-100}$ & 0.11 & 7.3$\pm$1.0 & 4.2e+29$\pm$2.1e+29 & -4.14$\pm$0.22 & -0.12$\pm$0.56 & -0.91$\pm$1.17\\
J000849.5+512514 & 2.20651 & 51.42067 & G & 268$^{+17}_{-15}$ & 0.07 & 205$\pm$4 & 2.1e+29$\pm$1.8e+28 & -3.97$\pm$0.04 & -0.11$\pm$0.09 & -0.77$\pm$0.11\\
J001051.6-120543 & 2.71524 & -12.09552 & - & 522$^{+70}_{-56}$ & 0.02 & 2.1$\pm$0.7 & 3.1e+29$\pm$1.2e+29 & -2.66$\pm$0.17 & 0.33$\pm$0.69 & 0.19$\pm$0.48\\
J001116.8-151526 & 2.82019 & -15.25745 & - & 375$^{+64}_{-48}$ & 0.02 & 3.8$\pm$0.8 & 2.4e+29$\pm$5.6e+28 & -2.12$\pm$0.1 & 0.33$\pm$0.69 & 0.19$\pm$0.48\\
J001123.9-151541 & 2.84977 & -15.26147 & Kd & 533$^{+15}_{-14}$ & 0.02 & 1.6$\pm$0.8 & 1.4e+29$\pm$4.6e+28 & -4.01$\pm$0.14 & -0.21$\pm$0.39 & -0.29$\pm$0.6\\
J001144.7+522859 & 2.93639 & 52.48312 & K & 393$^{+10}_{-9}$ & 0.12 & 24$\pm$2 & 3.1e+29$\pm$6.1e+28 & -3.04$\pm$0.08 & 0.28$\pm$0.21 & -0.66$\pm$0.23\\
J001145.8-285501 & 2.94100 & -28.91695 & M & 129$^{+2}_{-2}$ & 0.01 & 9.6$\pm$3.7 & 1.4e+28$\pm$6.3e+27 & -3.29$\pm$0.2 & -0.42$\pm$0.73 & 0.41$\pm$0.76\\
J001147.4-152319 & 2.94754 & -15.38880 & K & 374$^{+13}_{-13}$ & 0.01 & 14$\pm$3 & 2.3e+29$\pm$3.0e+28 & -3.08$\pm$0.06 & -0.33$\pm$0.14 & -0.53$\pm$0.23\\
J001152.5+523845 & 2.96884 & 52.64607 & F & 1303$^{+77}_{-69}$ & 0.15 & 4.1$\pm$1.4 & 2.5e+30$\pm$7.5e+29 & -3.31$\pm$0.13 & 0.36$\pm$0.33 & -0.72$\pm$0.42\\
\noalign{\smallskip}\hline
\end{tabular}
\end{center}
\tablecomments{The columns are:
(1) Object.
(2) Right ascension in degree.
(3) Declination in degree.
(4) Stellar classification.
(5) Distance from $Gaia$ DR2 data.
(6) $E(B-V)$ from PS1 3D map.
(7) Averaged unabsorbed X-ray flux in the 0.3--8 keV.
(8) X-ray luminosity in the 0.3--8 keV.
(9) X-ray-to-bolometric luminosity ratio.
(10) Hardness ratio $HR{\rm 1} = (M-S)/(M+S)$. S/M represents background-subtracted counts in soft (0.3-1 keV) and medium (1-2 keV) bands.
(11) Hardness ratio $HR{\rm 2} = (H-M)/(H+M)$. M/H represents background-subtracted counts in medium (1-2 keV) and hard (2-8 keV) bands.}
(This table is available in its entirety in machine-readable and Virtual Observatory (VO) forms in the online journal. A portion is shown here for guidance regarding its form and content.)
\end{table*}

\begin{figure*}[!t]
\center
\includegraphics[width=0.48\textwidth]{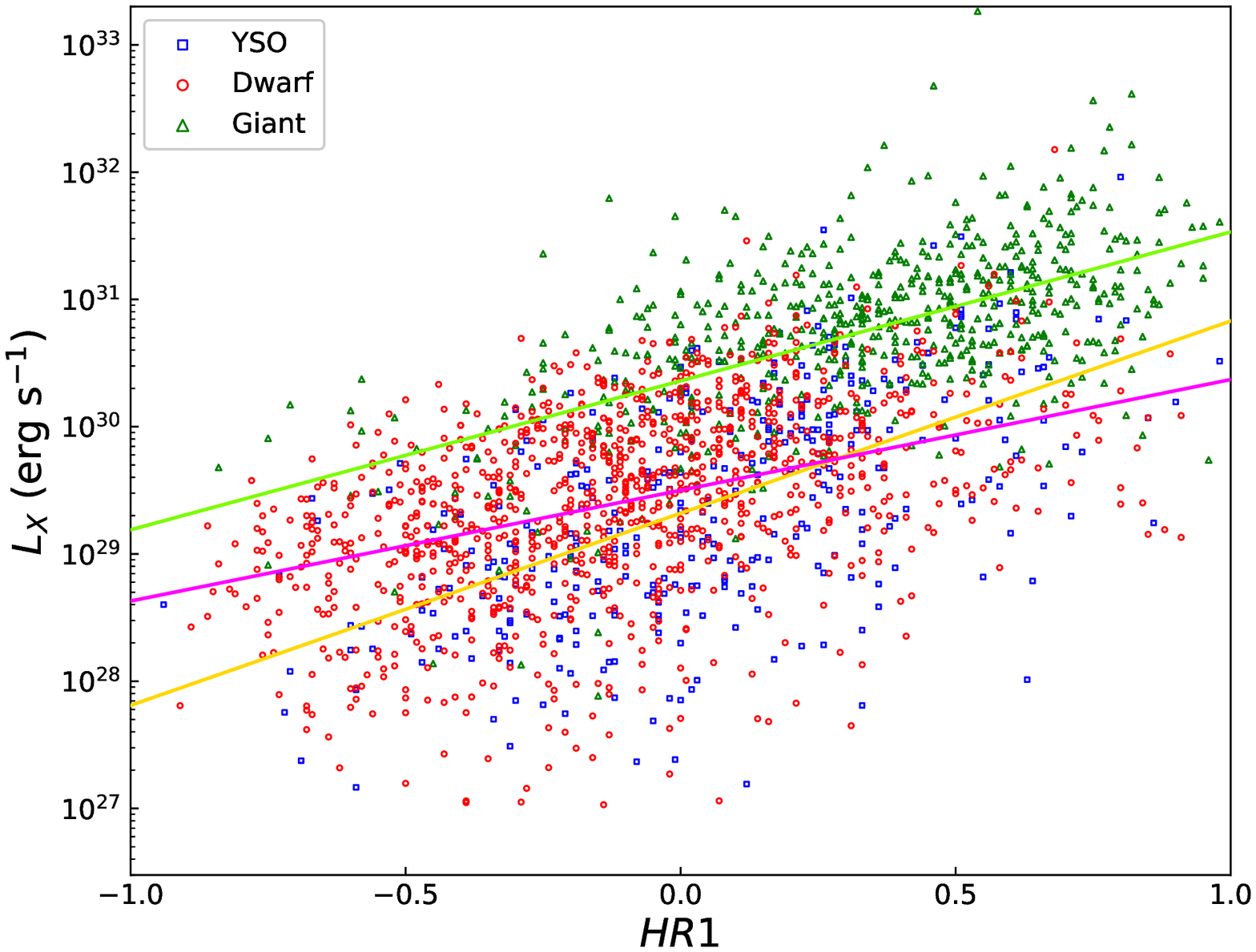}
\includegraphics[width=0.48\textwidth]{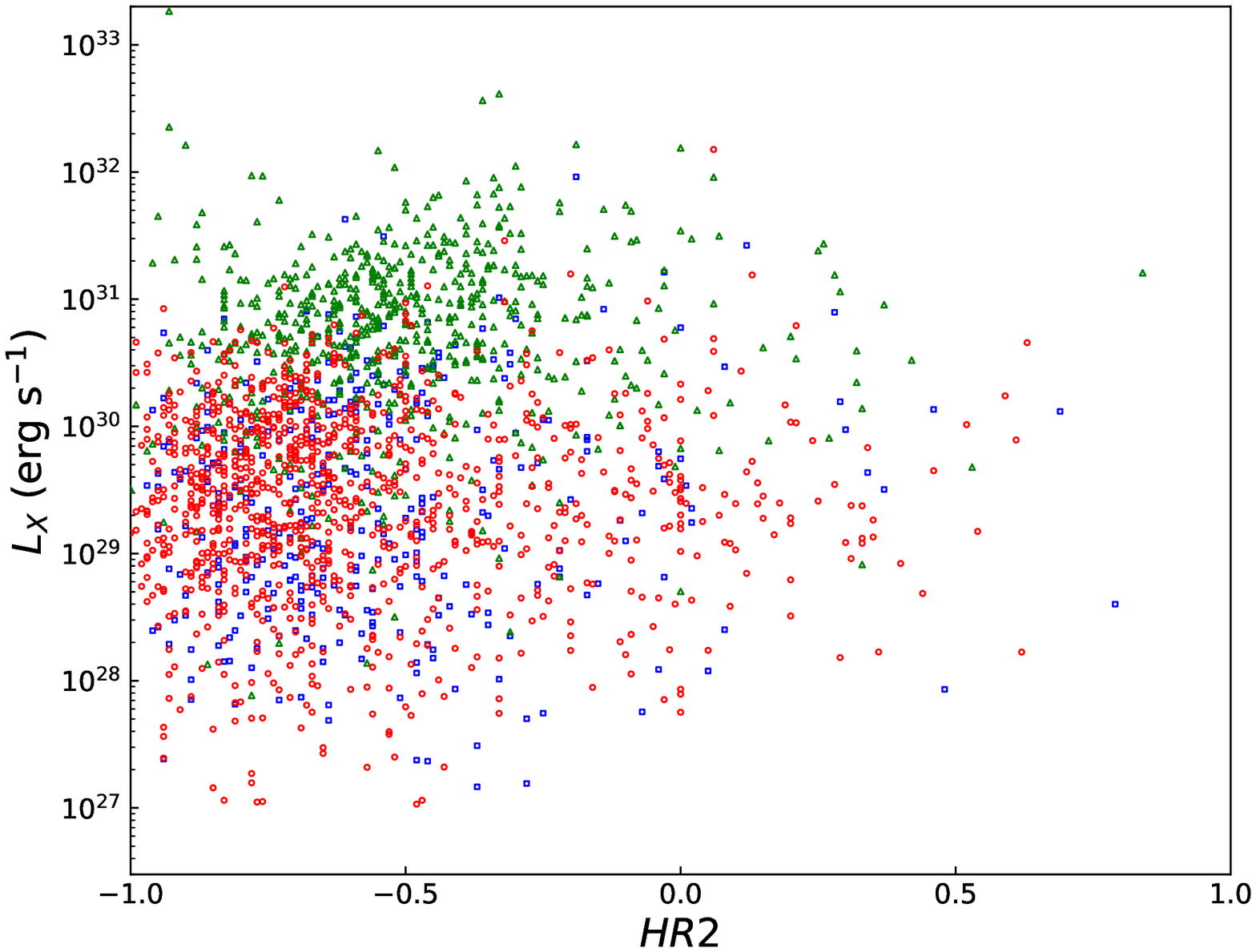}
\caption[]{Left Panel: $L_X$ versus $HR$1.
The positive correlation means stronger X-ray emitters (higher $L_X$) have higher coronal temperatures. The yellow, purple, and green lines are the fits for YSOs, dwarfs, and giants, respectively.
Right Panel: $L_X$ versus $HR$2.}
\label{lxhr.fig}
\end{figure*}

\begin{figure*}[!t]
\center
\includegraphics[width=0.48\textwidth]{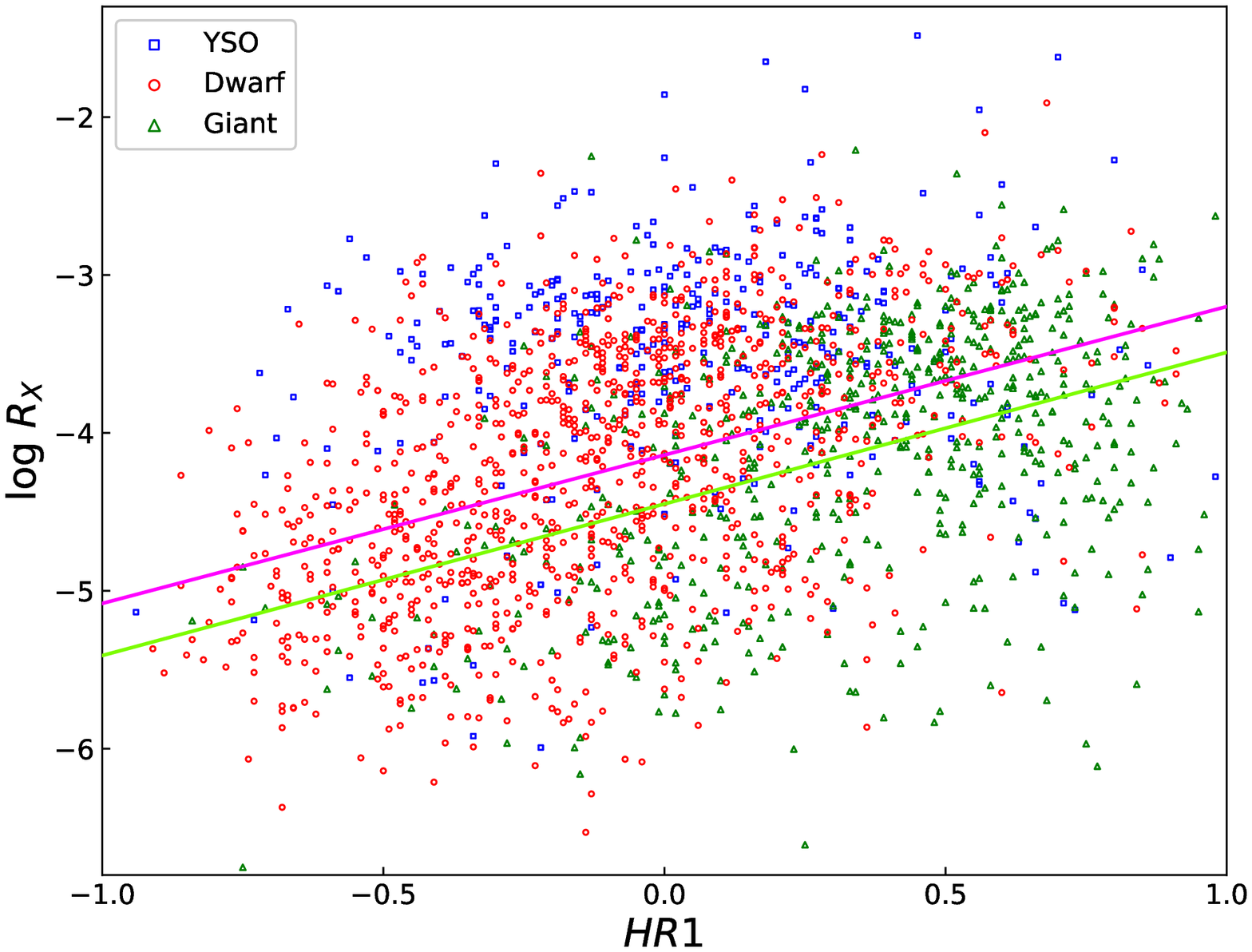}
\includegraphics[width=0.48\textwidth]{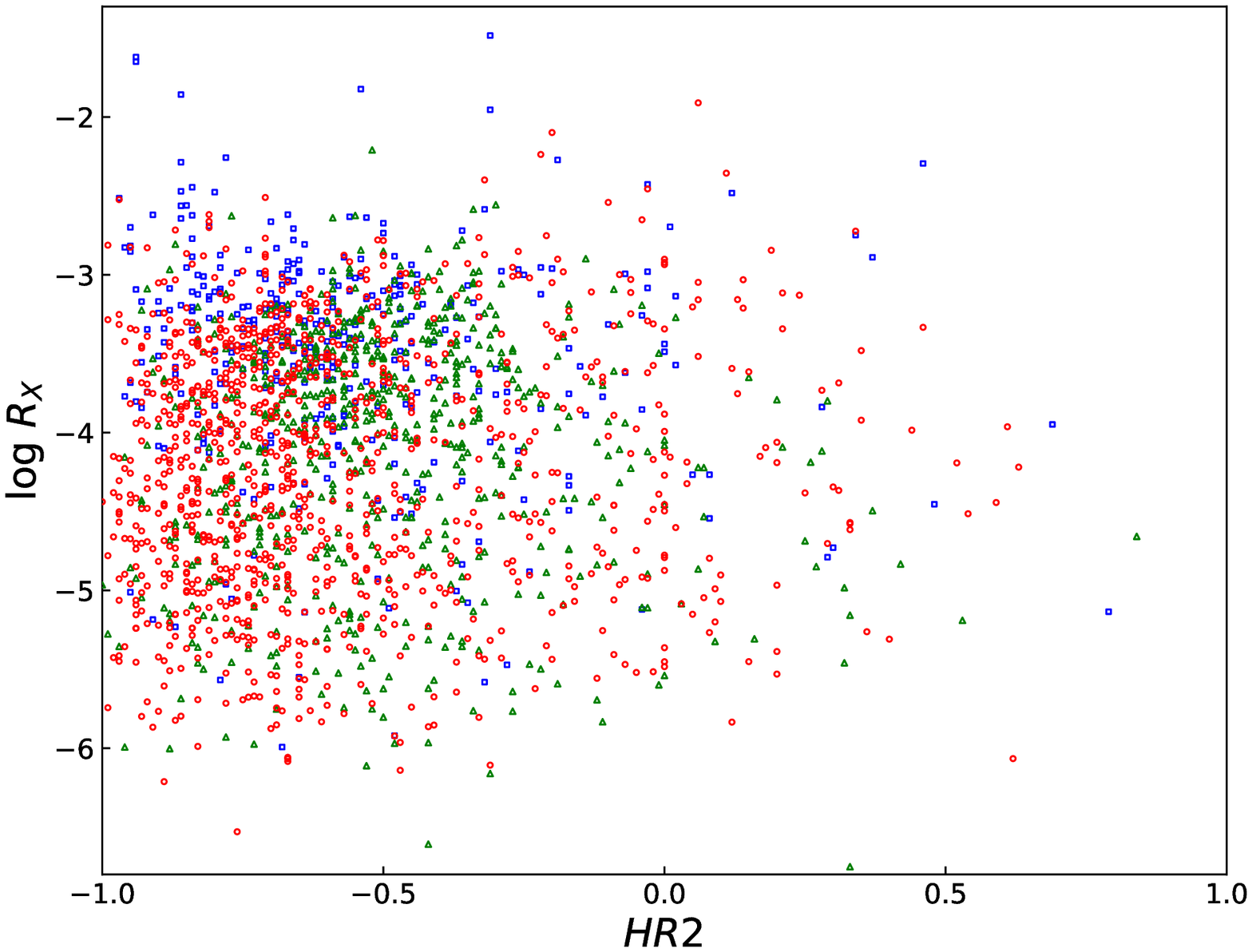}
\caption[]{Left Panel: $R_X$ versus $HR$1. The purple and green lines are the fits for dwarfs and giants, respectively.
Right Panel: $R_X$ versus $HR$2.
\\}
\label{rxhr.fig}
\end{figure*}

\subsection{X-ray Activity and Hardness Ratio}
\label{hr.sec}

X-ray emission of late-type stars is from solar-like corona, which suggests a relation between stellar X-ray activity and coronal activity.
The positive correlation between X-ray luminosity and coronal temperature has been explored \citep[e.g.,][]{Schmitt1995,Gudel2004}, which can be referred on the basis of a ``loop" model for stellar coronae \citep{Vaiana1983}.
Active stars with more efficient dynamo
have stronger magnetic fields in the corona, and thus higher rate of field line reconnections and flares.
This results in higher coronal temperatures and a larger plasma density of energetic electrons.
Here we take the X-ray $HR$ to track the coronal temperature for late-type stars.
There is a clear positive correlation between $L_X$ and $HR$1 (Figure \ref{lxhr.fig}), which means stronger X-ray emitters
have higher coronal temperatures.
Most YSOs are in the saturated regime, with a constant $R_X$ value (log$R_X \approx -3$) when $HR$1 varies (Figure \ref{rxhr.fig}).
We performed a Spearman correlation test ({\it scipy.stats.spearmanr}) for all these relations. Then linear regression fits were done for these relations with higher correlation coefficients (Table \ref{fits.tab}).

\begin{figure}[!htb]
\center
\includegraphics[width=0.49\textwidth]{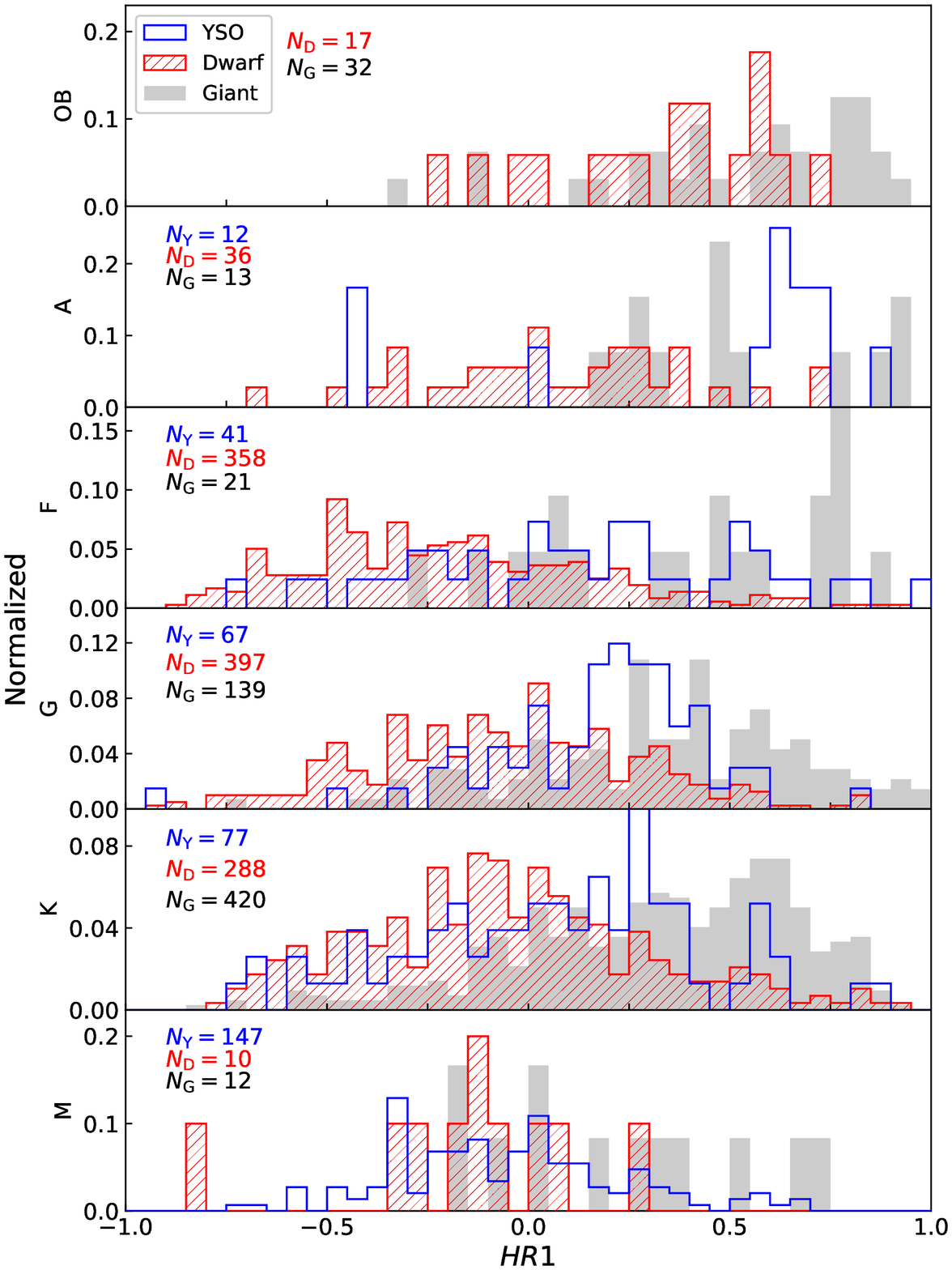}
\caption[]{Histograms showing the distributions of $HR$1 for stars from OB to M type.}
\label{hr1sub3.fig}
\end{figure}

The YSOs and dwarfs share similar $HR$1 distribution (Figure \ref{hr1sub3.fig}), while for all late stellar types (from F to K), there are giants showing very high $HR$1 values, which are good candidates possessing high-temperature corona.
A K-S test was also run for the $HR$1 distributions.
For YSOs and dwarfs, the $P_{\rm K-S}$ values are $\approx$2$\times$10$^{-4}$ (G stars) and $\approx$0.05 (K stars), respectively;
for giants and dwarfs, the $P_{\rm K-S}$ values are smaller than 10$^{-7}$ for both G and K stars.
Again, we remind that the samples of each stellar type are incomplete.

\begin{table}
\begin{center}
\renewcommand{\arraystretch}{1.5}
\caption[]{Fits to log($y$)$=a{\times}x+b$ in Logarithmic space.}
\label{fits.tab}
\begin{tabular}{lcccc}
\hline\noalign{\smallskip}
$y$  &   $x$    &  Spearman  & $a$   &  $b$   \\
 &      &  correlation coefficient &   &     \\
  (1)   &     (2)  &    (3)      &   (4)  &      (5)   \\
\hline\noalign{\smallskip}
\multicolumn{5}{c}{YSO}\\
\hline\noalign{\smallskip}
$L_X$ & $HR$1  & 0.96 & 1.51$\pm$0.17  & 29.32$\pm$0.06  \\
$R_X$ & $HR$1  & 0.48 & -  & - \\
$L_X$ & $HR$2  & 0.49 & - & - \\
$R_X$ & $HR$2  & 0.58 & - & - \\
\hline\noalign{\smallskip}
\multicolumn{5}{c}{dwarf}\\
\hline\noalign{\smallskip}
$L_X$ & $HR$1  & 0.94 &  0.87$\pm$0.08  & 29.50$\pm$0.04 \\
$R_X$ & $HR$1  & 0.94 & 0.98$\pm$0.08  & $-$4.14$\pm$0.04 \\
$L_X$ & $HR$2  & 0.16 & - & - \\
$R_X$ & $HR$2  & 0.36 & - & - \\
\hline\noalign{\smallskip}
\multicolumn{5}{c}{giant}\\
\hline\noalign{\smallskip}
$L_X$ & $HR$1  & 0.99 & 1.17$\pm$0.09  & 30.36$\pm$0.04 \\
$R_X$ & $HR$1  & 0.85 & 0.96$\pm$0.13  & $-$4.45$\pm$0.06 \\
$L_X$ & $HR$2  & 0.23 & - & - \\
$R_X$ & $HR$2  & 0.01 & - & - \\
\noalign{\smallskip}\hline
\end{tabular}
\end{center}
\end{table}

\subsection{X-ray Flux Variation}
\label{variation.sec}

Stellar X-ray emission is variable \citep{Soderblom2010,Stelzer2017}.
The most prominent signature of the variability is flares, which usually show sudden and intense brightness increase and decay.
The short X-ray variability can be also due to rotational modulation of the structural inhomogeneity of magnetic field \citep{Marino2003}.
Significant correlations have been found between the positions of active
longitudes in the activity diagnostics, including the X-ray light curve, star spot, and magnetic field maps \citep{Hussain2007}.
The long-term X-ray variability may reflect magnetic dynamo cycles \citep{Sanz-Forcada2013,Ayres2014}.
However, there are only few sources showing X-ray cycles,
since long-term X-ray monitoring is not easily feasible and it is hard to detect dynamo cycle in the X-ray band \citep{Stelzer2017}.
For YSOs, the variability can also be related to the shocks and absorption associated with accretion disk and protostellar jets \citep[e.g.,][]{Wolk2005,Flaccomio2006,Ustamujic2018,Guarcello2017}.

\begin{figure}[!t]
\center
\includegraphics[width=0.49\textwidth]{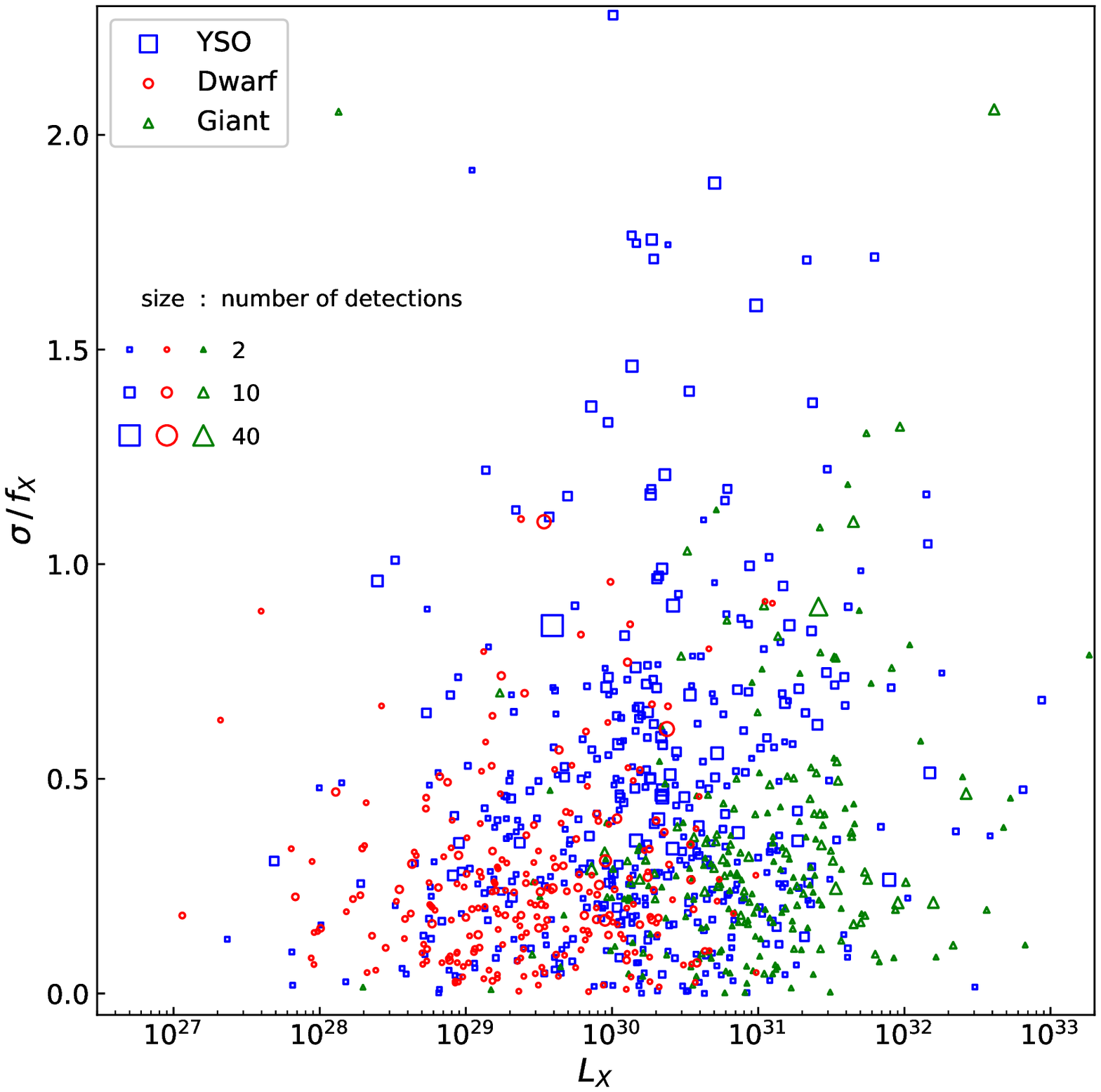}
\includegraphics[width=0.49\textwidth]{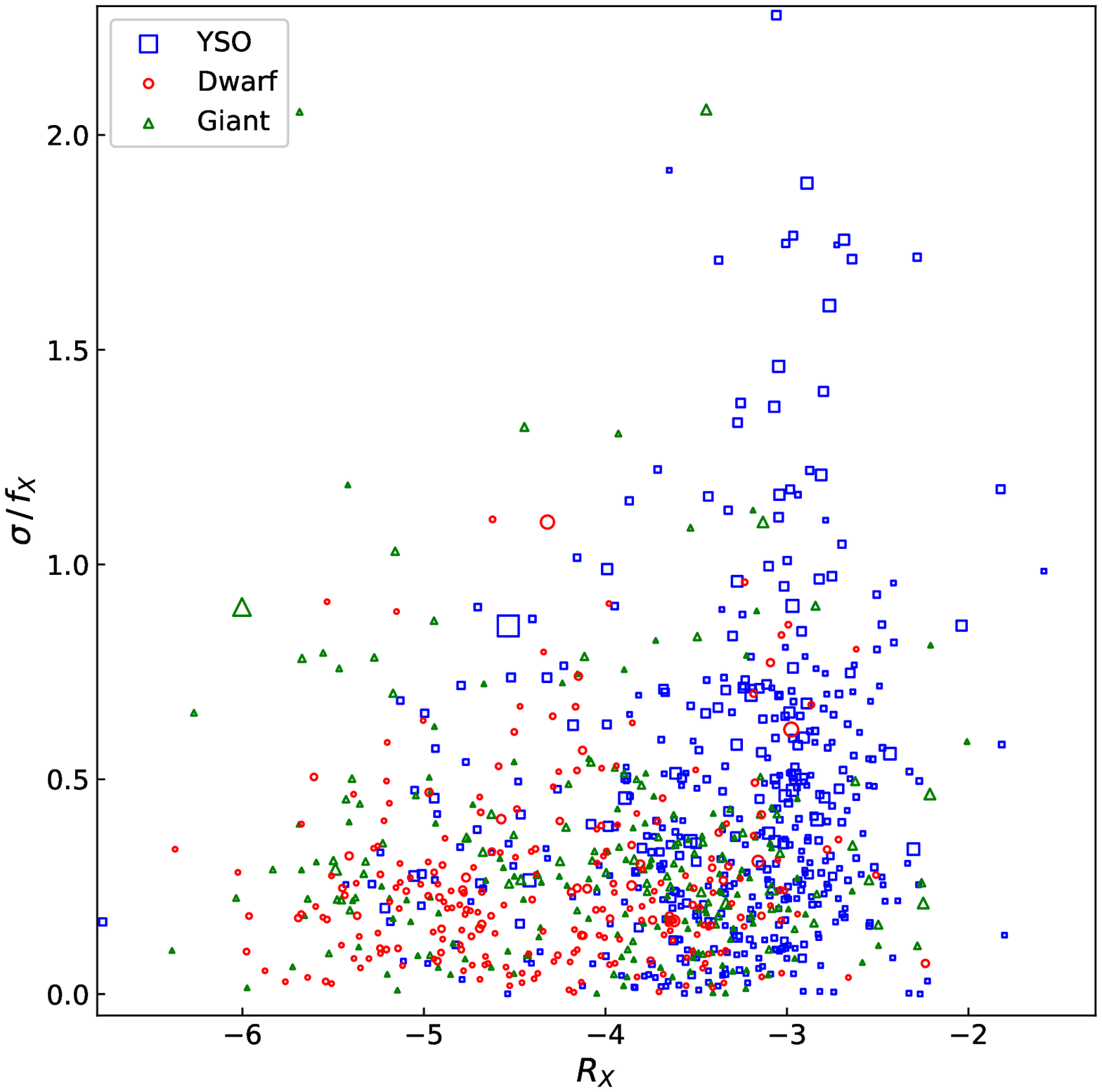}
\caption[]{Top Panel: relative flux variation $\sigma/f_X$ versus $L_X$. The symbol size represents the number of detections from $Chandra$ data. Bottom Panel: relative flux variation $\sigma/f_X$ versus $R_X$.}
\label{stdf.fig}
\end{figure}

\begin{figure}[!htbp]
\center
\includegraphics[width=0.49\textwidth]{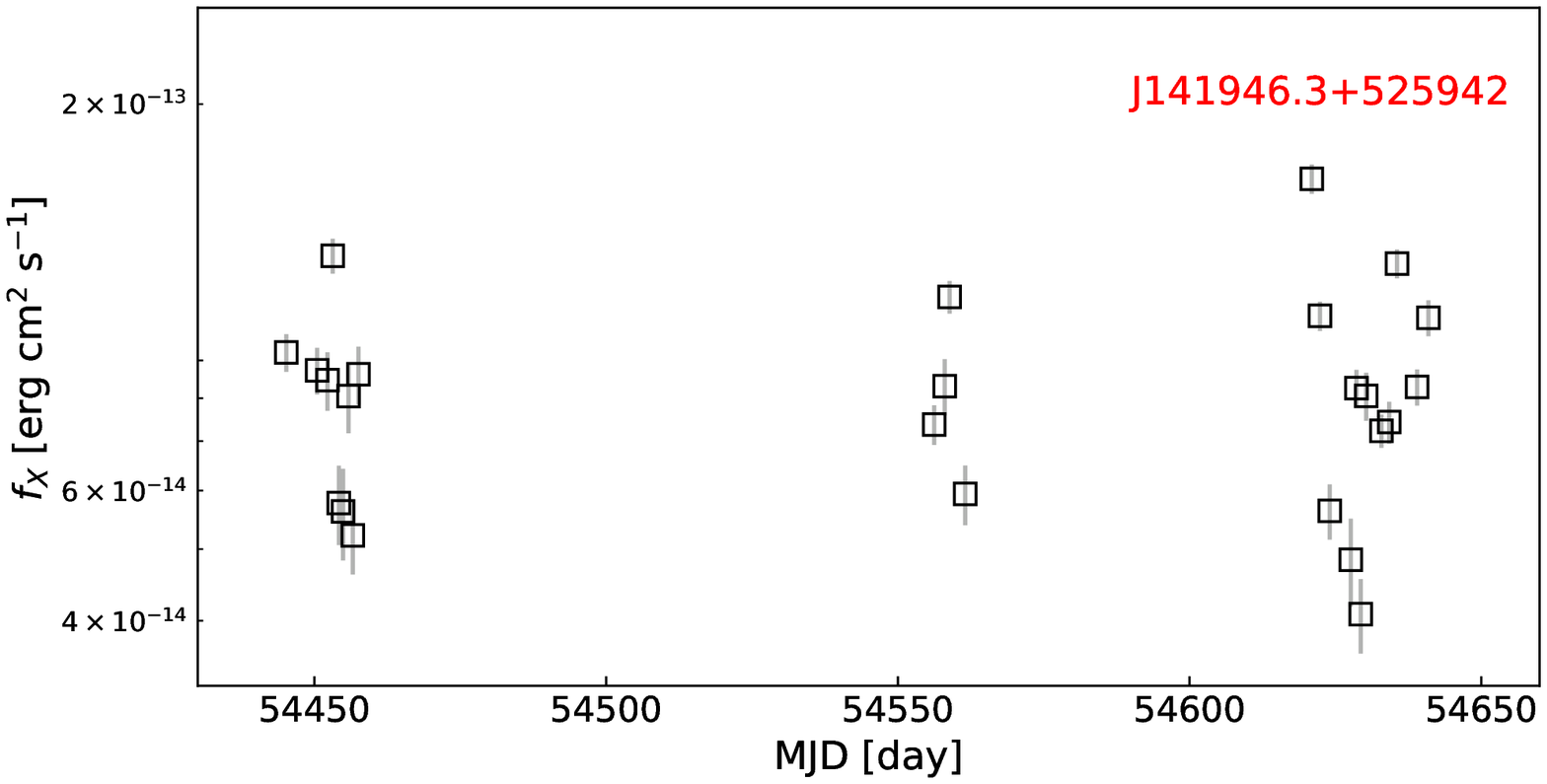}
\includegraphics[width=0.49\textwidth]{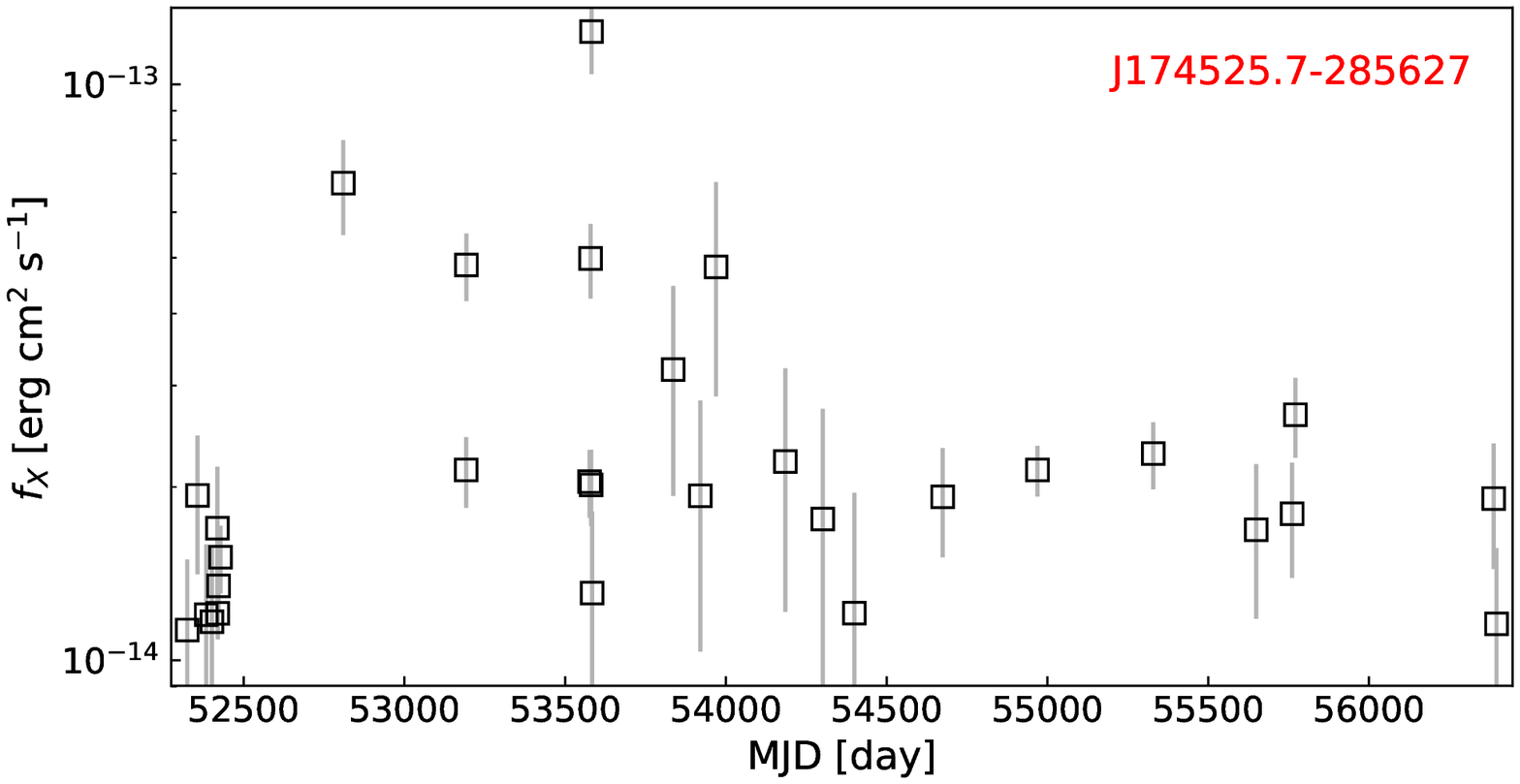}
\caption[]{Examples of long-term X-ray light curves.}
\label{long.fig}
\end{figure}

In our sample, about 1400 stars were observed more than once by {\it Chandra}.
We calculated the relative flux variation as the standard flux variation during non-flare observations divided by the averaged flux. Most stars have small relative flux variation ($\sigma$/$f_X < $0.5).
The $\sigma$/$f_X$ shows no clear relation with $R_X$ and $L_X$ (Figure \ref{stdf.fig}).
It also can be seen that the most variable sources are mainly YSOs.
Figure \ref{long.fig} shows some examples of long-term light curves

\section{ANALYSES AND SOME APPLICATIONS}
\label{discussion.sec}

In this work, we provided a large sample with X-ray activities estimated from a uniform procedure, which can help do detailed investigation of the magnetic dynamo for different type stars.
Here we give some possible scientific applications with this catalog. We remind that this catalog suffers from selection biases and is incomplete.

\subsection{The Activity--Rotation Relation}

Many studies have investigated the correlation between several coronal and chromospheric magnetic activity indicators and stellar rotation rate \citep[e.g.,][]{Pizzolato2003,Mamajek2008,Wright2011,Lehtinen2020}.
The Rossby number is used to trace the stellar rotation, which is defined as the ratio of the rotation period to the convective turnover time (Ro = $P/{\tau}$).

In order to derive stellar rotation periods from {\it Kepler} data, we have developed an online platform, {\it Kepler Data Integration Platform}\footnote{http://kepler.bao.ac.cn}, which integrates query, view, and period calculation on the whole {\it Kepler} and $K$2 data set \citep[see][for details]{Yang2019}.
In our sample, more than 900 stars were observed by the {\it Kepler} telescope, among which four are eclipsing binaries and about 100 ones show rotational modulation.
By using the baseline-detrended light curve, their periods were computed with the Lomb--Scargle periodogram \citep[see][for details]{Gao2016, Yang2019}.

The classical empirical estimate of $\tau$ from 
colors and effective temperatures works mainly for main-sequence dwarf \citep[e.g.,][]{Noyes1984,Wright2011}. Since our sample also includes YSOs and giants, we decided to derive the $\tau$ value using a grid of stellar evolution models from the YalePotsdam Stellar Isochrones (YaPSI) following \citet{Lehtinen2020}. We did a fitting, by using the log$T_{\rm eff}$ and log$g$ from LAMOST DR7 and \citet{Anders2019}, to the model evolutionary tracks. 
The YAPSI models include five initial metallicities, [Fe/H] = $+$0.3, 0.0, $-$0.5, $-$1.0, and $-$1.5. 
For each star, we obtained best-fit models for these metallicities, and calculated the final $\tau$ value (and stellar radius) by linear interpolation to the metallicity from LAMOST DR7 and \citet{Anders2019}.

\begin{figure}[!htbp]
\center
\includegraphics[width=0.49\textwidth]{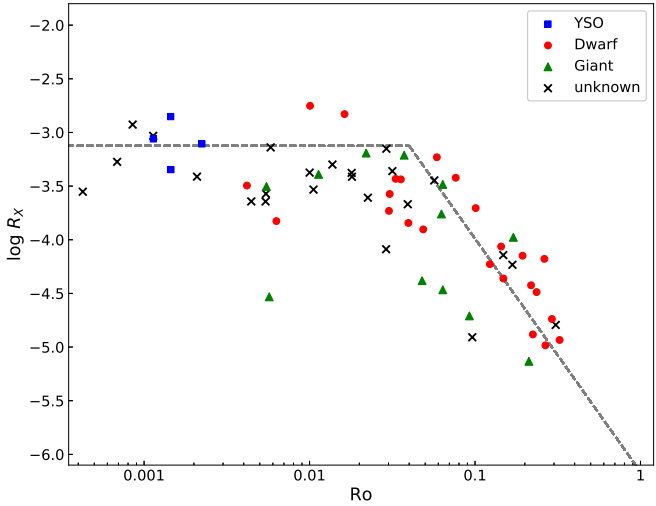}
\caption[]{Relation between $R_X$ and Ro for G, K, and M stars. The dashed line is from \citet{Wright2011}, but shifted to left with a ratio of Ro/3.}
\label{rxro.fig}
\end{figure}

\begin{figure}[!htbp]
\center
\includegraphics[width=0.23\textwidth]{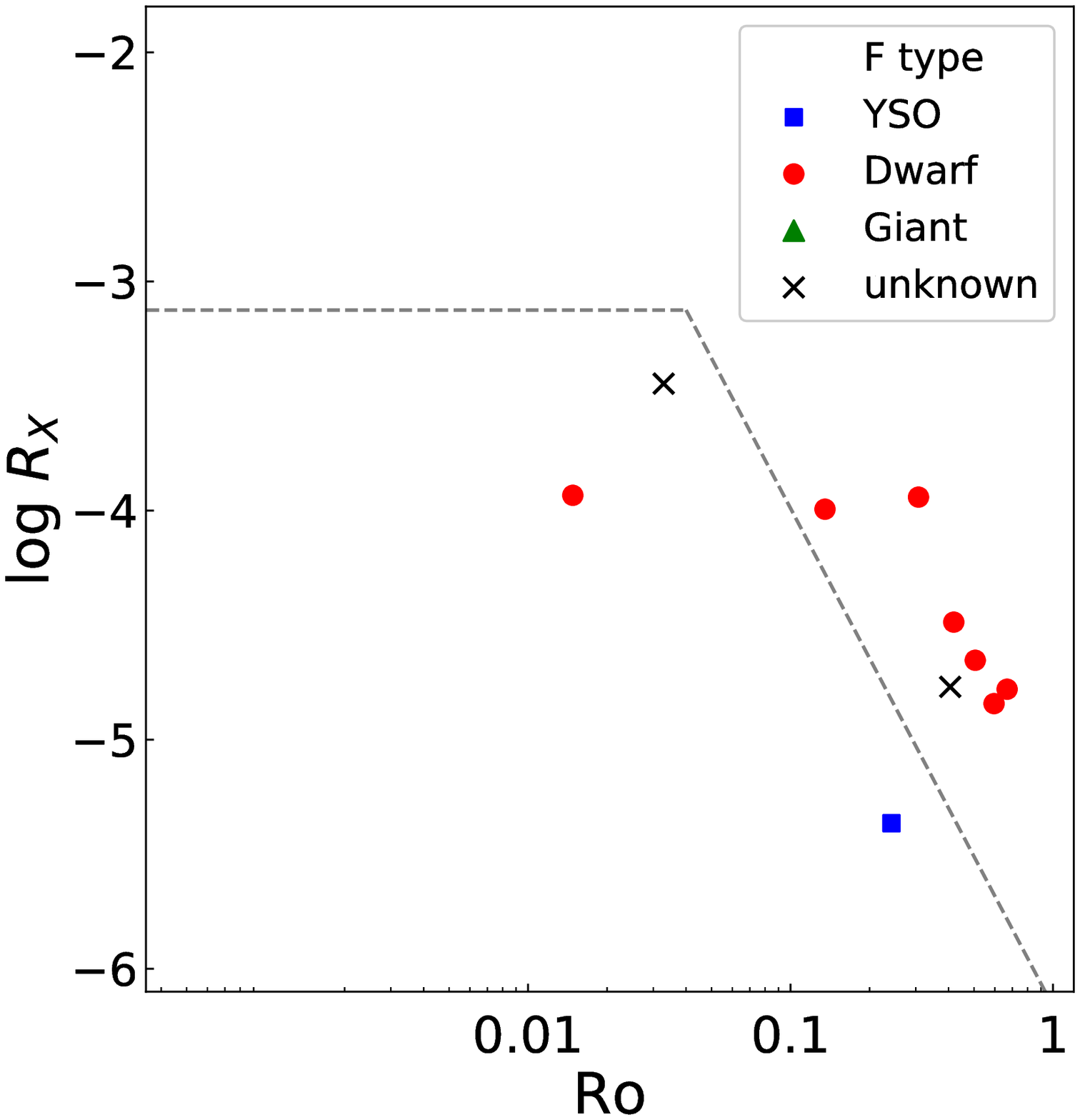}
\includegraphics[width=0.23\textwidth]{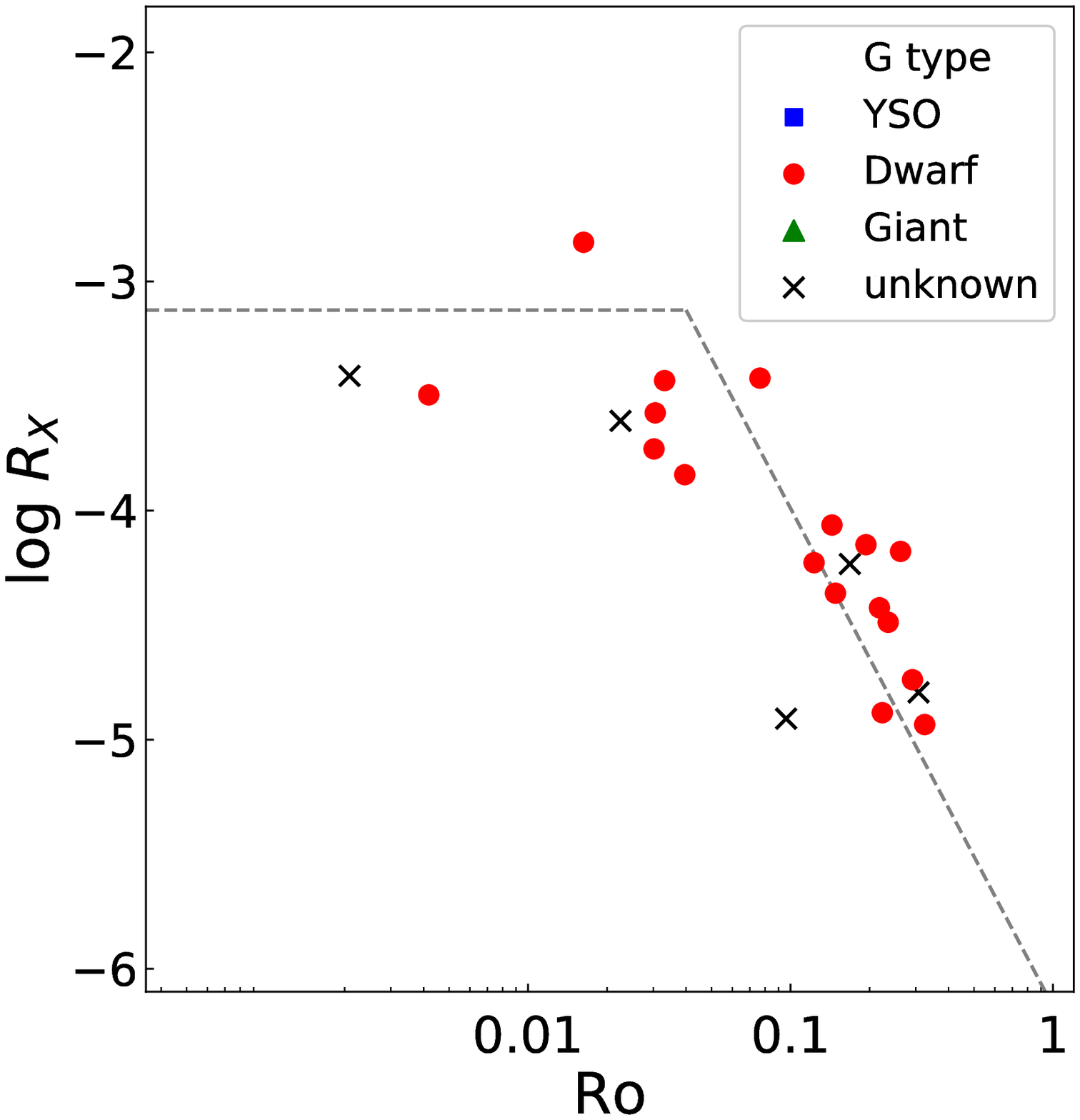}\\
\includegraphics[width=0.23\textwidth]{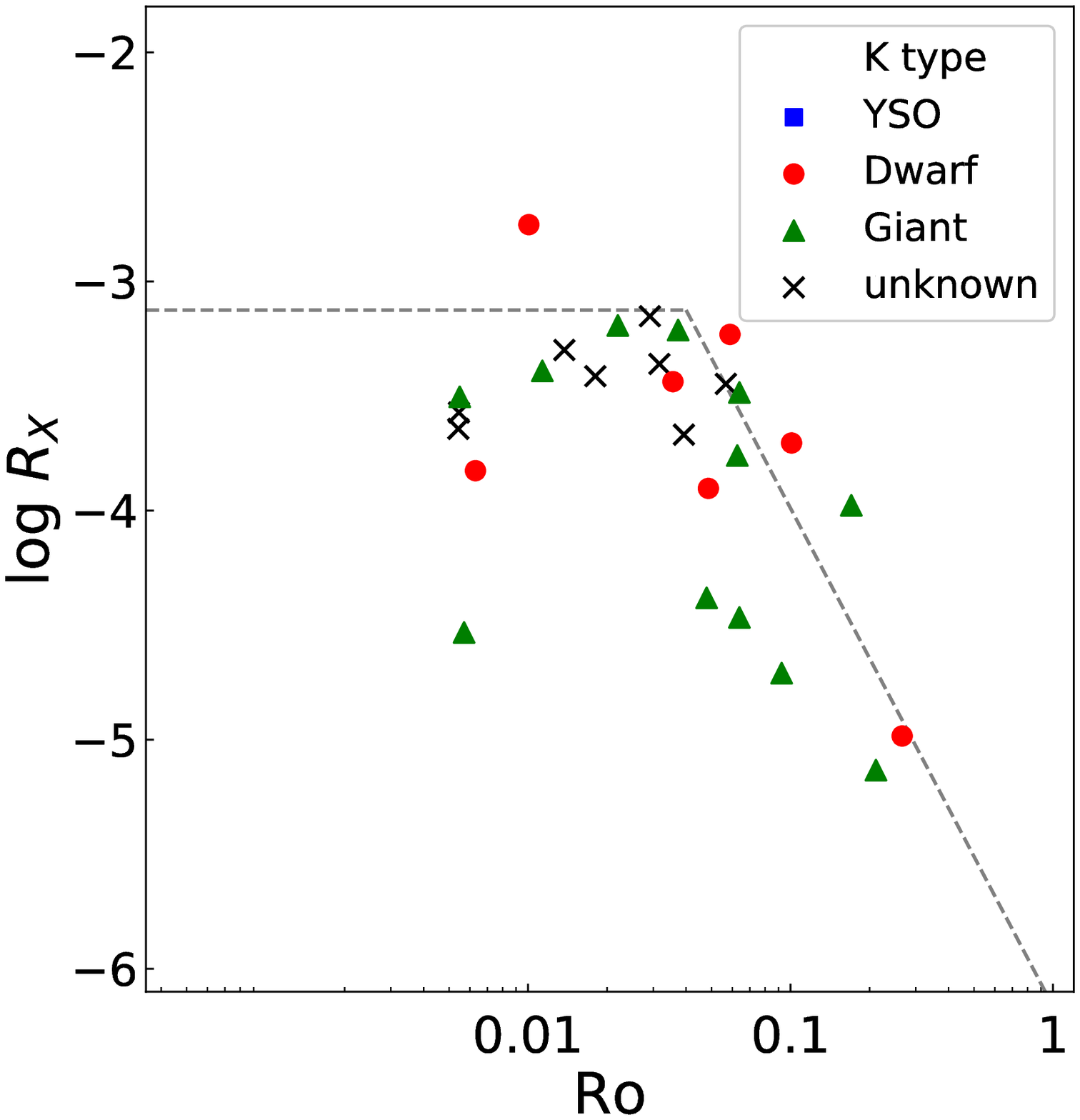}
\includegraphics[width=0.23\textwidth]{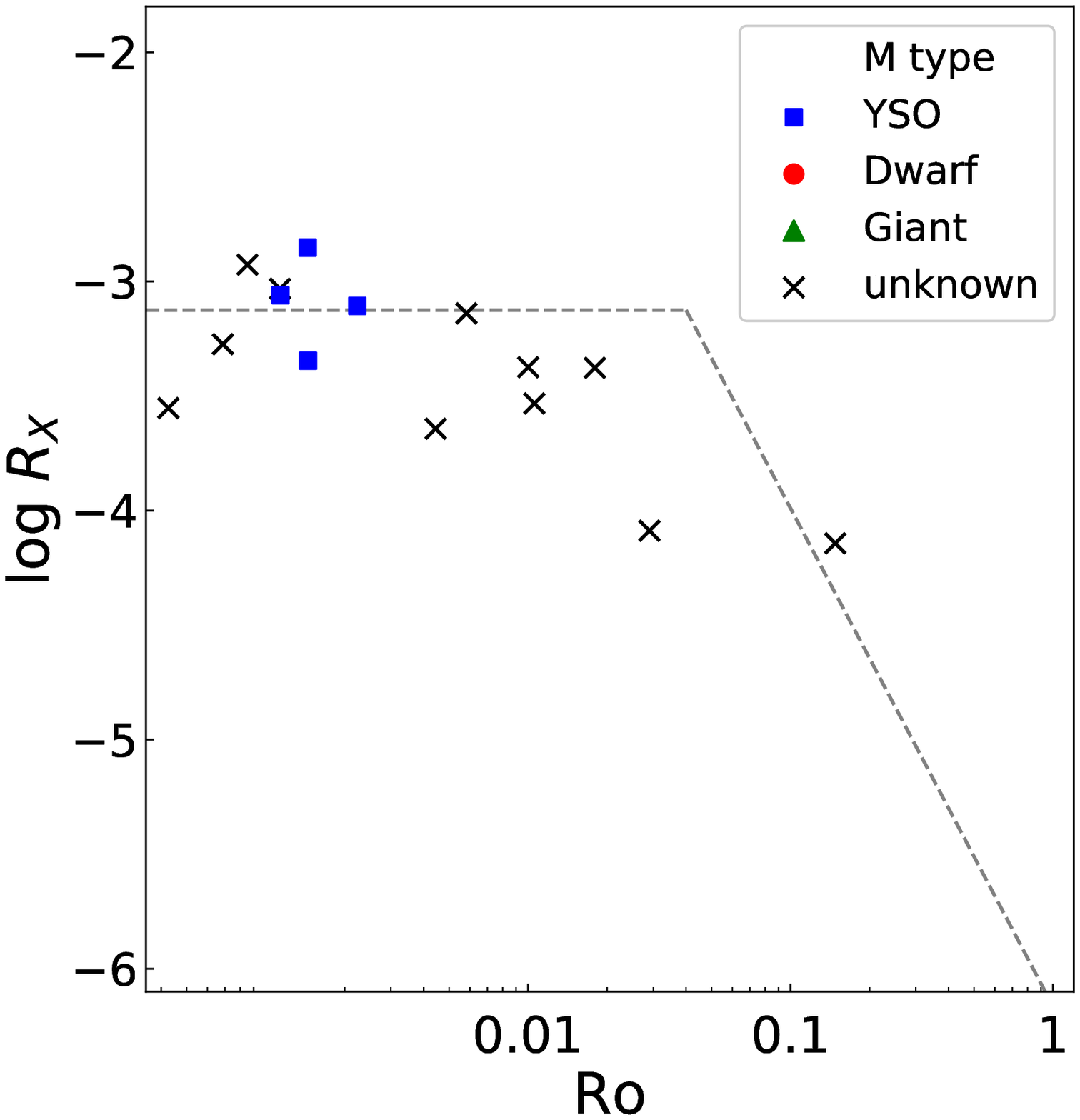}
\caption[]{Relation between $R_X$ and Ro for different type stars. The dashed line is from \citet{Wright2011}, but shifted to left with a ratio of Ro/3.}
\label{rxros.fig}
\end{figure}

Figure \ref{rxro.fig} shows $R_X$ as a function of Ro. The dashed line is from \citet{Wright2011}, but shifted to the left with a ratio of Ro/3, since we derived a ratio between the theoretical and empirical $\tau$ values around 3 (\citealt{Lehtinen2020} obtained a ratio of $\approx$2.6). For each stellar spectral type, the X-ray activity increases with decreasing rotation period, following a well-known activity--rotation relation (Figure \ref{rxros.fig}).
Most M stars are located in the saturated regime, which is usually explained by the magnetic/dynamo saturation \citep{Reiners2009} or a change in the dynamo configuration of the star\citep{Wright2011}. 

There are some theories for the saturation \citep[and super-saturation;][]{Prosser1996} of stellar X-ray emission: a saturation of the dynamo itself \citep[e.g.,][]{Gilman1983,Vilhu1987}, a saturation of the filling factor of active regions on the stellar surface \citep{Vilhu1984,Solanki1997,Stepien2001}, or centrifugal stripping of the corona \citep{Jardine1999}. 
The lack of observed saturation in chromospheric emission \citep{Cardini2007,Mamajek2008,Marsden2009,Jackson2010,
Argiroffi2016} are in contrast with the idea that the dynamo itself saturates. Analogously, the scenario of a saturation of the filling factor was argued against by some observations of rotational modulation of X-ray emission \citep{Marino2003} and small coronal filling factors \citep{Testa2004} in saturated/supersaturated stars. 
In addition, compact and dense coronal loops are found far below the corotation radius of the supersaturated star VW Cephei \citep{Huenemoerder2006}, indicating that saturation is not necessarily caused by coronal stripping \citep{Wright2011}.
These could suggest that saturation is due to a change in regime of the underlying dynamo.

\begin{figure*}[!htbp]
\center
\includegraphics[width=0.49\textwidth]{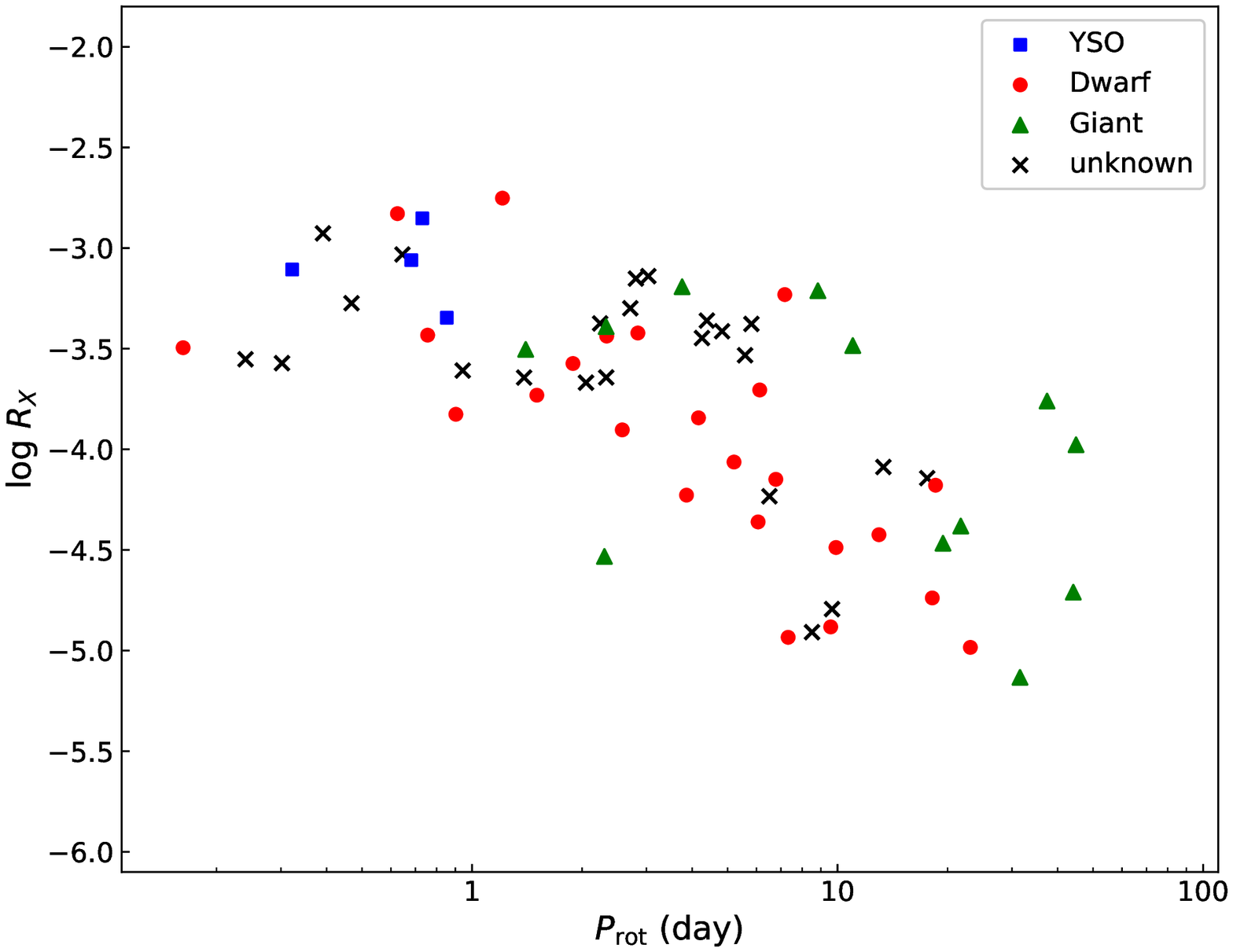}
\includegraphics[width=0.49\textwidth]{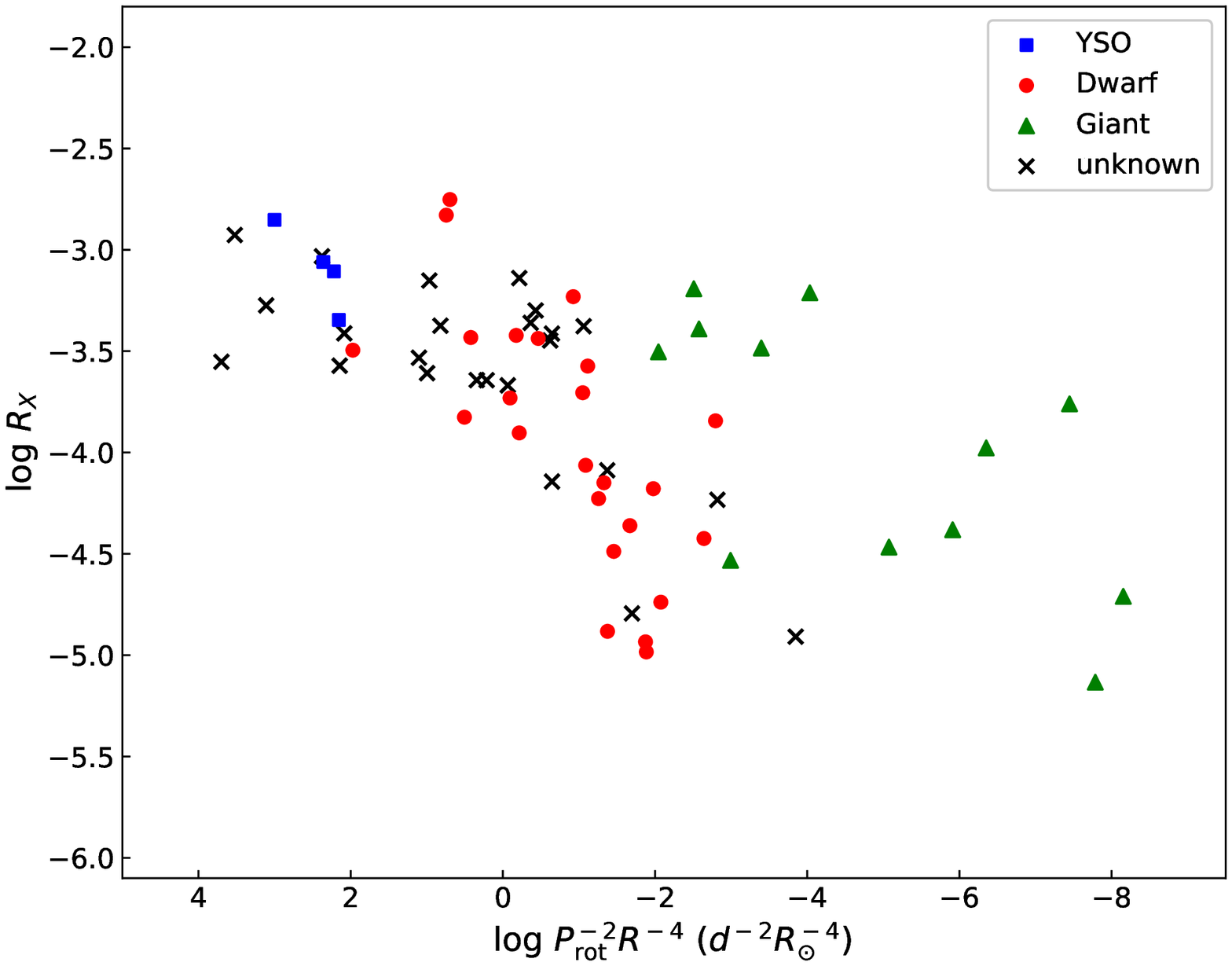}
\caption[]{Left Panel: relation between $R_X$ and rotational period for G, K, and M stars. Right Panel: relation between $R_X$ and $P_{\rm rot}^{-2}R^{-4}$ for G, K, and M stars.}
\label{rop.fig}
\end{figure*}

Some recent studies \citep[e.g.,][]{Reiners2014} reported that the rotation period alone suffices to determine the activity-rotation scaling, and the
relation $L_X$/$L_{\rm bol}$ $\propto$ $P_{\rm rot}^{-2}R^{-4}$ optimally describes the non-saturated fraction of stars.
However, \citet{Lehtinen2020} found that the dwarfs and giants   are located in a single sequence in the unsaturated regime in relation to Rossby numbers, while they are clearly separated in the relations $R'_{HK}$ versus $P_{\rm rot}$ and $R'_{HK}$ versus $P_{\rm rot}^{-2}R^{-4}$.
In our sample, the giants, with longer rotation periods, do not well follow the relation $R_X$ versus $P_{\rm rot}^{-2}R^{-4}$ for dwarfs (Figure \ref{rop.fig}).
More stars with accurate period estimations will help examine these relations.

\subsection{Comparison with UV activity}

A number of proxies have been used to study the activity occurring in the photosphere (star spots), chromosphere (Ca II HK; Mg II; H$\alpha$; optical flares; UV flux),
and corona (X-ray emission).
Observations have revealed good correlations between some of these proxies
\citep[e.g., ][]{Mamajek2008,Stelzer2013},
and the empirical scalings of these proxies with rotation period or Rossby number
\citep[e.g.,][]{Pallavicini1981,Pizzolato2003,Wright2011}.

To have a comparison with the chromospheric activity,
we calculated the UV activity index as following:
\begin{equation}
R^\prime_{\rm UV} = \frac{f_{\rm UV,exc}}{f_{\rm bol}} = \frac{f_{\rm UV,obs} - f_{\rm UV,ph}}{f_{\rm bol}},
\label{eq:act_index}
\end{equation}
where `UV' stands for the NUV and FUV bands, respectively.
Here $f_{\rm UV,exc}$ is the UV excess flux due to activity.
The observed UV flux $f_{\rm UV,obs}$ was estimated from the {\it GALEX} magnitude,
by using the transformation relations \footnote{https://asd.gsfc.nasa.gov/archive/galex/FAQ/counts\_backgr-
ound.html} as
\begin{equation}
f_{\rm FUV,obs} = 10^{0.4\times(18.82-M_{\rm FUV})}\times 1.40\times10^{-15} \times \delta\lambda_{\rm FUV}
\label{eq:fuv}
\end{equation}
and
\begin{equation}
f_{\rm NUV,obs} = 10^{0.4\times(20.08-M_{\rm NUV})}\times 2.06\times10^{-16} \times \delta\lambda_{\rm NUV}.
\label{eq:nuv}
\end{equation}
The $M_{\rm FUV}$ and $M_{\rm NUV}$ are absolute magnitudes, 
obtained from the observed magnitudes adopting the distance from $Gaia$ DR2 and the extinction from the PS1 3D extinction map (See Section \ref{sample.sec}). The extinction coefficients were calculated as 8.11 (FUV) and 8.71 (NUV) following \citet{Cardelli1989}.
The $\delta \lambda_{\rm FUV}$ and $\delta \lambda_{\rm NUV}$ are the
effective band width\footnote{http://galexgi.gsfc.nasa.gov/docs/galex/Documents/ERO\_
data\_description\_2.htm}
of the FUV and NUV filters (268 \AA~and 732 \AA), respectively.
The photospheric flux $f_{\rm UV,ph}$, which means
the photospheric contribution to the FUV and NUV emission,
were estimated with the PARSEC model.
Using the best fit model (see Appendix C), we derived the FUV and NUV magnitudes attributed to photospheric emission.
The PARSEC models present the {\it GALEX} absolute magnitudes in VEGA system, therefore we first converted them into AB magnitudes \citep{Bianchi2011} and then converted them into fluxes using Equations (\ref{eq:fuv}) and (\ref{eq:nuv}).
The bolometric flux $f_{\rm bol}$ was obtained with the bolometric luminosity from the best fit PARSEC model.
We calculated it using a distance of 10 pc.
Table \ref{uv.tab} lists the UV activity indexes.

\begin{figure}[!htbp]
\center
\includegraphics[width=0.23\textwidth]{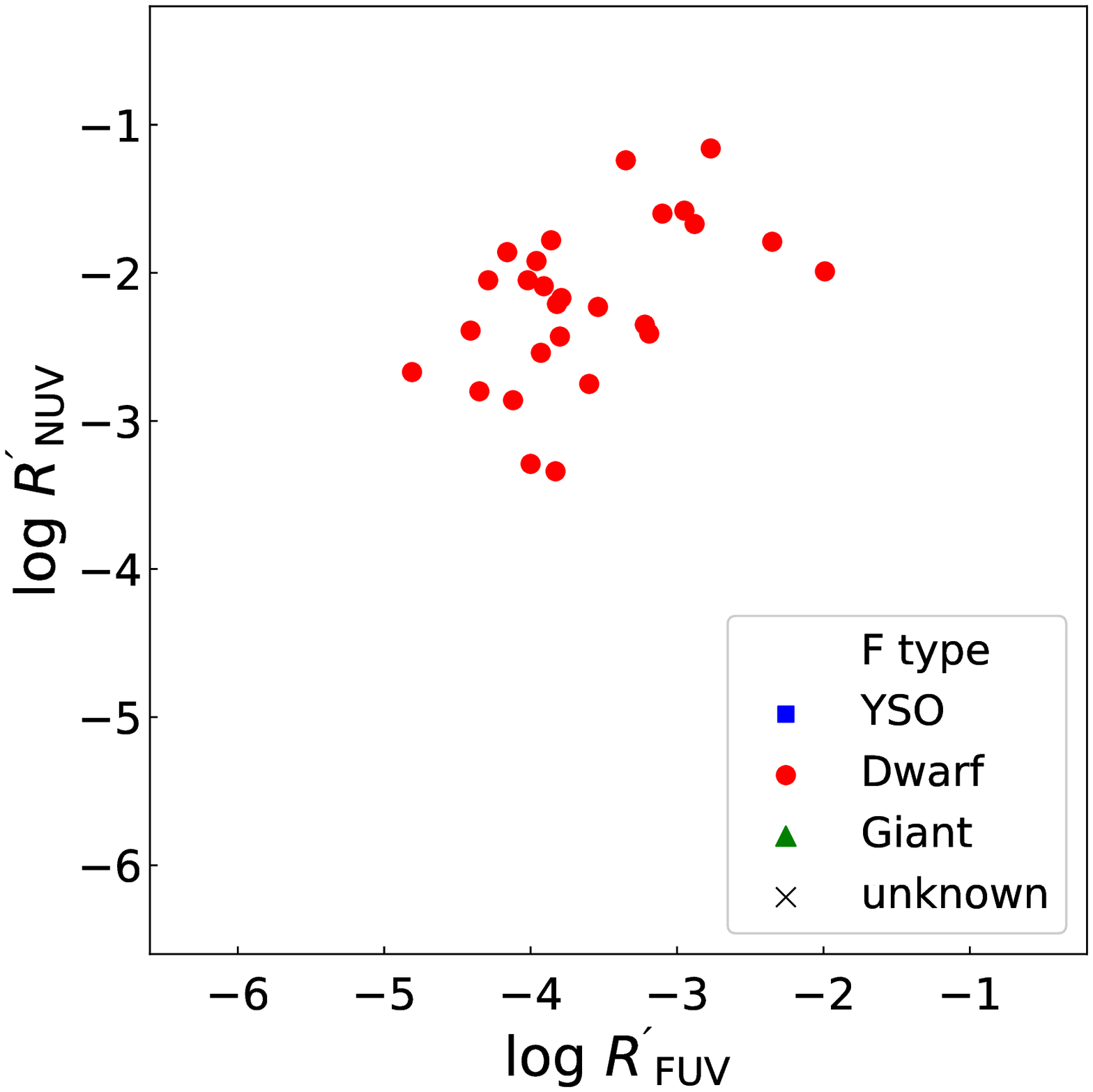}
\includegraphics[width=0.23\textwidth]{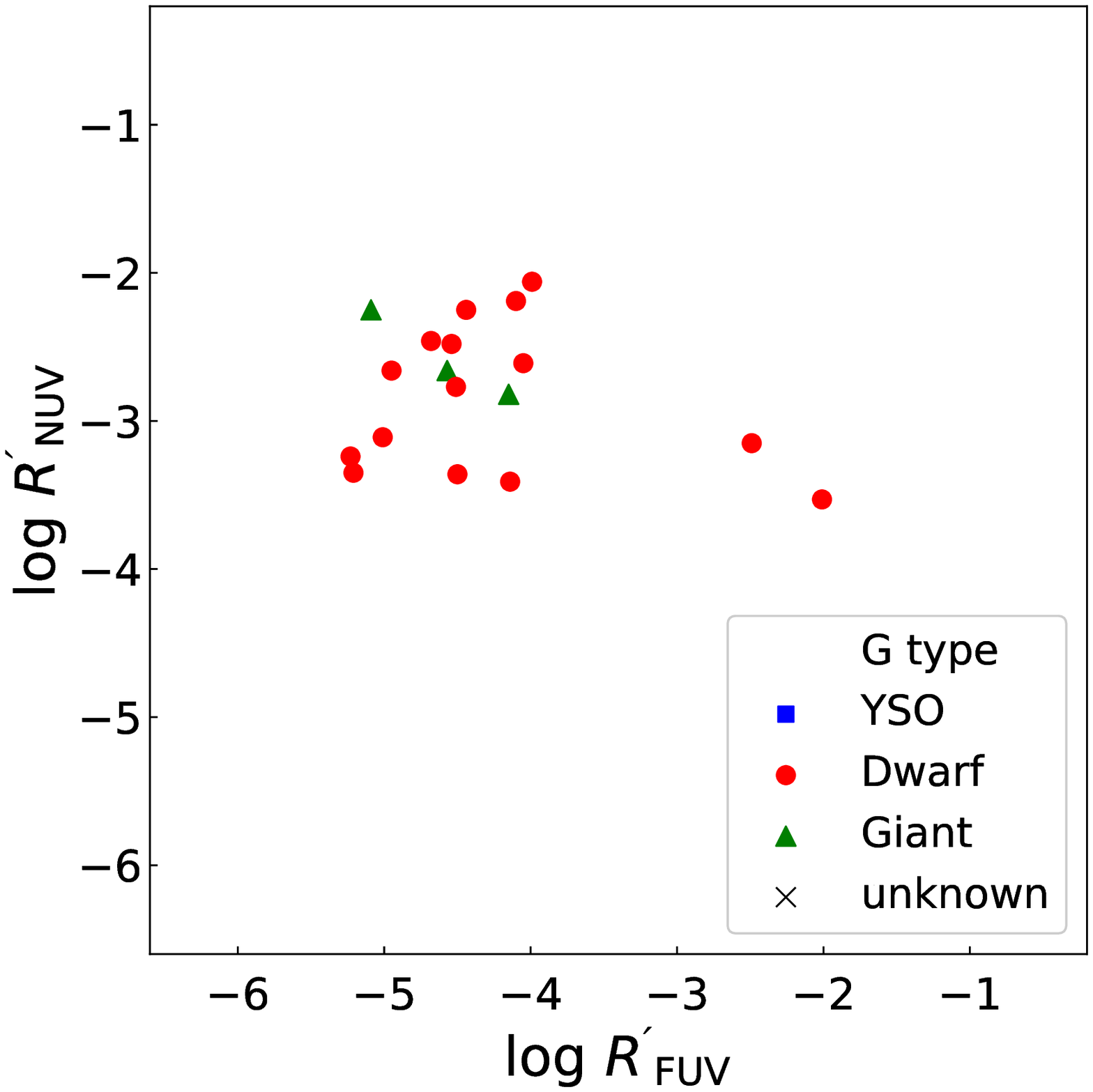}\\
\includegraphics[width=0.23\textwidth]{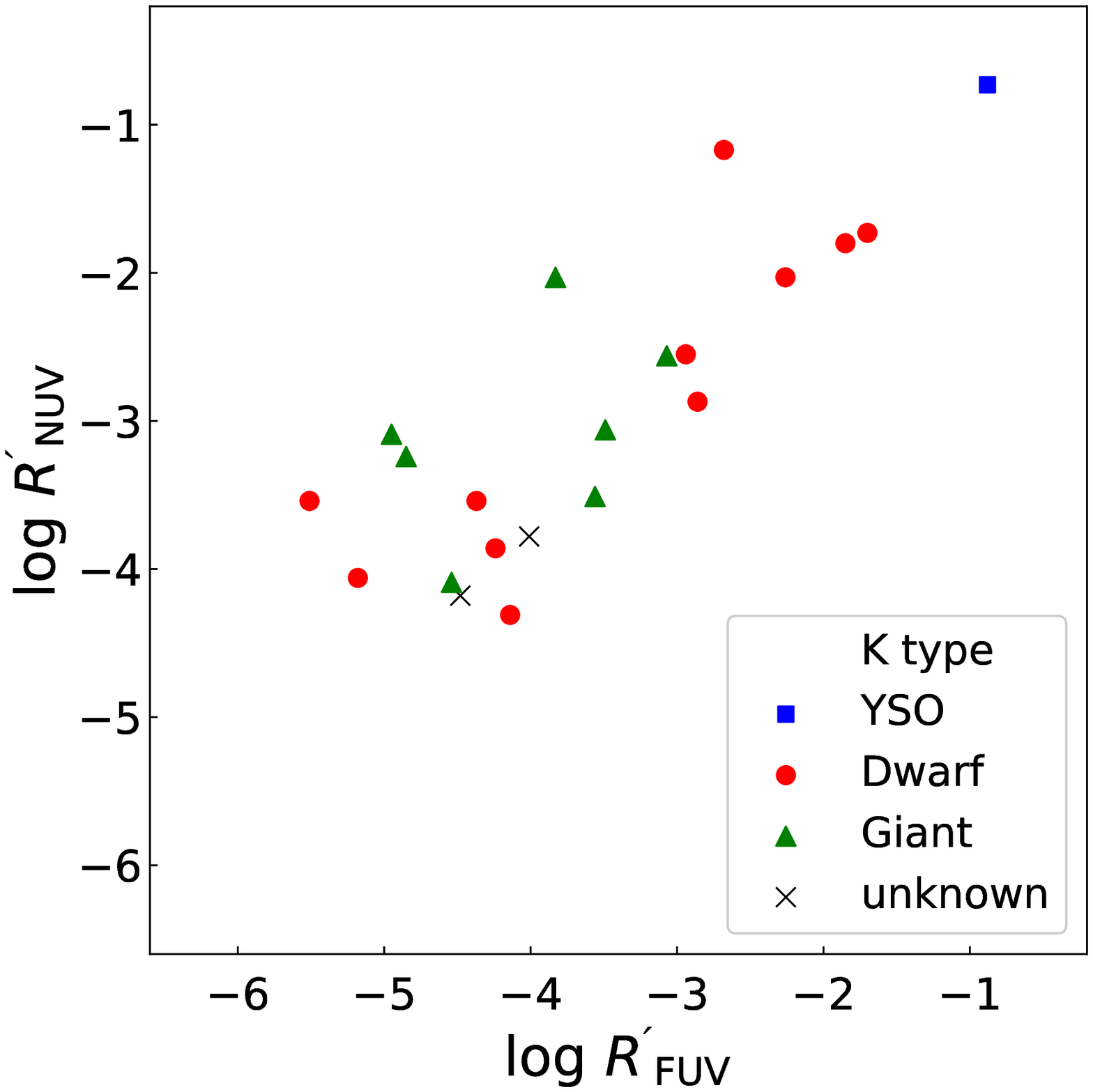}
\includegraphics[width=0.23\textwidth]{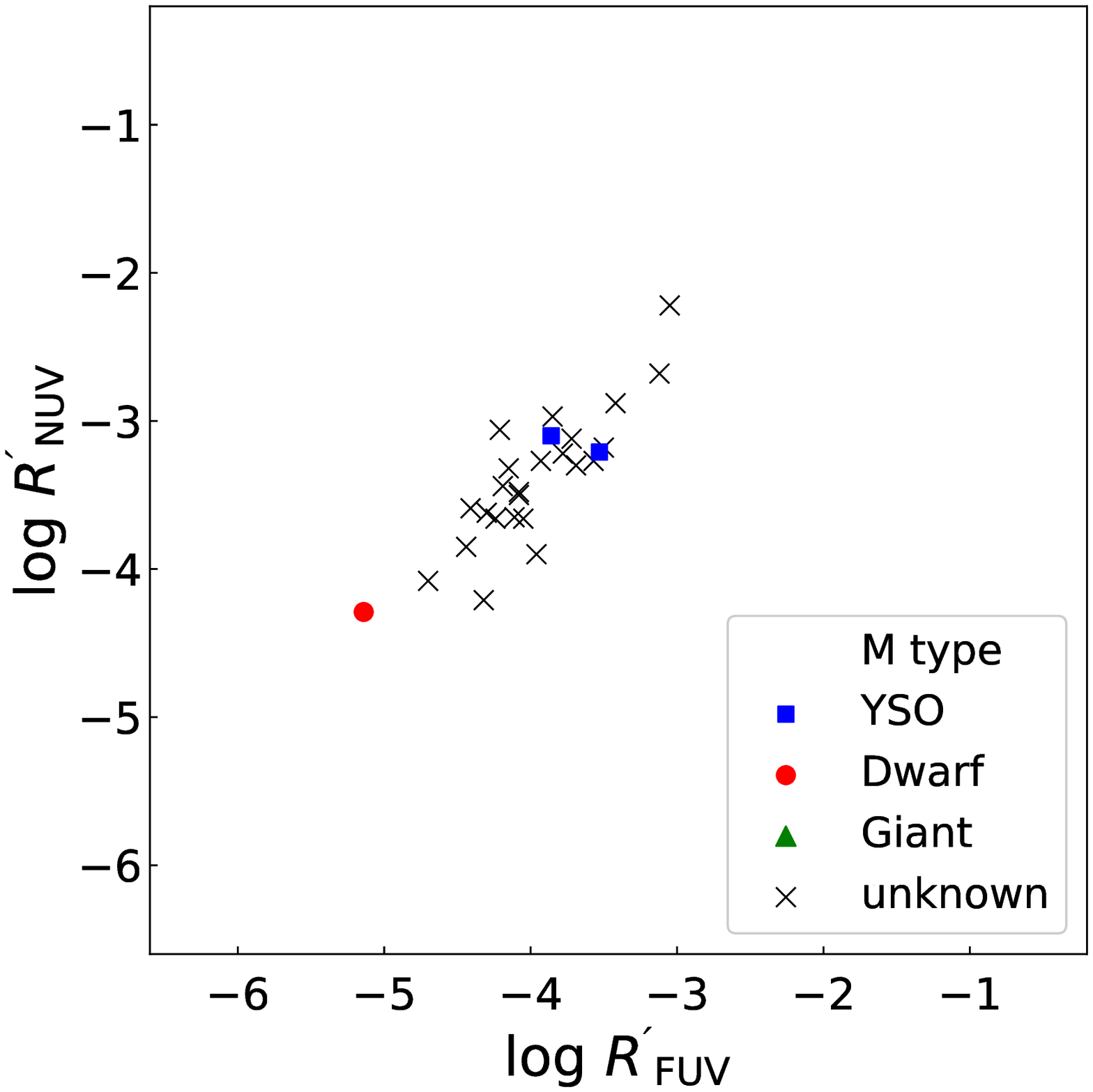}
\caption[]{Relation between $R^\prime_{\rm NUV}$ and $R^\prime_{\rm FUV}$ for stars of different spectral types.}
\label{FUVNUV.fig}
\end{figure}

\begin{figure}[!htbp]
\center
\includegraphics[width=0.49\textwidth]{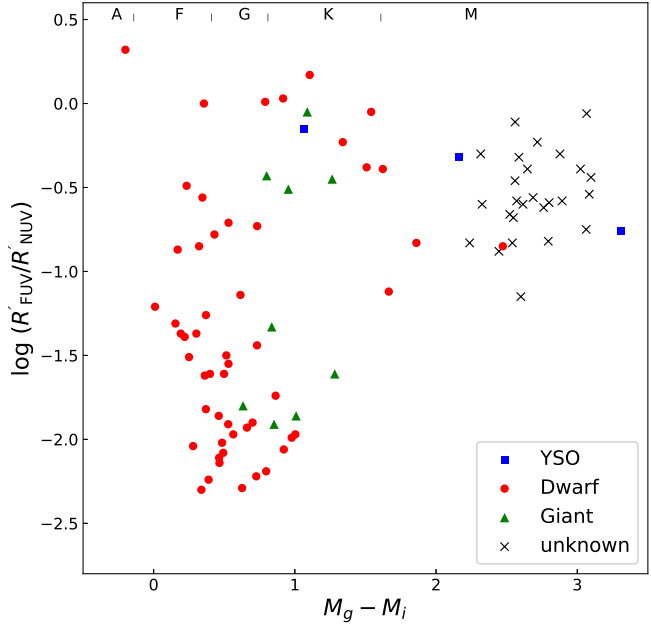}
\caption[]{Ratio of FUV and NUV activities as a function of color $M_g - M_i$.}
\label{fuv2nuv.fig}
\end{figure}

\begin{figure}[!htbp]
\center
\includegraphics[width=0.23\textwidth]{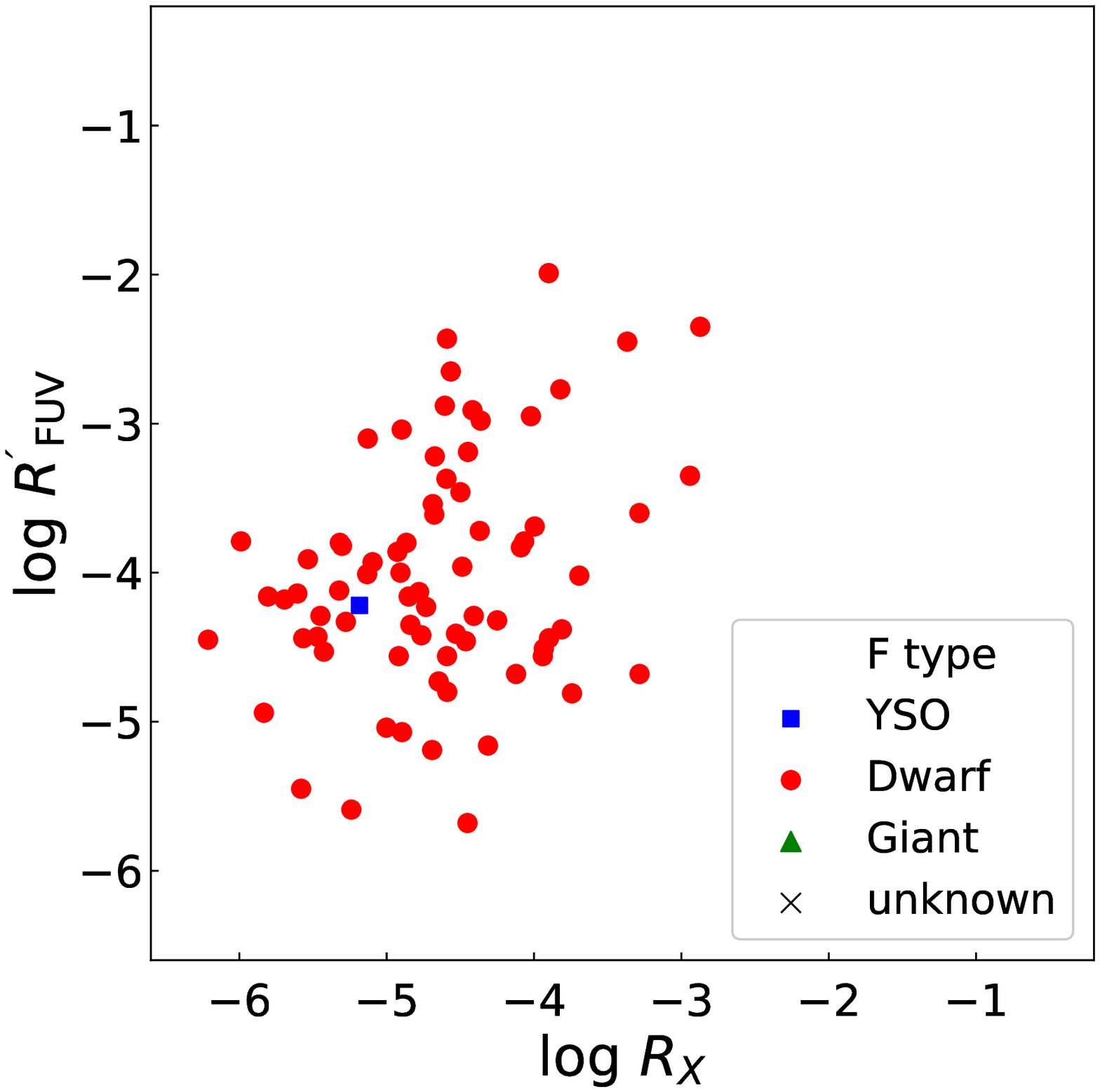}
\includegraphics[width=0.23\textwidth]{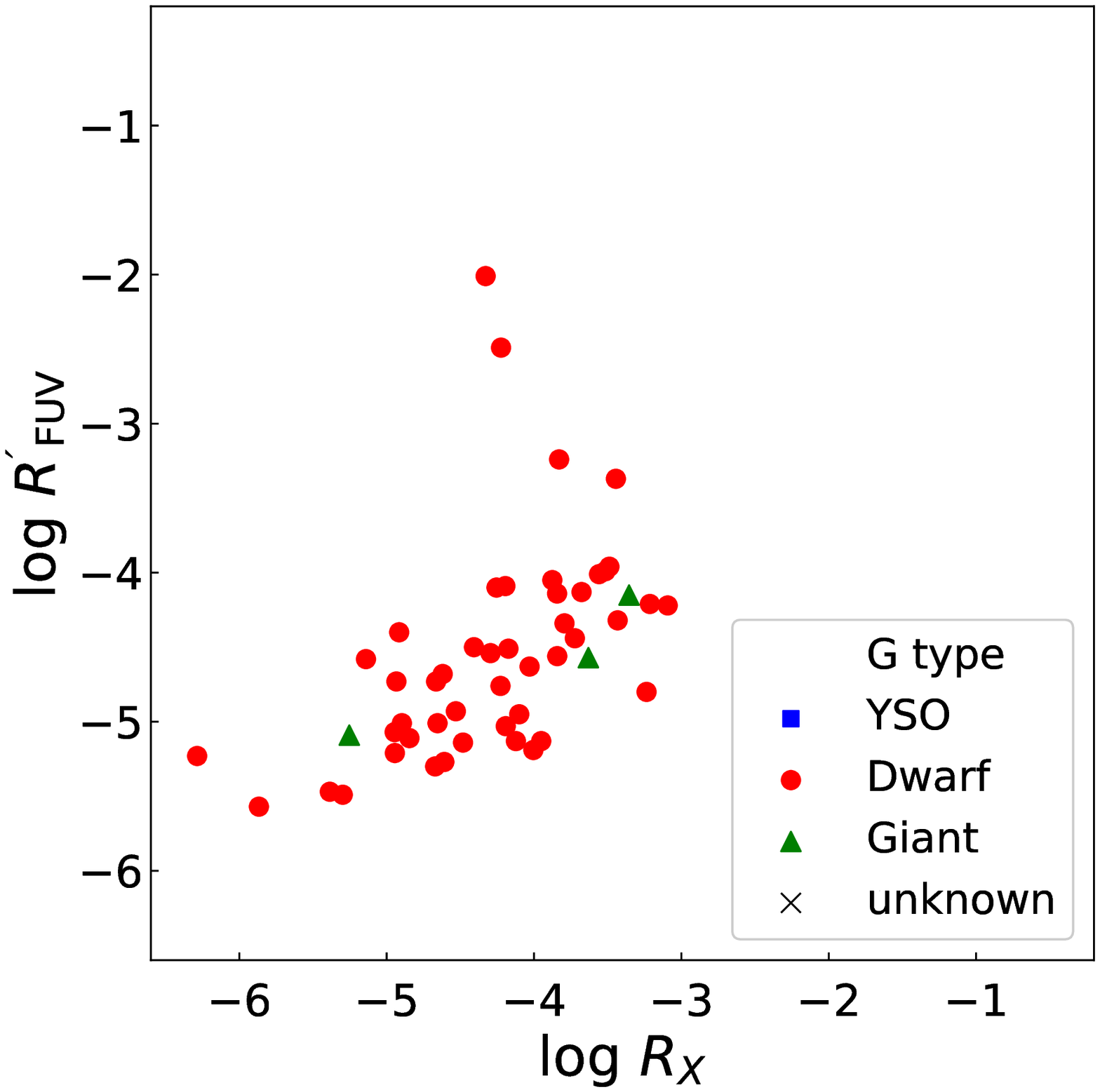}\\
\includegraphics[width=0.23\textwidth]{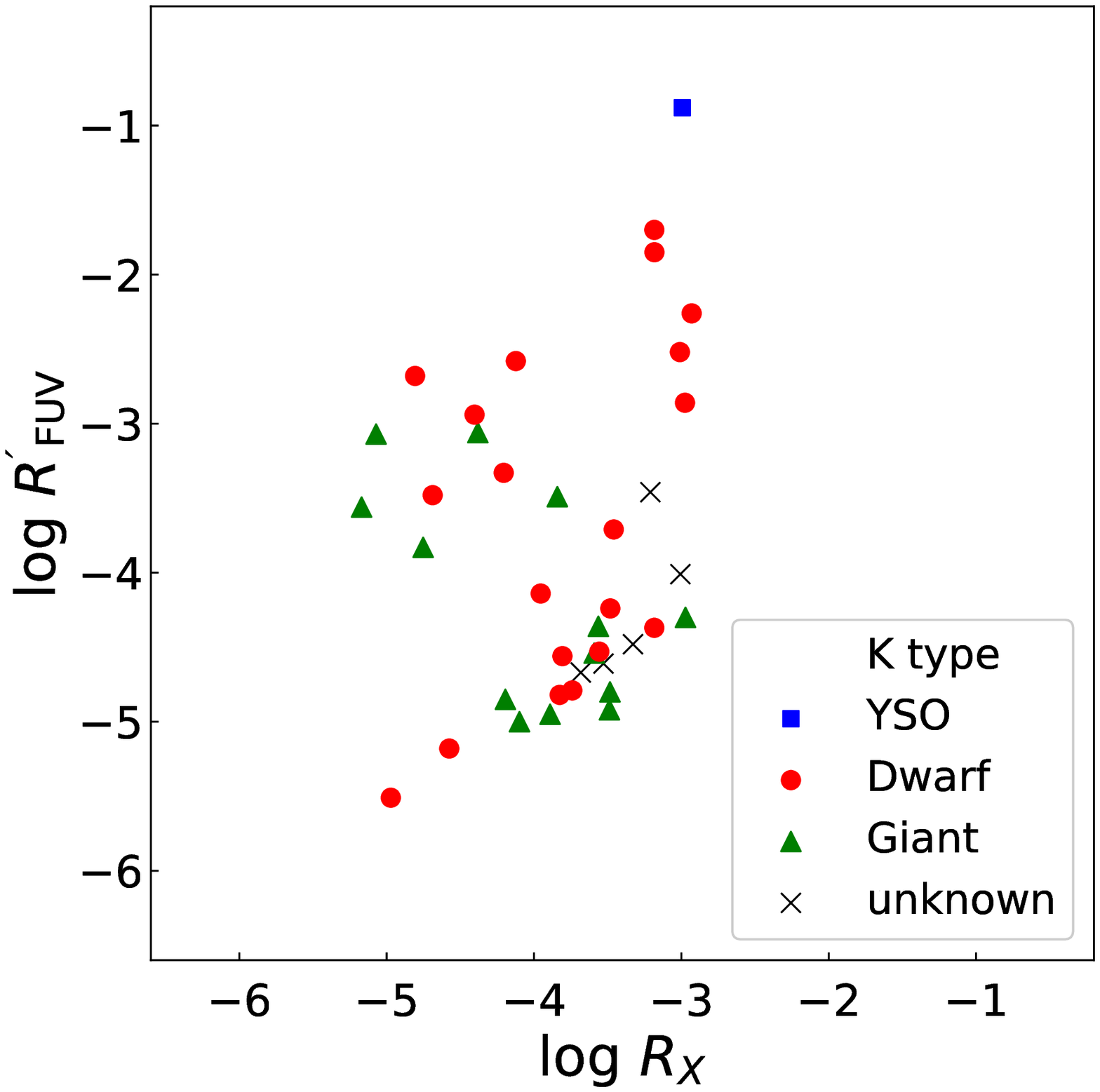}
\includegraphics[width=0.23\textwidth]{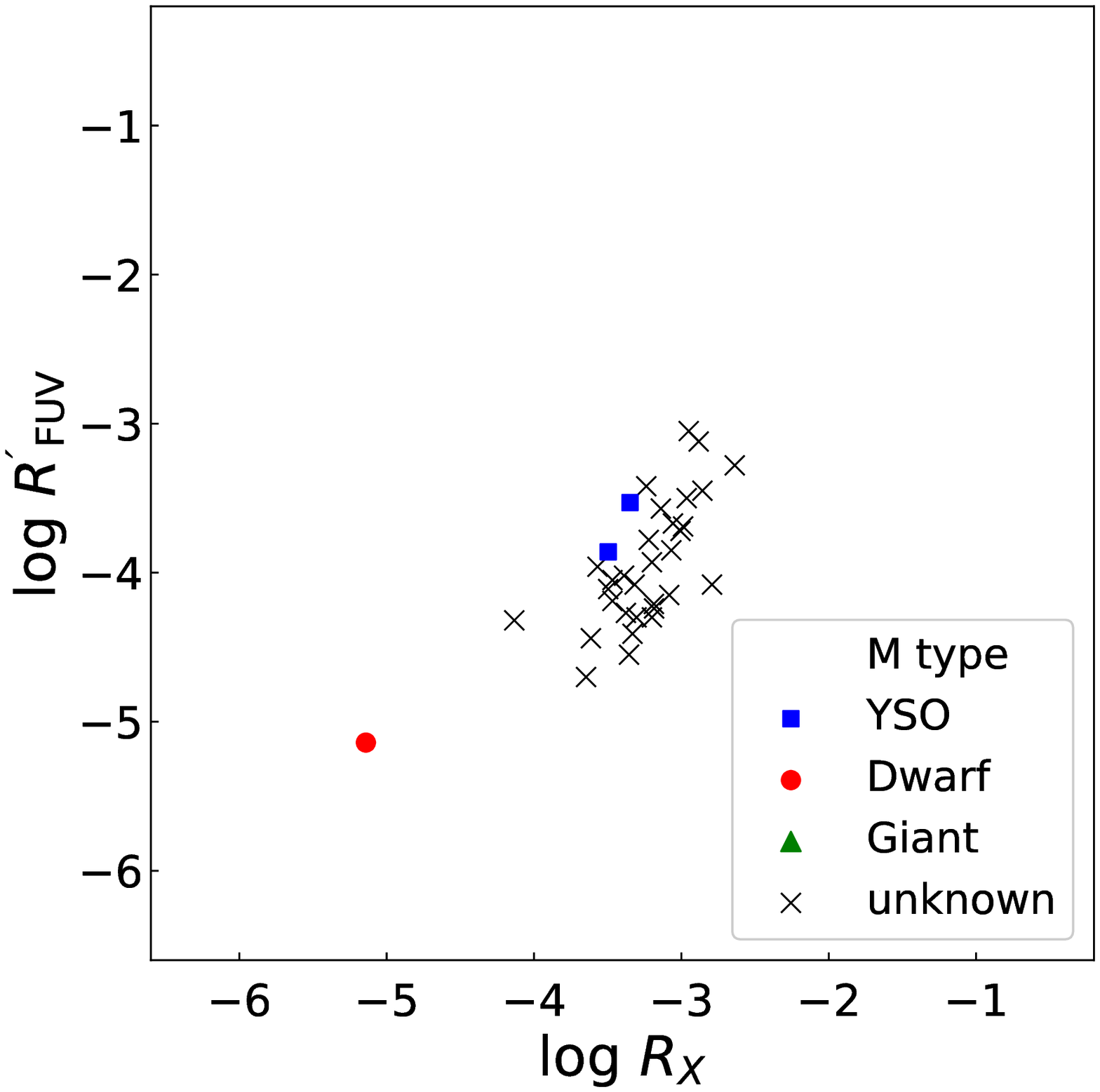}
\caption[]{Relation between $R^\prime_{\rm FUV}$ and $R_X$ for stars of different spectral types.}
\label{RxFUV.fig}
\end{figure}

\begin{figure}[!htbp]
\center
\includegraphics[width=0.23\textwidth]{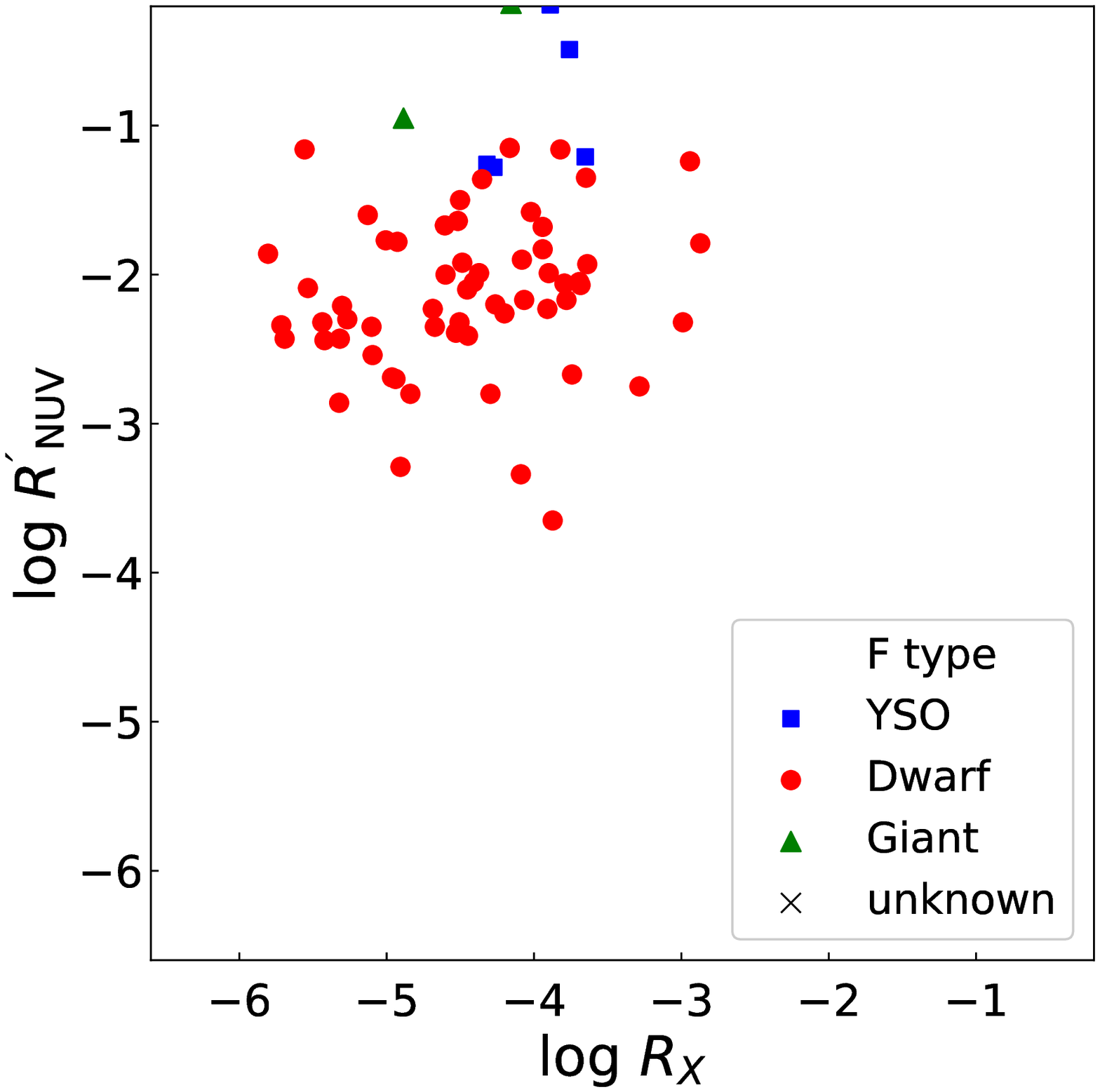}
\includegraphics[width=0.23\textwidth]{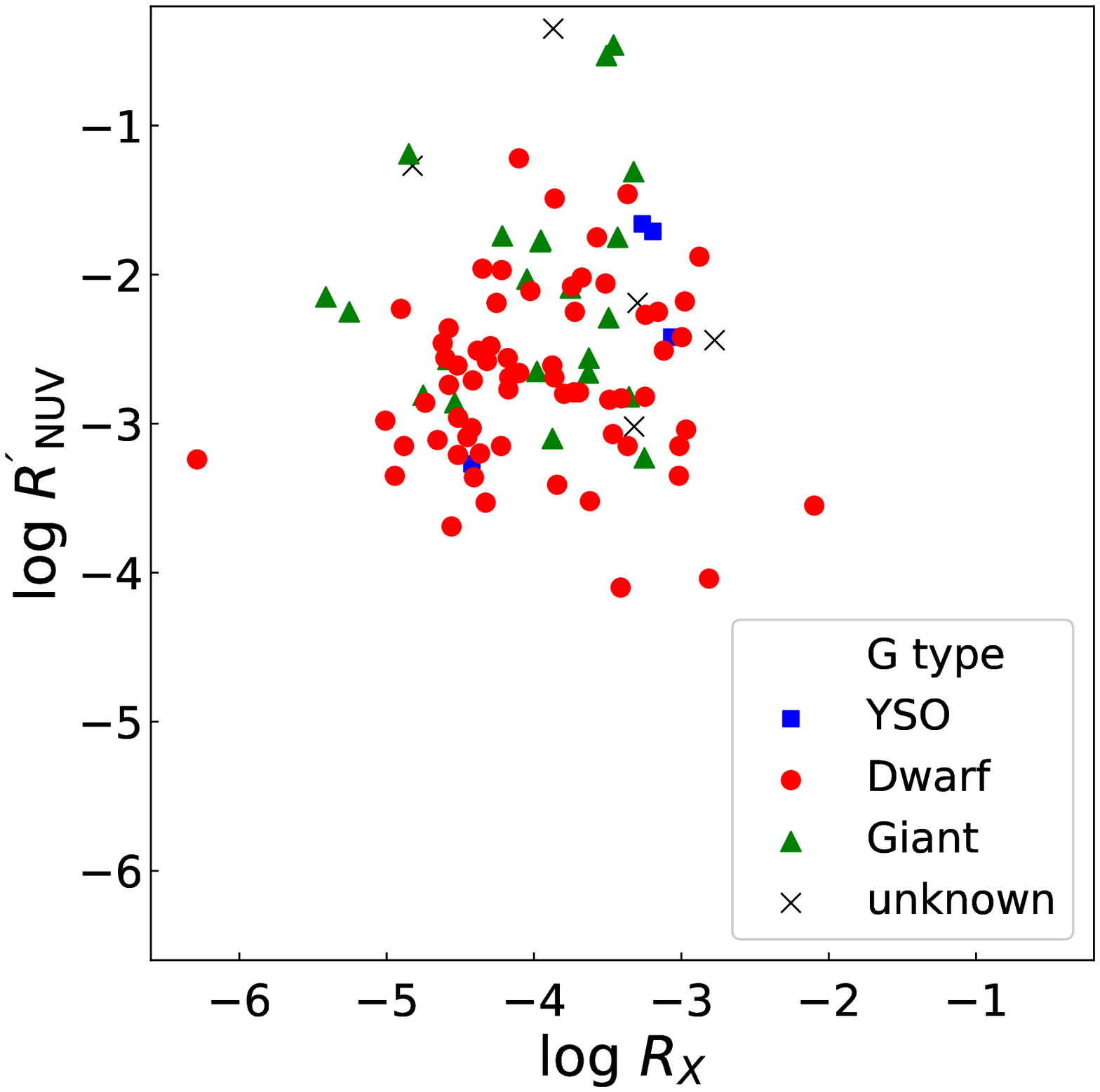}\\
\includegraphics[width=0.23\textwidth]{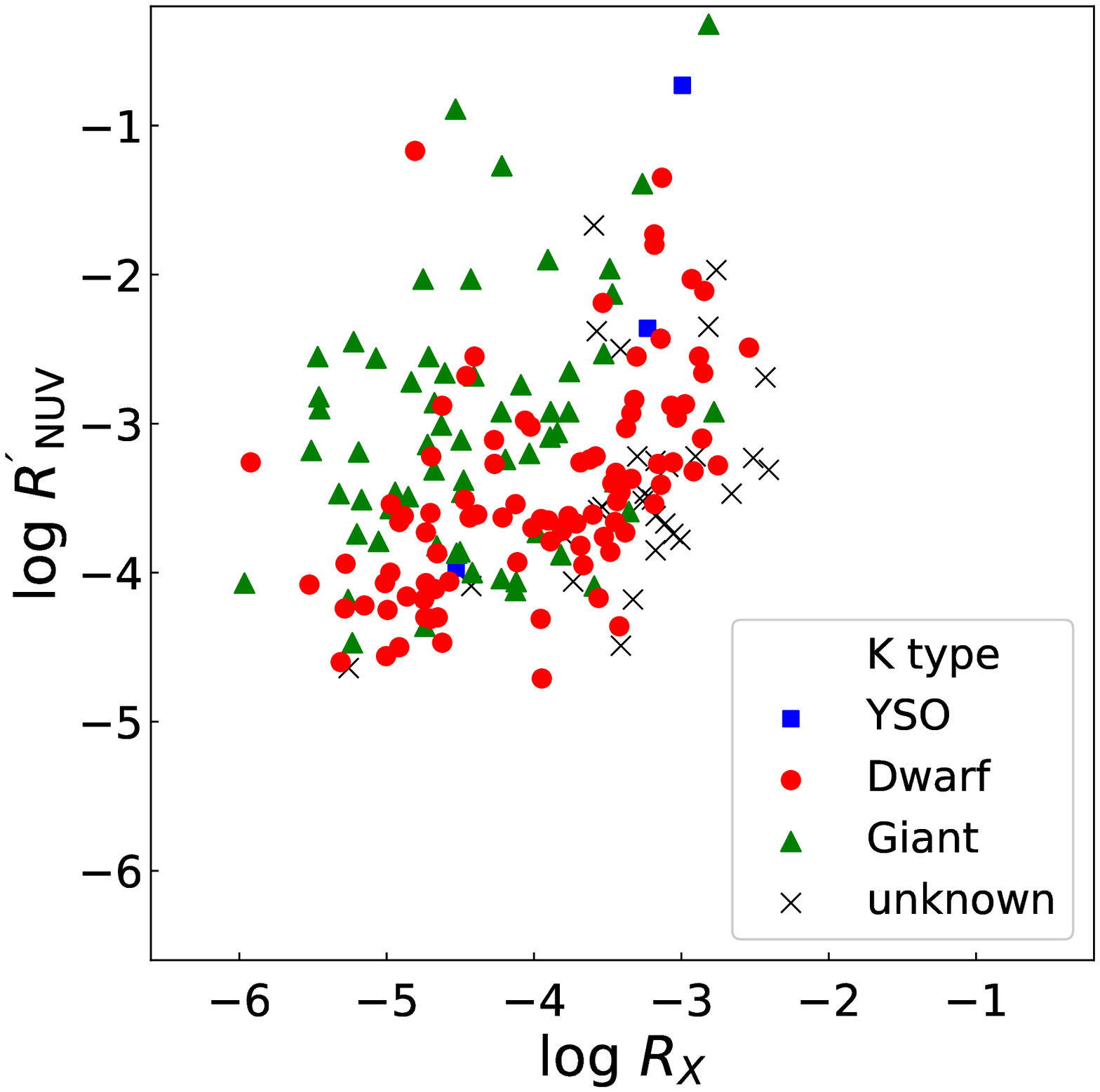}
\includegraphics[width=0.23\textwidth]{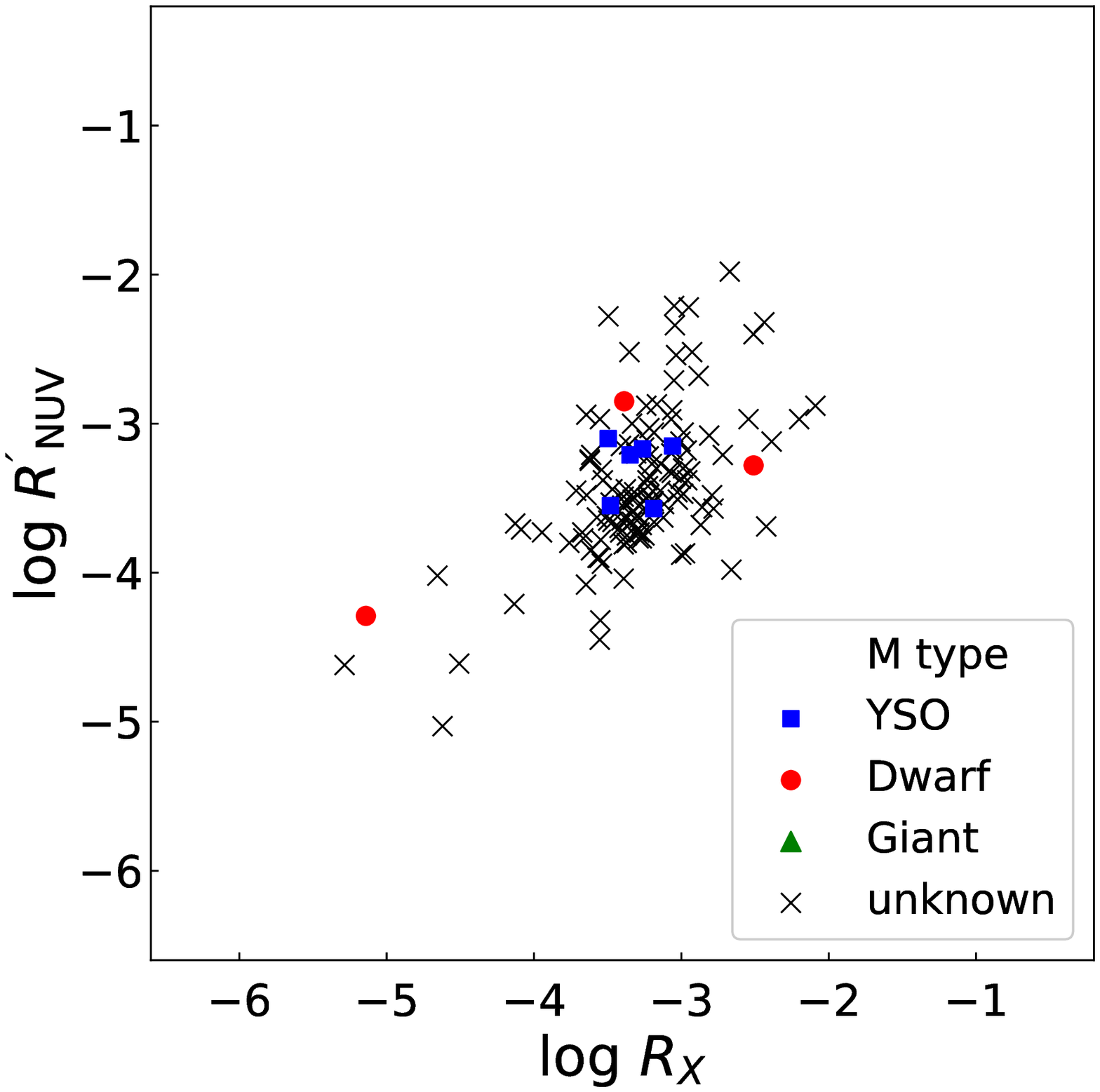}
\caption[]{Relation between $R^\prime_{\rm NUV}$ and $R_X$ stars of different spectral types.}
\label{RxNUV.fig}
\end{figure}

We searched for the relations between FUV and NUV activities for different type stars (Figure \ref{FUVNUV.fig}).
For F- and G-type stars, there is no clear relation, while
for K and M stars, the FUV and NUV fluxes show an obvious positive relation \citep{Stelzer2013}.
By using different stellar models (i.e., BT-Cond grid), \citet{Bai2018} calculated the UV excesses emission for millions of stars, and their sample showed similar trends.
The lack of clear relation for F and G stars may be explained as the mixed populations with different ages.
As can be seen in Figure \ref{fuv2nuv.fig}, the F and G stars with lower FUV/NUV ratio may be older ones, while the young population show higher ratio of log(${R^{\prime}}_{\rm FUV}/{R^{\prime}}_{\rm NUV}$) $\gtrsim$ $-$1
\citep[see also Figure 12 in][]{Richey-Yowell2019}.
Similar trends can be found in the comparisons between X-ray and UV activities (Figure \ref{RxFUV.fig} and \ref{RxNUV.fig}). For G, K, and M stars, a positive relation can be seen between $R'_X$ and $R'_{\rm FUV}$, while only M stars show an obvious relation between $R'_X$ and $R'_{\rm NUV}$.

\begin{table}
\begin{center}
\renewcommand{\arraystretch}{1.5}
\scriptsize
\setlength{\tabcolsep}{1.5pt}
\caption[]{Stars with UV emission in our sample.}
\label{uv.tab}
\begin{tabular}{lcccc}
\hline\noalign{\smallskip}
Object  &   log$f_{\rm FUV,exc}$    &   log$f_{\rm NUV,exc}$   &   log$R^\prime_{\rm FUV}$ & log$R^\prime_{\rm NUV}$  \\
    & (erg cm$^{-2}$  &  (erg cm$^{-2}$   &      &          \\
       & s$^{-1}$)  &  s$^{-1}$)   &      &          \\
  (1)   &     (2)  &    (3)      &         (4)                        &      (5)        \\
\hline\noalign{\smallskip}
J000136.1+130639 & 6.12$\pm$0.73 & 8.25$\pm$0.07 & -4.81$\pm$0.73 & -2.67$\pm$0.07\\
J000238.8+255219 & ... & 7.03$\pm$0.96 & ... & -3.47$\pm$0.96\\
J000833.8+512412 & 8.81$\pm$0.04 & ... & -2.66$\pm$0.04 & ...\\
J000835.3+512142 & 7.62$\pm$0.08 & 9.13$\pm$0.01 & -2.68$\pm$0.08 & -1.17$\pm$0.01\\
J001325.4+791537 & 6.60$\pm$0.16 & 8.14$\pm$0.05 & -4.35$\pm$0.16 & -2.80$\pm$0.05\\
J001507.9-302342 & ... & 8.34$\pm$0.44 & ... & -2.44$\pm$0.44\\
J001801.8+300816 & ... & 7.77$\pm$0.05 & ... & -2.86$\pm$0.05\\
J001944.6+591349 & ... & 10.71$\pm$0.38 & ... & -0.49$\pm$0.38\\
J002023.5+591444 & ... & 8.54$\pm$0.05 & ... & -2.43$\pm$0.05\\
J002101.6+591518 & ... & 8.40$\pm$0.41 & ... & -2.03$\pm$0.41\\
J002111.3-084140 & 6.53$\pm$0.06 & 6.98$\pm$0.03 & -3.12$\pm$0.06 & -2.68$\pm$0.03\\
J002344.4+641111 & ... & 10.90$\pm$0.28 & ... & -0.19$\pm$0.28\\
J002410.3-020127 & ... & 6.26$\pm$0.49 & ... & -3.95$\pm$0.49\\
J002438.4+641122 & ... & 12.04$\pm$0.02 & ... & -0.05$\pm$0.02\\
J002519.7-123303 & 5.83$\pm$0.47 & 6.66$\pm$0.14 & -4.37$\pm$0.47 & -3.54$\pm$0.14\\
J002537.0-121450 & ... & 6.02$\pm$0.35 & ... & -4.31$\pm$0.35\\
J002609.9+171559 & ... & 6.19$\pm$0.24 & ... & -3.93$\pm$0.24\\
J002611.2+171234 & 5.54$\pm$0.29 & ... & -5.09$\pm$0.29 & ...\\
J002756.6+261651 & ... & 7.56$\pm$0.11 & ... & -3.07$\pm$0.11\\
J003151.5+003233 & 6.02$\pm$0.23 & 6.41$\pm$0.13 & -3.69$\pm$0.23 & -3.30$\pm$0.13\\
\noalign{\smallskip}\hline
\end{tabular}
\end{center}
(This table is available in its entirety in machine-readable and Virtual Observatory (VO) forms in the online journal. A portion is shown here for guidance regarding its form and content.)
\end{table}

\section{SUMMARY}
\label{conclusion.sec}

We are carrying out systematic studies of stellar magnetic activities using a uniformly processed X-ray data set.
By using the $Chandra$ and $Gaia$ DR2 data, we first presented a catalog of X-ray emitted stars and studied the X-ray activities of different type stars.

We used a machine learning method to select and exclude QSOs and galaxies from the initial sample, and divided the stellar sample into YSOs, dwarfs, and giants in different spectral types.
We calculated the unabsorbed X-ray flux from count rate  (taken from the {\it Chandra} point source catalog), using PIMMS with an APEC model.
X-ray flares were detected with the Bayesian block analysis, and the entire observations with flares were removed.
An exposure-weighted averaged flux was then calculated for each star, and the X-ray luminosity ($L_X$) was estimated adopting the $Gaia$ DR2 distance.
Finally, we calculated the X-ray-to-bolometric luminosity ($R_X$) as the X-ray activity index.

We studied the X-ray activities for different stellar types, by using a well-selected but incomplete sample.
The $L_X$ of late-type stars ranges from $10^{27}$ to $10^{32}$ ergs s$^{-1}$.
For each stellar type, giants are much brighter than dwarfs and YSOs, while YSOs have higher $R_X$ values than giants and dwarfs.
In addition, many giants have very high hardness ratio $HR$1, indicating a high coronal temperature.
The $L_X$ and $R_X$ are positively correlated with $HR$1.

This catalog can be used to explore some interesting scientific topics.
The activity--rotation relation provides fundamental information on stellar dynamos and angular momentum evolution. With a selected sample, we found that the YSOs, dwarfs, and giants follow
a single sequence in the relation $R_X$ versus Ro, while the giants do not follow the relation $R_X$ versus $P_{\rm rot}^{-2}R^{-4}$ for dwarfs.
More stars with period estimations are needed to review these relations.
In the future, the TESS mission will cover most of the $Chandra$ fields.
Combining with the rotational periods revealed by TESS,
this catalog can help to understand how the magnetic fields and potential dynamo depend on their rotation.

Stars with unexpected high activities are worthy of follow-up detailed studies.
A-type stars were expected to have weak large-scale magnetic fields due to their shallow convective envelopes. However, many studies have observed activity of normal A stars though chromospheric emission lines \citep{Simon1991, Simon1997, Simon2002} and photospheric stellar spots and flares \citep{Balona2012, Balona2013, Balona2017}.
Our work shows that many A stars have clear X-ray activity, with log$R_X$ ranging from $\approx$ $-$6 to $\approx$ $-$3.5.
Ordinary late-type giants are thought to harbor mainly weak surface magnetic fields, and thus weak stellar activity, due to their large radii and slow rotation.
In our sample, more than 700 giants show X-ray emission, and some have high X-ray activity (log$R_X \approx -$3) and high $HR$ values.
This raise the question that whether
the solar-like dynamo operates in these stars.
Some studies proposed that the X-ray emission of A-type stars and giants are from an unresolved low-mass companion.
We are carrying out a campaign of high-resolution spectral observations of some candidates, in order to confirm whether there is a cool companion.

\begin{acknowledgements}
We sincerely thank the anonymous referee for the very helpful constructive comments and suggestions, which have significantly improved this article.
We thank Dr. Yang H. Q. and Cui K. M. for developing the {\it Kepler Data Integration Platform}.
This work has made use of data obtained from the {\it Chandra} Data Archive, and software provided by the {\it Chandra} X-ray Center (CXC) in the application packages CIAO.
This work presents results from the European Space Agency (ESA) space mission $Gaia$. $Gaia$ data are being processed by the $Gaia$ Data Processing and Analysis Consortium (DPAC). Funding for the DPAC is provided by national institutions, in particular the institutions participating in the Gaia MultiLateral Agreement (MLA).
This publication makes use of data products from the Pan-STARRS1 Surveys (PS1) and the PS1 public science archive,
the Two Micron All Sky Survey, and the Wide-field Infrared Survey Explorer.
We acknowledge use of the SIMBAD database and the VizieR catalogue access tool, operated at CDS, Strasbourg, France, and of Astropy, a community-developed core Python package for Astronomy (Astropy Collaboration, 2013). This work was supported by the National Key Research and Development Program of China (NKRDPC) under grant numbers 2019YFA0405000, 2019YFA0405504, and 2016YFA0400804, and the B-type Strategic Priority Program of the Chinese Academy of Sciences under grant number XDB41000000.
It was also supported by the National Science Foundation of China (NSFC) under grant numbers 11988101 and 11425313. \end{acknowledgements}

\section*{Appendix A \\
Soft excess of a few sources}

There are about 450 detections with $HR$1 values lower than $-$0.5. About 60 detections have vignetting-corrected net counts lower than 10; about 300 detections have counts between 20 and 50; about 90 detections have counts ranging from 50 to 650. 

\begin{figure*}[!]
\center
\includegraphics[width=0.49\textwidth]{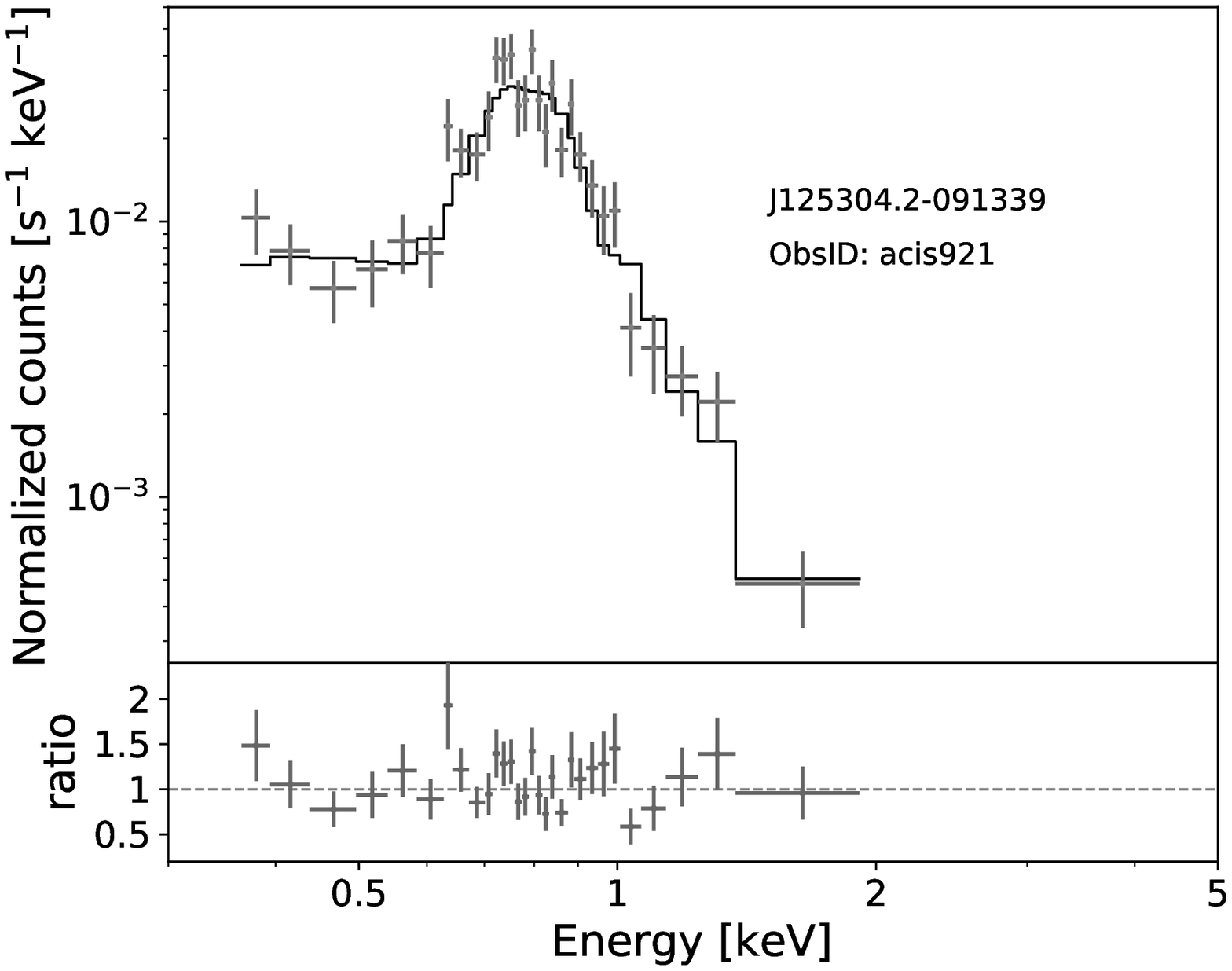}
\includegraphics[width=0.49\textwidth]{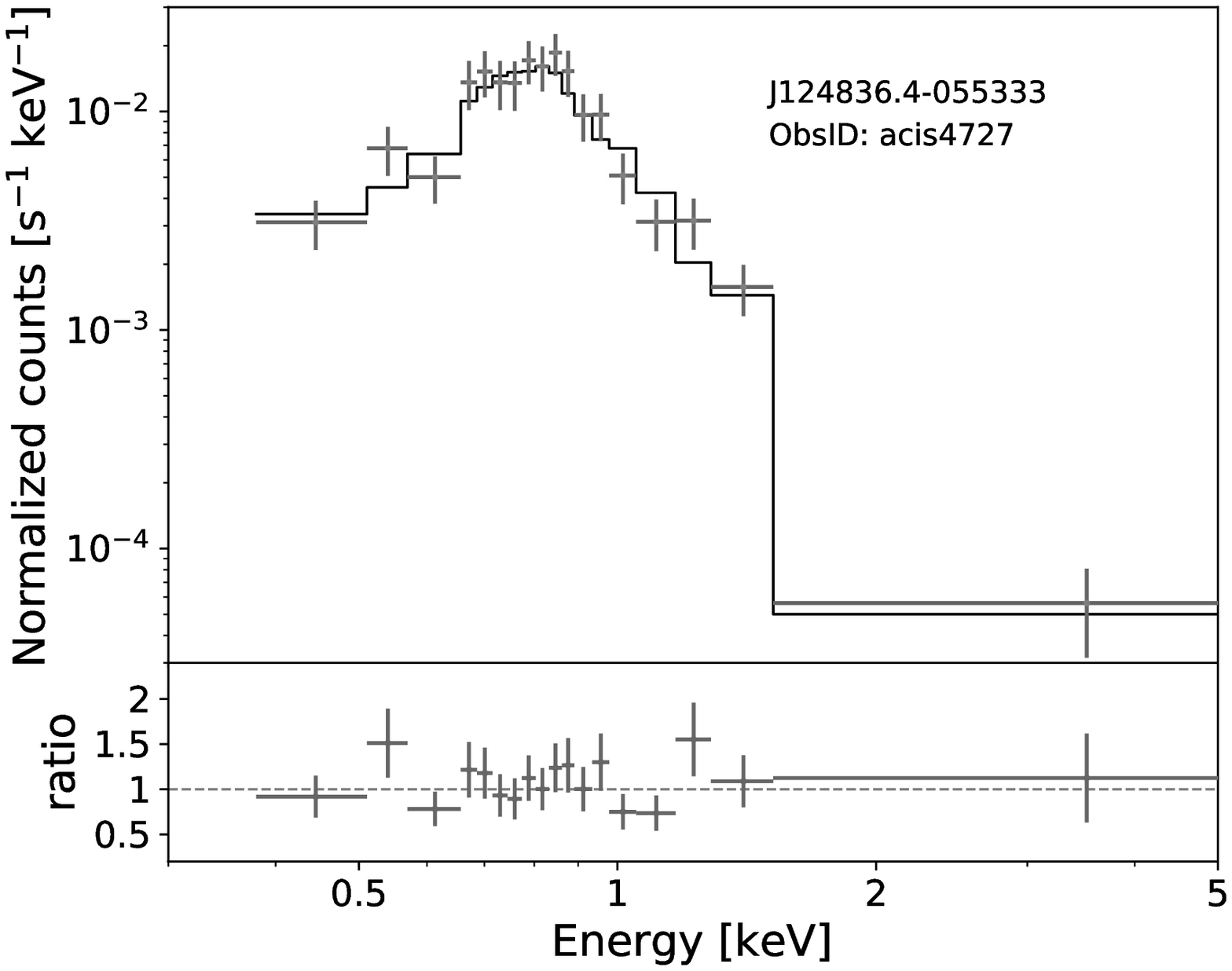}
\caption[]{Left panel: best-fit model (Blackbody plus APEC; $kT_{bb} < kT_{apec}$) for J125304.2-091339. Data points and model values are plotted in the top subpanel; data/model  ratios in the  bottom subpanel.  Right panel: best-fit model (Blackbody plus APEC; $kT_{bb} < kT_{apec}$) for J124836.4-055333.}
\label{soft.fig}
\end{figure*}

The low $HR$ values are mostly caused by statistical fluctuation of the photons.
We also found some sources, with the most counts, show a soft excess in their spectra.
This low-temperature thermal component can also help reduce $HR$1 values.
Figure \ref{soft.fig} shows that the spectra of two sources (J125304.2-091339 and J124836.4-055333) can be best fitted with a two-component model (Blackbody plus APEC).

\section*{Appendix B \\
YSO classification}

\begin{figure*}[!]
\center
\includegraphics[width=0.49\textwidth]{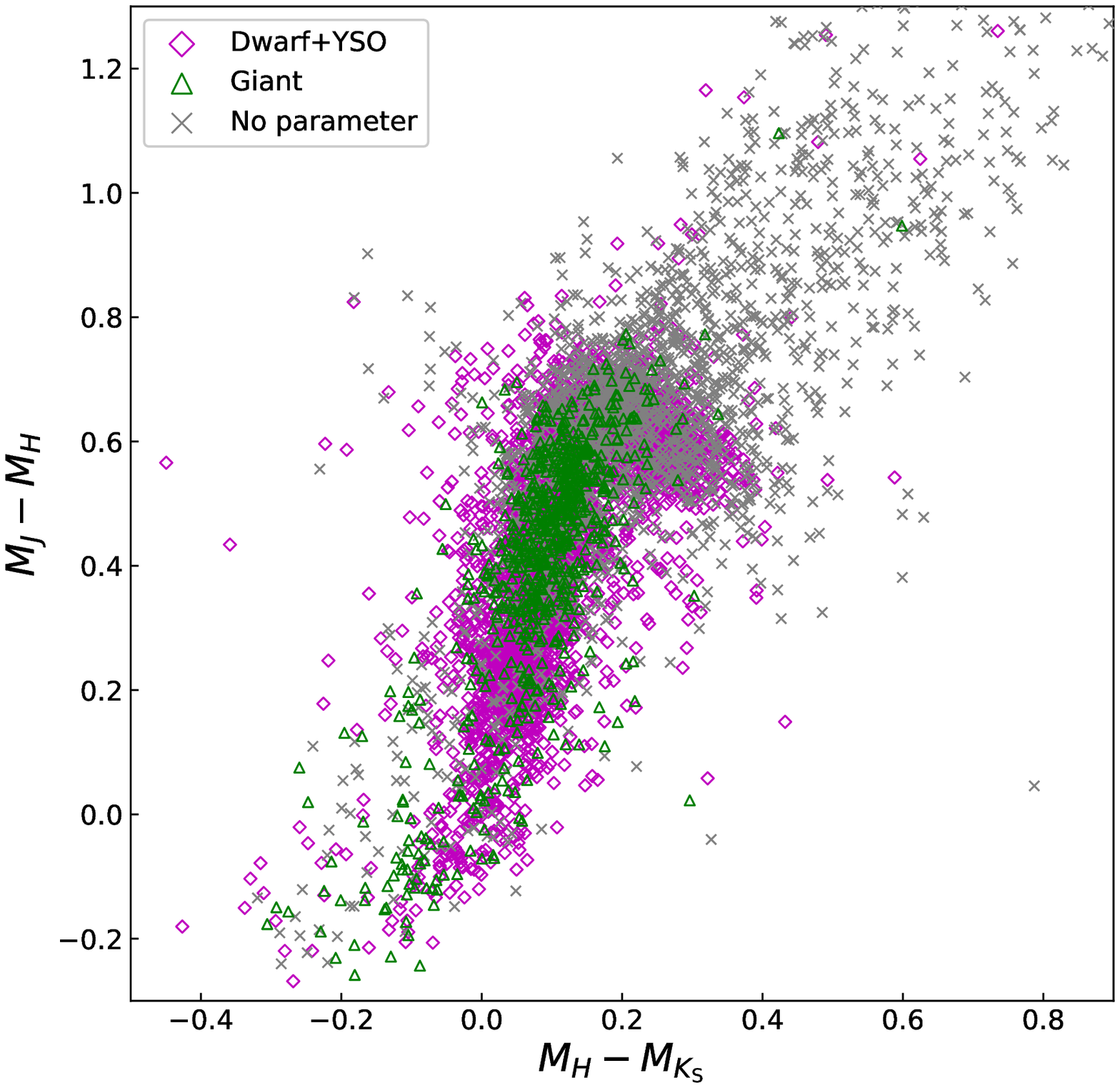}
\includegraphics[width=0.49\textwidth]{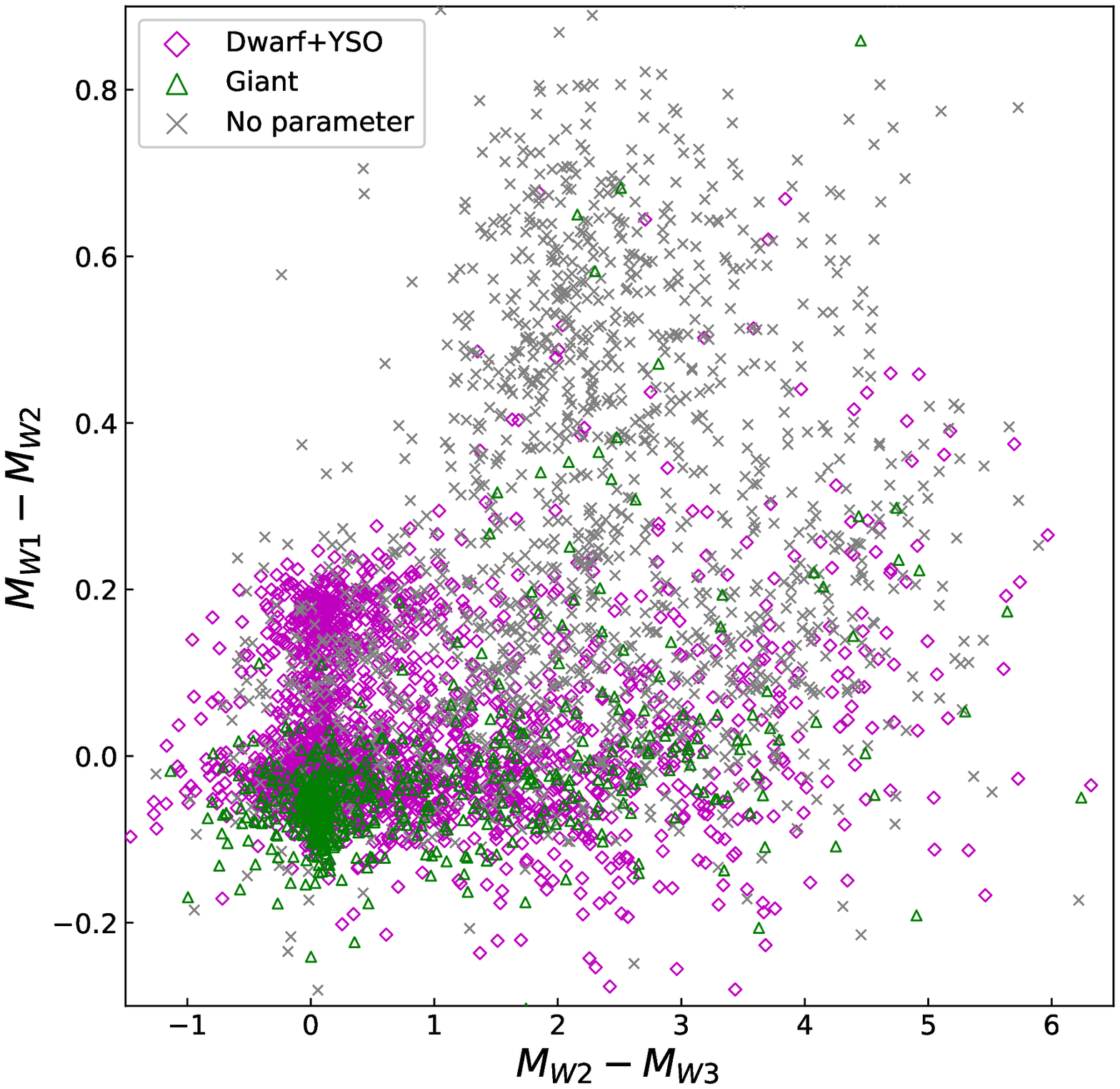}
\caption[]{Extinction corrected color-color diagram. The stellar classification is from LAMOST DR7 catalog and \citet{Anders2019}, both of which do not separate the main-sequence dwarfs and YSOs. Other objects marked with crosses are from the ``{\it non-parameter}" sample.}
\label{color3u.fig}
\end{figure*}

By cross-matching our sample with LAMOST DR7 catalog and \citet{Anders2019},
685 stars in our sample are classified as giants and 3312 ones are dwarfs and YSOs (see Section \ref{classify.sec}).
For the remaining sources without classification (``{\it non-parameter"} sample), many stars with infra-red (IR) excess can be YSO candidates (Figure \ref{color3u.fig}).

\begin{figure*}[!]
\center
\includegraphics[width=0.49\textwidth]{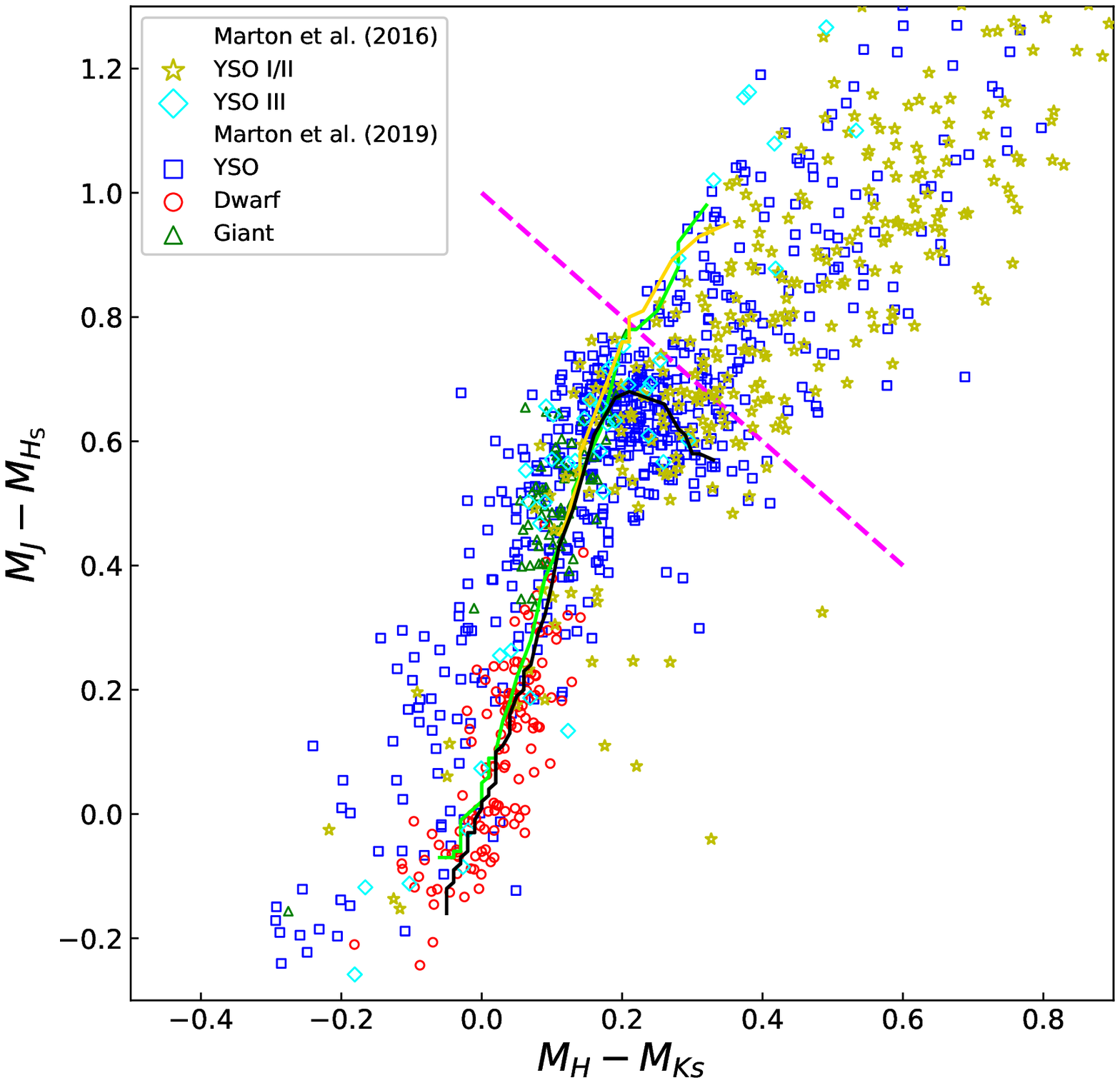}
\includegraphics[width=0.49\textwidth]{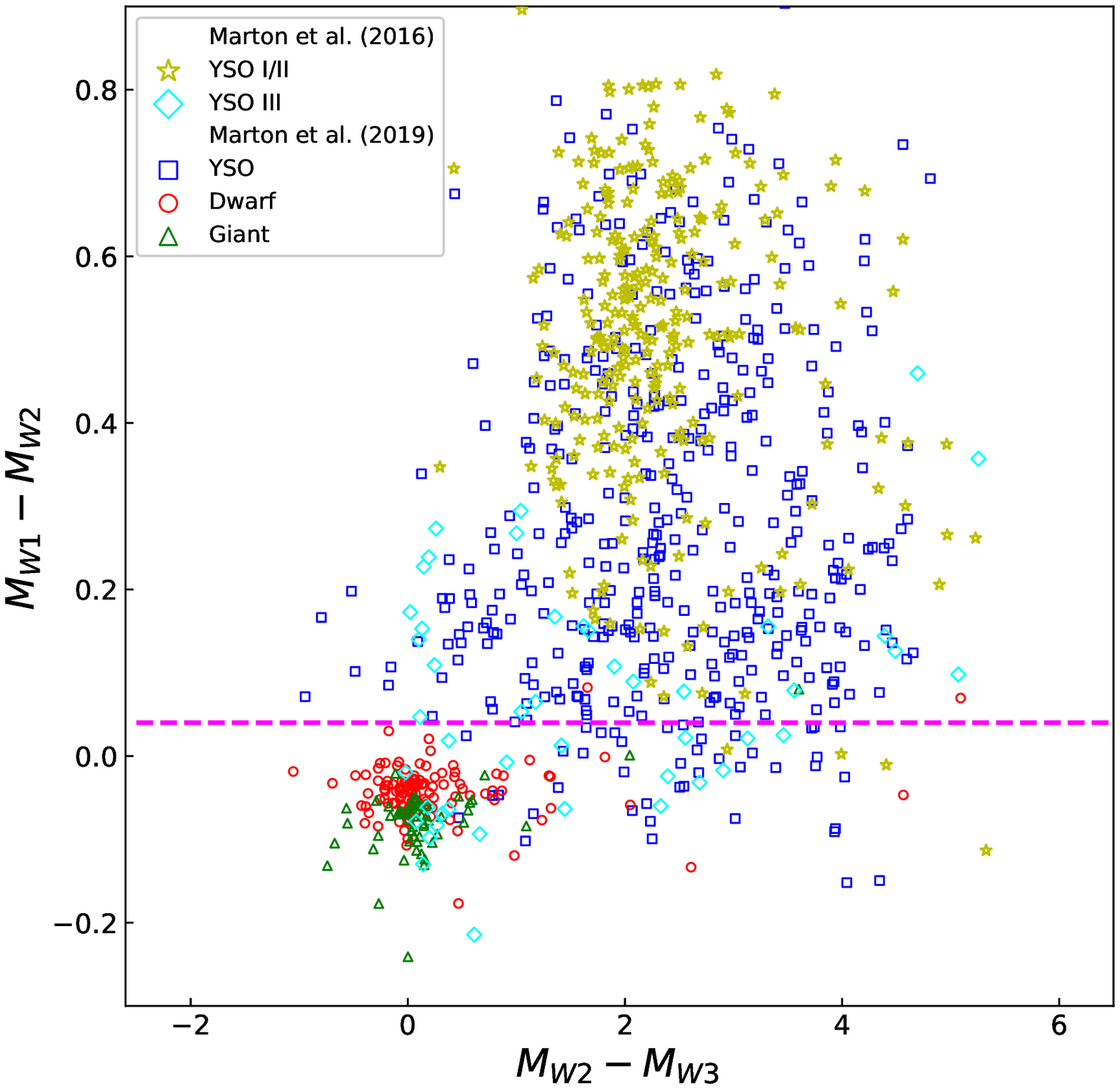}
\caption[]{Extinction corrected color-color diagram. The stellar classification is from the catalogs in \citet{Marton2016, Marton2019}.
The black, yellow, green lines show the colors of main-sequence stars, giants, and supergiants, respectively \citep{Koornneef1983}. The magenta dashed lines indicate $J-H = 1 - (H-K_{\rm S})$ (left panel) and $W1-W2 = $ 0.04 (right panel).
}
\label{color3p.fig}
\end{figure*}

We cross-matched our sample with the catalogs in \citet{Marton2016, Marton2019}.
There are 1100 objects in common between our sample and these catalogs, including 68 giants, 900 YSOs, and 132 dwarfs. 
Using these sources, we found that there is no dwarf or giant with color $J-H > 1 - (H-K_{\rm S})$ (left panel in Figure \ref{color3p.fig}),
while there are only two dwarfs and one giant showing color $W1-W2 > $ 0.04 (right panel in Figure \ref{color3p.fig}).
Thus, we will use these two colors as one criterion of classifying YSOs.
In addition, as \citet{Marton2019} reported, 99\% of the known YSOs are located in the regions where the dust opacity value is higher than 1.3$\times$10$^{-5}$. 
Therefore we picked out YSO candidates if one star meets the requirements: (1) $\tau >$ 1.3$\times$10$^{-5}$ and (2)
$J-H > 1 - (H-K_{\rm S})$ or $W1-W2 > $ 0.04.

However, many AGB stars have color $W2-W3 < $1 \citep{Koenig2014}. 
For the ``{\it non-parameter}" sample, in which the giants can not be recognized, we add one criterion $W2-W3 \geq $1 to select new YSOs.

\begin{figure*}[!]
\center
\includegraphics[width=0.49\textwidth]{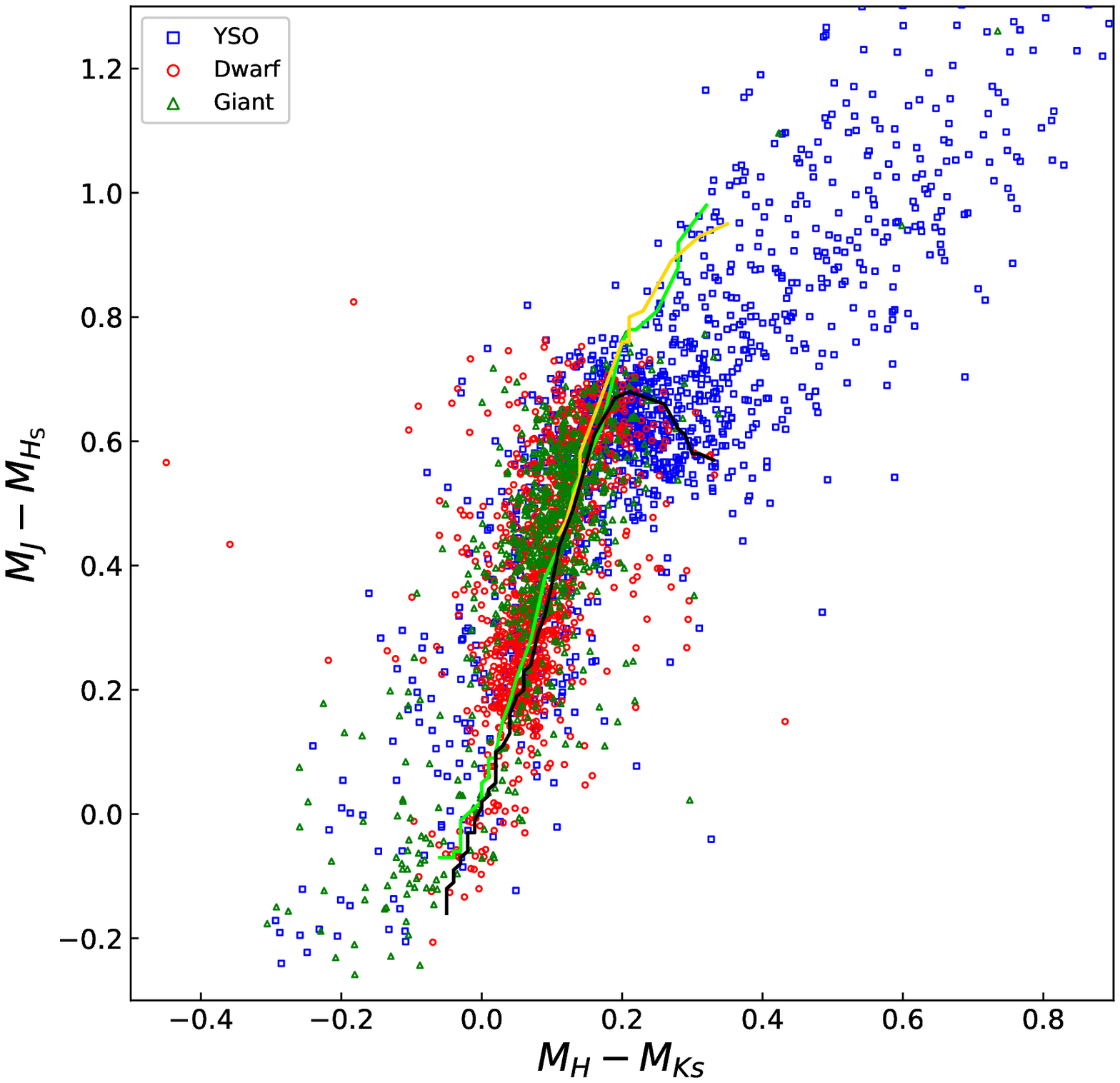}
\includegraphics[width=0.49\textwidth]{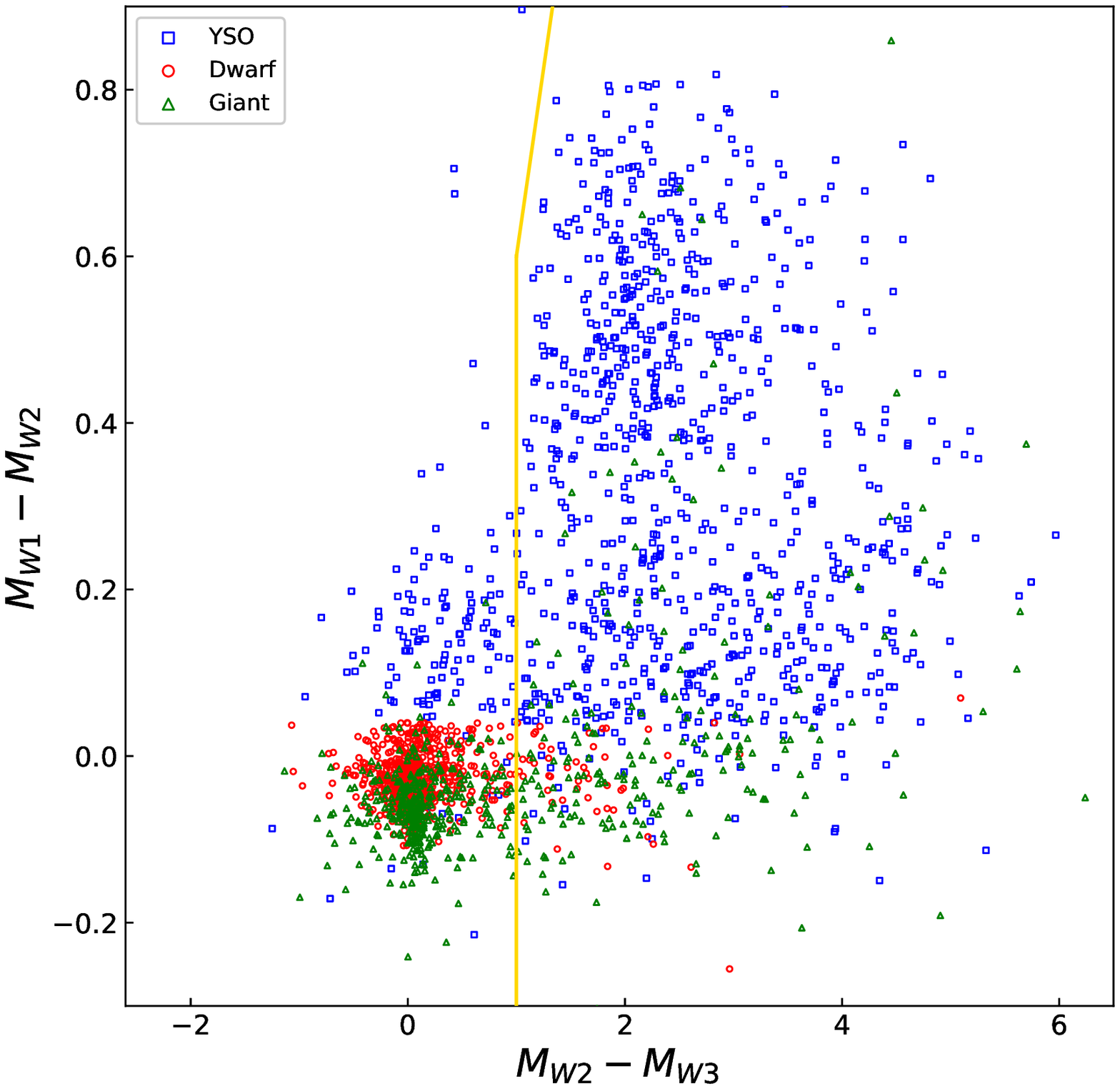}
\caption[]{Extinction corrected color-color diagram. Final stellar classification of 3005 sources, including 1196 YSOs, 1112 dwarfs, and 697 giants. The black, yellow, green lines show the colors of main-sequence stars, giants, and supergiants, respectively \citep{Koornneef1983}.}
\label{color3new.fig}
\end{figure*}

Finally, with the constraints of colors and dust opacity,
we classified additional $\approx$300 stars to be YSOs.
We totally determined stellar classes for 3005 sources, including 1196 YSOs, 1112 dwarfs, and 697 giants (Figure \ref{color3new.fig}).

\section*{Appendix C \\
Bolometric luminosity}

We used the PARSEC theoretical models to determine the bolometric luminosity.
The PARSEC isochrones\footnote{http://stev.oapd.inaf.it/cgi-bin/cmd\_3.1} were downloaded in eight metallicities ($Z =$ 0.0001, 0.0005, 0.001, 0.0021, 0.0043, 0.0085, 0.017, 0.034), with stellar ages log($t/{\rm yr}$) ranging from 6.6 to 10.13 at steps of $\Delta$log($t/{\rm yr}$) $=$ 0.02.
We divided our sample into four subsamples:
(1) 2872 sources with both parameter estimations and stellar classifications;
(2) 1125 sources with parameter estimations but no classification;
(3) 1549 sources with classifications but no parameter estimation;
(4) 360 sources with neither parameter estimation nor classification.

For subsamples (1) and (2), we first selected the models with closest metallicity, and then extracted the best model by comparing the observed and theoretical $T_{\rm eff}$ and log$g$ values. Because the PARSEC models divide stars into different evolution stages (e.g., pre-main sequence, main sequence, subgiant, red giant), the models in the same stellar stage were used for the fitting for subsample (1).

For subsamples (3) and (4), we constructed their SEDs ($g,$ $r$, $i$, $J$, $H$, and $K_{\rm }$ magnitudes) and compared them with the PARSEC models.
We used a $\chi^2$ minimization test to determine which PARSEC models are most compatible with the observed SEDs, following
\begin{equation}
\chi^2=\sum_{i=1}^{n}{\frac{[m_{i}^{\rm abs}-m_{i}^{\rm mod}(Z,t)]^2}{\sigma_{i}^{2}}},
\end{equation}
where $m_{i}^{\rm mod}(Z,t)$ is the magnitude in the $i{\rm th}$ filter of a model at metallicity $Z$ and age $t$, $m_{i}^{\rm abs}$ represents the absolute observed magnitude in the same filter, $\sigma_{i}$ is the observational uncertainty for the $i{\rm th}$ filter, and $n$ is the number of the filters used for fitting.
For subsample (3), we only used the models in the same stellar stage.

\end{document}